\documentclass[aps,twocolumn,amsmath,amssymb,bbm,nofootinbib,preprintnumbers,showpacs]{revtex4}
\usepackage{epsfig}
\usepackage{mathrsfs,graphicx,bm,amsmath}

\newcommand{\Tr}{\mathrm{Tr}}

\newcommand{\sigex}{\tilde{\sigma}}

\newcommand{\rex}{\tilde{r}}
\newcommand{\hpn}{\tilde{h}_\phi}
\newcommand{\Tn}{\tilde{T}}
\newcommand{\qn}{\tilde{q}}

\newcommand{\nun}{\tilde{\nu}}
\newcommand{\hF}{^{(F)}}
\newcommand{\hB}{^{(B)}}
\newcommand{\lit}{^6\mathrm{Li}}
\newcommand{\kal}{^{40}\mathrm{K}}
\newcommand{\epm}{\epsilon_M}

\newcommand{\hpb}{\bar{h}_\phi}
\newcommand{\lpb}{\bar{\lambda}_\phi}

\newcommand{\rhob}{\bar{\rho}}
\newcommand{\mpb}{\bar{m}_\phi^2}
\newcommand{\mpbF}{\bar{m}_\phi^{(F)\,2}}

\newcommand{\nB}{\bar{n}_B}
\newcommand{\nF}{\bar{n}_F}
\newcommand{\nC}{\bar{n}_C}
\newcommand{\nM}{\bar{n}_M}
\newcommand{\OmB}{\bar{\Omega}_B}
\newcommand{\OmF}{\bar{\Omega}_F}
\newcommand{\OmC}{\bar{\Omega}_C}
\newcommand{\OmM}{\bar{\Omega}_M}

\newcommand{\rhooR}{\rho_0}
\newcommand{\rhoR}{\rho}

\newcommand{\Apb}{\bar{A}_\phi}
\newcommand{\Apn}{\tilde{A}_\phi}
\newcommand{\ApR}{A_\phi}
\newcommand{\ZpR}{Z_\phi}

\newcommand{\hpR}{h_\phi}
\newcommand{\lpR}{\lambda_\phi}

\newcommand{\mpR}{m_\phi^2}

\newcommand{\nRF}{n_F}
\newcommand{\nRC}{n_C}
\newcommand{\nRM}{n_M}

\newcommand{\OmRF}{\Omega_F}
\newcommand{\OmRC}{\Omega_C}
\newcommand{\OmRM}{\Omega_M}

\newcommand{\MatPpb}{\bar{\mathcal{P}}_\phi}

\begin{document}

\title{Functional Integral for Ultracold Fermionic Atoms}

\author{S. Diehl}
\email{S.Diehl@thphys.uni-heidelberg.de}
\author{C. Wetterich}
\email{C.Wetterich@thphys.uni-heidelberg.de}

\address{
Institut f{\"u}r Theoretische Physik,
Philosophenweg 16, 69120 Heidelberg, Germany}


\begin{abstract}
We develop a functional integral formalism for ultracold gases of fermionic atoms. It describes the BEC - BCS crossover
and involves both atom and molecule fields. Beyond mean field theory we include the fluctuations of the molecule field by
the solution of gap equations. In the BEC limit, we find that the low temperature behavior is described by a Bogoliubov
theory for bosons. For a narrow Feshbach resonance these bosons can be associated with microscopic molecules. In contrast,
for a broad resonance the interaction between the atoms is approximately pointlike and microscopic molecules are irrelevant.
The bosons represent now correlated atom pairs or composite ``dressed molecules''. The low temperature results agree with
quantum Monte Carlo simulations. Our formalism can treat with general inhomogeneous situations in a trap. For not too
strong inhomogeneities the detailed properties of the trap are not needed for the computation of the fluctuation effects -
they enter only in the solutions of the field equations.
\end{abstract}

\pacs{03.75.Ss; 05.30.Fk } 

\maketitle



\section{Introduction}
\label{sec:intro}

Ultracold fermionic atoms can exhibit both the phenomena of a Bose-Einstein condensate (BEC) \cite{Einstein24,Einstein25}
of molecules and the condensation of correlated atom pairs similar to BCS-superconductivity \cite{ACooper56,BBCS57}. Recent
experimental progress \cite{Jin04,Ketterle04,ZGrimm04,Partridge05} in the crossover region reveals the universality of the
condensation phenomenon, as anticipated theoretically \cite{ALeggett80,BNozieres85,CMelo93,DStoof96,ECombescot99,FPethick00}.

In this paper we develop a systematic functional integral formulation for the treatment of the equilibrium state
\footnote{We do not touch in this work on the highly interesting non-equilibrium physics of the atom gas.} of ultracold
fermionic atoms. We discuss in detail how to arrive at a formulation that treats the fermionic fluctuations of unbound atoms
and the bosonic fluctuations of the molecule or di-atom field on equal footing. An approach based on a Hubbard-Stratonovich
transformation is an ideal starting point for a unified inclusion of fluctuations of molecules or ``Cooper pairs''.
We show how this formalism can be implemented in practice in a self-consistent approximation scheme.
We carefully discuss the renormalization procedure that is needed to absorb the ultraviolet divergences in this
nonrelativistic quantum field theory. The result is an effective low energy formulation which is insensitive to the
microphysical cutoff scale $\Lambda$.

We concentrate on dimensionless quantities by measuring all quantities in units of the Fermi momentum $k_F$ or the Fermi
energy $\epsilon_F = k_F^2/2M$. The inverse Fermi momentum is the most important length scale in the problem, related to the
total number density of atoms by $k_F = (3\pi^2 n)^{1/3}$. It measures the typical
interparticle spacing. We show that the atom density does not appear as an independent parameter in the computations which
can be performed in terms of dimensionless ratios. This renders our formalism highly universal, since the results of
experiments with different atoms and densities can be related by simple scaling laws.

For this purpose, we introduce three dimensionless parameters that characterize the crossover problem efficiently:
First, the ``concentration'' $c$ describes the ratio between the in-medium scattering length and the average distance between
two unbound atoms or molecules. Its inverse, $c^{-1}$, smoothly connects the weakly coupling BCS regime ($c<0, |c| \ll 1$)
with the BEC regime ($c>0, |c| \ll 1$). The perhaps most interesting region is the ``crossover regime'' in between,
$|c^{-1}|\lesssim 1$. Second, a Yukawa or Feshbach coupling $\hpn$ characterizes the interaction between atoms and
molecules. The third parameter is the temperature in units of the Fermi energy,
$\Tn = T/\epsilon_F$. No further details of the ``microphysics'' are needed for the macroscopic quantities. In this sense
the description becomes universal. In the language of quantum field theory or critical phenomena the parameters $c$ and
$\hpn$ describe relevant couplings for the long distance physics.

Our formalism is well suited to describe all regimes of coupling and temperature, including the superfluid phase with
broken symmetry.  As a first application we compute the phase diagram for the crossover problem as described by the
dependence of the critical temperature on $c$. The function $\Tn(c^{-1})$ only depends on the value of $\hpn$ and shows
universal limits for large $|c^{-1}|$, i.e. in the BCS or BEC regime, respectively. In the crossover region the dependence
on the Yukawa coupling $\hpn$ is strongest. Nevertheless, we find that for the two limits $\hpn \to 0$ (narrow resonance)
and $\hpn\to \infty$ (broad resonance) the crossover becomes independent of $\hpn$. For broad Feshbach resonances the
Yukawa coupling becomes an ``irrelevant'' parameter.

In the narrow resonance limit for $\hpn\to 0$ the bosonic degrees of freedom can be associated with microscopic molecules.
In this limit the molecule fluctuations can be neglected
and mean field theory becomes valid. In contrast, for $\hpn \to \infty$ the microscopic molecule
degrees of freedom play no role and the model is equivalent to a purely fermionic model with pointlike
interactions. Nevertheless, effective molecular bound states (``dressed molecules'') become a crucial feature of the
crossover physics.

For an actual comparison with experimental results one needs to relate the ``universal parameters'' to experimental
observables, in particular
the strength of the magnetic field. This is done in \cite{Diehl:2005an}, where we connect the
scattering properties of atoms with the values of our universal parameters. The present paper may therefore be viewed as
the theoretical basis for the more detailed comparison with experiment in \cite{Diehl:2005an}. Here we concentrate
on the conceptual and formal developments.

A further aspect of universality concerns the geometry of the trap. We will present a systematic ``derivative expansion''
which computes the effects of fluctuations and interactions independently of the shape of the trap. The details of the trap
only enter at the end through the solution of effective field equations in presence of a trap potential.
The systematic character of the derivative
expansion allows for a quantitative estimate of the reliability of simple approximations, like the ``local density'' resp.
``Thomas-Fermi approximation'' or the ``local condensate approximation'' frequently used in literature
\cite{DDKokkelmans,FFGriffin,BBStrinati,CCStrinati,ZStrinati}. The effective field equation for the condensate
has the same status as the (time independent) Gross-Pitaevskii equation \cite{Pitaevskii61,Gross61,Gross63} for a
Bose-Einstein condensate.

Our approach basically relies on two ingredients: The functional integral and the Hubbard-Stratonovich transformation or
partial bosonization. The partial bosonization permits us to formulate the problem microscopically as a Yukawa theory,
thereby allowing to deal with nonlocal interactions. This route  is also taken in
\cite{BBKokkelmans,CCKokkelmans,DDKokkelmans,EEGriffin,FFGriffin,AATimmermans,GGChen,HHChenReview,WWStoofBos,XXStoofBos,YYStoof,ZZStoof}.
The power of the functional integral techniques, however, is so far only marginally used. A first attempt in this direction
was made by Randeria \cite{CMelo93,CMelo97}. Later, other approaches employed this concept more as an argumentative
tool than as a method for concrete calculations \cite{DDStrinati,CCKokkelmans,WWStoofBos,XXStoofBos} \footnote{The present
paper covers part of a longer first version of \cite{Diehl:2005an}. A publication using functional integral computations has
appeared more recently \cite{ZZVivas05}.} .

Beyond the systematic functional integral formulation and the emphasis on the universal aspects of the phase
transition our work extends previous results by the systematic inclusion of the molecule fluctuations. These fluctuations
are important for the quantitative understanding of the phase transition for a broad Feshbach resonance with a large
dimensionless Yukawa coupling $\hpn$, as relevant for the present experiments in $\lit$ and $\kal$. For zero temperature
our calculations agree well with Quantum Monte Carlo simulations \cite{Carlson03} at the resonance.

This paper is organized as follows:

In sect. \ref{sec:metastabledilutegas}
we investigate the Feshbach resonance and introduce the important molecule degrees of freedom in terms of a di-atom or molecule
field $\hat{\phi} (x)$. This allows us to cover the whole range of temperature and the crossover. The Bose-Einstein
condensate or the superfluid order parameter corresponds to a nonvanishing expectation value $\langle \hat{\phi}\rangle$.
We establish the equivalence of our formulation with a purely fermionic formulation for which the effective interaction
between the atoms contains a nonlocal piece reflecting the molecule exchange. Only in the broad resonance limit $\hpn\to
\infty$ this interaction becomes pointlike.

Our functional integral formulation in terms of an independent field $\hat{\phi}(x)$ is particularly well adapted to the
crossover from a BEC to a BCS condensate: Molecules and Cooper pairs are described by the same field. Of course, the
dynamical properties depend strongly on the BEC or BCS regime. In particular, we compute in sect. \ref{sec:EffActMFT} the
gradient coefficient $\Apb$ which determines the gradient contribution to the free energy for a spatially varying molecule
field. For the BEC regime, $\Apb$ is dominated by the ``classical value'', corresponding to the dominance
of the tightly bound molecules. In contrast, for the BCS regime the fluctuation effects dominate $\Apb$. In this case
$\hat{\phi}(x)$ can be associated to a collective degree of freedom (Cooper pair). One could omit the classical contribution to
$\Apb$ such that $\hat{\phi}$ becomes on the microscopic level an auxiliary field. Indeed the presence of a molecular bound
state is no longer crucial in the BCS regime. One may work in a purely fermionic setting with a local
interaction which depends on a magnetic field $B$. For a broad Feshbach resonance this feature holds for the entire
crossover region.

We discuss the derivative expansion of the effective action in sect. \ref{sec:EffActMFT}. In particular,
sect. \ref{sec:renormalization} addresses the issue of additive renormalization of the detuning and sect.
\ref{sec:WFR} computes the wave function renormalization $Z_\phi$ which distinguishes the ``renormalized field''
for the dressed molecules and the field for the microscopic of ``bare'' molecules. We
turn in sect. \ref{sec:Relevant} to the discussion of the relevant parameters that describe the
universal aspects of ultracold atoms. In terms of the dimensionless concentration $c$ and Yukawa coupling $\hpn$ the
system becomes independent of the detailed short distance properties.

Sect. \ref{sec:MolFrac} discusses the fraction of atoms bound in molecules. It is crucial to distinguish between the ``bare'' or
microscopic molecules and the dressed molecules \cite{Stoof05,IIChen05}. Their numbers are related by the
multiplicative wave function renormalization $\ZpR$. In order to achieve a complete symmetry between the
fermionic fluctuations of unbound atoms and the bosonic fluctuations of molecules we adapt our functional integral
setting in sect. \ref{EffAtDens}. In sect. \ref{sec:renconstZR} we turn to the BEC limit. For a
broad Feshbach resonance (large $\hpn$) one finds very large $\ZpR$, such that the condensate fraction (condensed dressed
molecules) exceeds by far the number of microscopic molecules. The latter becomes completely negligible for $\hpn\to \infty$.
Nevertheless, we find a Bogoliubov theory for bosons in the low temperature BEC regime for all values of $\hpn$. Finally,
we include
the molecule fluctuations (or collective fluctuations of di-atom states) in sect. \ref{sec:beyond} in the form of new
bosonic gap equations. We draw conclusions in sect. \ref{sec:conclusions}.

While the main part of this paper deals with a homogeneous situation our formalism can be extended to cover the
inhomogeneous situation in a trap of atoms if the inhomogeneity is not too large. Since the main part is independent of
the discussion of inhomogeneities we display the formalism for inhomogeneous situations in appendix \ref{sec:partial}.

We introduce a general formalism for a functional integral which applies to arbitrary
fermionic systems and is easily generalized to systems with bosons, far beyond the particular case of a Feshbach
resonance (where di-atom states play a role). In addition to the (Grassmann) field variables $\hat\psi (x)$ for the fermionic
atoms we employ a bosonic field variable $\sigma (x)$. It corresponds to a varying effective chemical potential which
is associated to the density field $n(x)$. This procedure allows computations for the inhomogeneous setting of atoms in a
trap beyond the small density approximation or beyond the Thomas-Fermi approximation. The bosonic field variable
$\sigma$ is introduced by partial bosonization. We formulate the effective action $\Gamma [\sigma]$ and establish the
exact formal relations between $\sigma (x)$, $n(x)$, the chemical potential $\mu$ and the local trap potential $V_l(x)$.

\section{dilute gas of ultracold atoms}
\label{sec:metastabledilutegas}

The ultracold gas of fermionic atoms in the vicinity of a Feshbach resonance can be treated in the idealization of
two stable atomic states denoted by a two component spinor $\hat\psi$. (For the example of $^6\mathrm{Li}$ these states
may be associated with the two lowest hyperfine states $|1\rangle$, $|2\rangle$.) The molecular state responsible for
the Feshbach resonance can be treated as a bosonic particle. In our idealization it is stable for negative binding
energy and can decay into a pair of fermionic atoms for positive binding energy.

For a realistic description our formalism has to be capable to describe the following
phenomena: (i) Condensates of atom pairs may form at low temperature, similar to the BCS description of superconductivity.
(ii) Molecules of two atoms can be exchanged between the single atoms, thus contributing to the interaction. Also these
molecules may form a Bose-Einstein condensate at low temperature.
In our formalism both effects find a \emph{unified} description as will be discussed in detail in this paper.

We work with a microscopic action which explicitly includes a bosonic field $\hat\phi$ with atom number two
\cite{CCKokkelmans},
\begin{eqnarray}\label{PhiAction}
 S_B&=& \int dx\Big\{\hat\psi^\dagger(\partial_\tau - \frac{1}{2M}\triangle -\sigma) \hat\psi\nonumber\\
&& \qquad+\hat{\phi}^*(\partial_\tau - \frac{\triangle}{4M} + \bar{\nu}_\Lambda-2\mu) \hat{\phi} \nonumber\\
          && \qquad - \frac{1}{2}\hpb \Big(\hat{\phi}^*\hat\psi^T \epsilon\hat\psi -
          \hat{\phi}\hat\psi^\dagger\epsilon\hat\psi^*\Big)
 \Big\}.
\end{eqnarray}
Here we employ the Matsubara formalism where the Euclidean time $\tau$ is wrapped on a torus with circumference
$\beta = 1/T$ with conventions
\begin{eqnarray}
x = (\tau,\textbf{x}), \quad \int dx  = \int_0^{\beta} d\tau \int d^3x. 
\end{eqnarray}
(Our units are $\hbar = c = k_B = 1$.) We will shortly see how this microscopic model relates to a purely fermionic
setting by the means of a partial bosonization or Hubbard-Stratonovich transformation.

The complex two-component spinors $\hat\psi(x)$ are anticommuting
Grassmann variables. We assume an equal mixture of the two atomic states. In this case the chemical potential
associated to the difference in the number of atoms in the ``up'' and ``down'' states precisely cancels the
energy difference between the two states such that both can be omitted. (For unequal mixtures the action contains an additional
term $\propto \hat\psi^\dagger \tau_3 \hat\psi$.)
The bosonic molecules are described by a complex bosonic field $\hat{\phi}$. The
propagator for these ``bare molecules'' is obtained from simple symmetry considerations, i.e. we assume a mass $2M$,
leading to a nonrelativistic kinetic energy $q^2/4M$. The quadratic term $\sim\hat{\phi}^* \hat{\phi}$ involves the
``bare'' binding energy or detuning ($\bar{\nu}_\Lambda$) which typically depends on the magnetic field. In order to make
contact to physical observables, $\bar\nu_\Lambda$ has to be additively renormalized, which is implemented in sect.
\ref{sec:renormalization}.

We use here two different chemical potentials $\sigma$ and $\mu$
for the fermionic atoms and bare molecules. This is useful if we want to obtain separately the densities of fermionic atoms
or bare molecules by differentiation of the free energy with respect to $\sigma$ or $\mu$. Since only the total number
of atoms is conserved, one has to set $\sigma =\mu$ at the end of the computations. The distinction between $\sigma$ and
$\mu$ is appropriate if one wants to understand explicitly the role of the microscopic (or bare) molecules. In sect.
\ref{EffAtDens}
we will drop this distinction in favor of a more unified approach. There we will set $\mu =\sigma$ from the outset such
that $\sigma$ will be conjugate to the total density of atoms irrespective of a microscopic distinction between unbound
atoms and molecules. In the main part of this paper we will treat $\sigma$ and $\mu$ as constant classical source terms.
However, we stress that this
source term can be straightforwardly promoted to a fluctuating field, $\sigma \to \hat\sigma(x)$. This issue, and its
use for the description of inhomogeneities beyond the usual Local Density Approximation, will be discussed in the
appendix \ref{sec:partial}.

The Yukawa or Feshbach coupling $\hpb$ describes the coupling between the single atoms and molecules, $\epsilon_{\alpha\beta}=
-\epsilon_{\beta\alpha}$, $\epsilon_{12}=1$. For $\hpb\to 0$ the molecular states decouple. However, in an
appropriately performed ``narrow resonance limit'' $\hpb\to 0$ which keeps the scattering length fixed, an exact solution
of the many-body problem becomes feasible above the critical temperature. A detailed analysis of
this limit is given in \cite{Diehl:2005an}. Broad Feshbach resonances correspond to large $\bar h_\phi$,  and we will
see that the limit $\bar h_\phi\to \infty$ describes a purely fermionic theory with pointlike interactions where
microscopic molecules can be neglected.

The thermodynamic equilibrium situation is described by the partition function. The basic ingredient for our formalism
is the representation of this object in terms of a functional integral with weight factor $e^{-S_B}$, with $S_B$ the
Euclidean action (\ref{PhiAction})
\begin{eqnarray}\label{1}
Z[\eta,j]&=&\int {\cal D}\hat\psi{\cal D}\hat\phi \exp \Big\{-S_B[\hat\psi,\hat\phi]\\\nonumber
&&+\int\hspace{-0.12cm} dx\,\, \eta(x)\hat\psi^\dagger(x) + \eta^\dagger\hat\psi(x) \\\nonumber
&& + j(x) \hat\phi^*(x) + j^*(x) \hat\phi(x)\Big\}.
\end{eqnarray}
This formulates the full quantum theory in terms of the sources $\eta$ and $j$ for the fermion and the boson fields.
All one particle irreducible $n$-point functions, including the order parameter and the correlation functions,
can be directly extracted from the effective action $\Gamma$, which obtains by a Legendre transform,
\begin{eqnarray}\label{GammaLeg}
\Gamma [\psi, \bar\phi] &=& - \ln Z + \int\hspace{-0.12cm} dx\,\, \eta(x)\psi^\dagger(x) + \eta^\dagger(x)\psi(x)\\\nonumber
&& \qquad \qquad + j(x) \bar\phi^*(x) + j^*(x) \bar\phi(x).
\end{eqnarray}
The effective action is a functional of the ``classical'' fields (or field expectation values)
$\psi = \langle \hat\psi\rangle , \bar\phi =\langle \hat\phi\rangle$ in the presence of sources. They are defined as
\begin{eqnarray}
\bar\phi (x) = \langle \hat\phi\rangle(x) = \frac{\delta \ln Z}{\delta j^*(x)}
\end{eqnarray}
and analogous for the fermion fields. Of course, due to Pauli's principle, the fermion field cannot acquire a nonvanishing
expectation value for $\eta = \eta^\dagger =0$ such that the physical value is simply $\psi =0$. It is often convenient
to write $\Gamma$ as an implicit functional integral over fluctuations $\delta \phi, \delta\psi$ around ``background
fields'' $\bar\phi, \psi$
\begin{eqnarray}\label{GammaFuncInt}
\Gamma[\psi ,\bar\phi] =-\ln \int \mathcal{D}\delta\psi\mathcal D \delta\phi \exp\big( -
S  [\psi + \delta \psi, \bar\phi + \delta \phi ]+\nonumber\\
 \int \big( j^*\delta\phi+  \eta^\dagger\delta\psi + \mathrm{h.c.}\big),
\end{eqnarray}
with $j^*(x) = \delta \Gamma /\delta \bar\phi(x)$. This form is particularly useful for the
construction of the equation of state, i.e. the explicit expression for the total atom number density.

The total number density of atoms $n$ includes those from unbound or ``open channel'' atoms and the ones arising from the
bare molecules or ``closed channel'' atoms. It obeys
\begin{eqnarray}\label{TotDens}
n (x)=  \nF(x) + \nB(x) = \langle\hat\psi^\dagger(x) \hat\psi(x)\rangle   + 2\langle \hat{\phi}^*(x)\hat{\phi}(x)\rangle.
\end{eqnarray}
Indeed, the action is invariant under $U(1)$ phase transformations of the fermions and bosons,
\begin{eqnarray}
\hat\psi \to e^{\mathrm{i} \theta }\hat \psi, \quad \hat\phi \to e^{2\mathrm{i} \theta} \hat \phi
\end{eqnarray}
and the corresponding Noether charge is the total atom number $N= \int d^3 x n(x)$. We emphasize that eq. (\ref{TotDens})
is no ``ad hoc'' assumption - it is an exact expression for the particle number and directly follows from the microscopic
formulation. More technically speaking, $\langle\hat\psi^\dagger \hat\psi\rangle$ and $2\langle \hat{\phi}^*
\hat{\phi}\rangle$ represent the full two-point correlation functions of the ``bare fields'' which appear in the
microscopic action (\ref{PhiAction}) and are quantized by means of the functional integral. In a homogeneous situation,
the conserved particle number can be replaced by a fixed constant particle density, $n =N/V$.

The bosonic part $\nB$ counts the total number of atoms contained in the microscopic or ``bare'' molecules. This number receives a
contribution from free molecules and from the condensate as discussed in more detail in sect. \ref{sec:MolFrac}.
In the language often used for a Feshbach resonance, $\nB$ measures the ``closed channel'' microscopic atoms.
Using the formalism of the present paper we have computed $\bar n_B$ as a function of the magnetic field $B$ in
\cite{Diehl:2005an}. We find very good agreement with observation \cite{Partridge05} over several orders of
magnitude in $\bar n_B$.

Since the action (\ref{PhiAction}) contains
only terms quadratic or linear in $\hat{\phi}$ it is straightforward to express the expectation value $\bar{\phi}_0$ in terms of
a local fermion-bilinear. It obeys (for vanishing source for $\hat{\phi}$)
\begin{eqnarray}
\big( \bar{\nu}_\Lambda- 2\mu -\frac{\triangle}{4M} + \partial_\tau \big)\bar{\phi}_0 =
\frac{\hpb}{2}\langle \hat\psi^T\epsilon\hat\psi\rangle.
\end{eqnarray}
In particular, for constant $\bar{\phi}_0$ we find
\begin{eqnarray}
\bar{\phi}_0 = \frac{\hpb}{2\bar{\nu}_\Lambda - 4\mu}\langle \hat\psi^T\epsilon\hat\psi\rangle.
\end{eqnarray}
This demonstrates directly that our formalism makes no difference between a ``condensate of molecules'' $\bar{\phi}$ and a
``condensate of atom pairs'' $\langle\hat\psi^T\epsilon\hat\psi\rangle$ - they are simply related by a multiplicative
constant.

We finally show the equivalence of our formalism with a model containing only fermionic atoms and no microscopic
molecules. The interaction in this fermionic description is, in general, not local. It becomes local, however, in the
limit of a broad Feshbach resonance for $\bar h_\phi\to \infty$. For this purpose we use again
the quadratic form of the bosonic part of the microscopic action (\ref{PhiAction}), which allows us to integrate out
the $\hat{\phi}$ field. Expressed only in terms of fermions our model contains now a momentum dependent four-fermion
interaction ($Q_4=Q_1+Q_2-Q_3$; $Q = (\omega_n, \vec q)$ with discrete bosonic Matsubara frequencies $\omega_n = 2\pi n T$
at finite temperature)
\begin{eqnarray}\label{Mom4Fermion}
S_{int} &=& - \frac{1}{2}\int\limits_{Q_1,Q_2,Q_3}\big(\hat\psi^\dagger(-Q_1)\hat\psi(Q_2)\big)\big(\hat\psi^\dagger(Q_4)
\hat\psi(-Q_3)\big)\nonumber\\
&&\Big\{\frac{\hpb^2}{ \bar{\nu}_\Lambda - 2\mu + 2\pi\mathrm{i}(n_1-n_4)T+(\vec{q}_1-\vec{q}_4)^2/4M } \Big\}.\nonumber\\
\end{eqnarray}
We emphasize that there is no difference between the Yukawa type model described by the action (\ref{PhiAction}) and a
purely fermionic model with interaction (\ref{Mom4Fermion}). All physical observables can be computed in either one or the
other of the two formulations. However, eq. (\ref{Mom4Fermion}) reveals that our model describes a setting beyond
pointlike interactions via the classical frequency and momentum dependence of the four-fermion interaction. The momentum
structure of (\ref{Mom4Fermion}) is compatible with interactions in the $\hat\psi\hat\psi$ -- channel. The action
(\ref{PhiAction}) hence models a nonlocal coupling between the fermionic constituents. Reversing the logic, eq.
(\ref{PhiAction}) could also be obtained by starting from eq. (\ref{Mom4Fermion}) and performing a Hubbard-Stratonovich
transform or partial bosonization \cite{Hubbard59,Stratonovich}. Finally, we note that we could also choose classical
``gradient coefficients'' $\bar A_\phi^{(cl)}$ (see below) different from $1/(4M)$ in order to model an experimentally
determined effective range.

In the pointlike limit the momentum dependence and the dependence on $\mu$ can be neglected and the coupling term in eq. (\ref{Mom4Fermion})
is replaced by the ``local interaction approximation''
\begin{eqnarray}\label{BosonCond2}
\bar{\lambda}_\Lambda=-\frac{\hpb^2}{\bar{\nu}_\Lambda}.
\end{eqnarray}
This limit obtains formally for $\bar h_\phi^2 \to \infty$, $\bar\nu_\Lambda \to \infty$, while keeping $\bar \lambda$
fixed. It is relevant for broad Feshbach resonances, as discussed in detail in \cite{Diehl:2005an}.

\section{Derivative Expansion for the Effective Action}
\label{sec:EffActMFT}
The condensation of atoms pairs or molecules is signalled by a non-vanishing expectation value $\langle\hat{\phi}\rangle
=\bar{\phi}_0$. The associated symmetry breaking of the global continuous symmetry of phase rotations of $\hat\psi$ and $\hat{\phi}$
(related to the conservation of the number of atoms) induces a massless Goldstone boson. This is the origin of
superfluidity. In this section, we show how to describe this phenomenon in the effective action formalism. For this
conceptual issue, it is sufficient to work in the mean field approximation or a simple extension thereof (extended MFT).
The more sophisticated approximation schemes beyond mean field are presented in sects. \ref{EffAtDens}, \ref{sec:beyond}.

In this work we will treat the effective action in a derivative expansion, i.e. we write
\begin{eqnarray}\label{GammaPosSpace}
\Gamma[\bar \phi] = \int dx\Big\{ U(\bar\phi) + Z_\phi \bar{\phi}^*\partial_\tau\bar{\phi} +
\Apb\vec{\nabla}\bar{\phi}^*\vec{\nabla}\bar{\phi}  + ...\Big\},
\end{eqnarray}
and compute the ``wave function renormalization'' $Z_\phi$, the ``gradient coefficient'' $\bar A_\phi$ and the effective
potential $U(\bar\phi)$. In the present paper we do not compute the corrections to the part of the effective action
involving fermions -- for this part we have simply taken the classical action. (See \cite{Diehl:2005an} for the
renormalization of the Yukawa coupling in presence of a ``background'' four-fermion interaction.) We therefore omit the
fermionic part of the effective action from now on. For the concrete calculation we work in momentum space. We
emphasize, however, that the above expression can be used for the investigation of weak inhomogeneities as encountered
in atom traps. The fluctuation problem, i.e. the computation of $Z_\phi, A_\phi, U$ in the above truncation, can then
be solved in momentum space, while the effects of weak inhomogeneities can be investigated by solving the classical
field equations derived from (\ref{GammaPosSpace}). This reaches substantially beyond the
usual local density approximation, which ignores the kinetic terms in eq. (\ref{GammaPosSpace}). We discuss the
implementation of an external trapping potential in our functional integral formalism in app. \ref{sec:partial}.
Here, however, we focus on the homogeneous situation.

In its most general form, the effective action depends on the parameters $T, \sigma$ and $\mu$ and on the classical field
$\bar\phi$. The dependence on the chemical potentials $\sigma$ and $\mu$ can already be inferred from the
partition function -- they are only spectators w.r.t. the Legendre transform. Following the thermodynamic
construction, the total particle density in a homogeneous setting is obtained as
\begin{eqnarray}\label{TheDensEq}
n = - \frac{\partial U}{\partial \sigma}\Big|_{\mu} \,\, - \,\, \frac{\partial U}{\partial \mu}\Big|_{\sigma}.
\end{eqnarray}
This prescription precisely reproduces eq. (\ref{TotDens}). Here $U$ has to be taken at its minimum.

In the absence of sources the field equation for $\bar\phi (x)$ reads
\begin{eqnarray}
\frac{\delta \Gamma }{\delta \bar\phi(x)} \stackrel{!}{=} 0.
\end{eqnarray}
For a homogeneous situation the stable solution corresponds to a minimum of the effective potential
\begin{eqnarray}
\frac{\partial U }{\partial \bar\phi} &=& U' \cdot \bar\phi^* = 0,
\end{eqnarray}
with
\begin{eqnarray}
U' &=& \frac{\partial U }{\partial \bar\rho}, \quad \bar\rho = \bar\phi^*\bar\phi.
\end{eqnarray}
Here we have introduced the $U(1)$ invariant $\bar\rho = \bar\phi^*\bar\phi$ --
the effective potential only depends on this combination. This simple equation
can be used to classify the thermodynamic phases of the system,
\begin{eqnarray}\label{CharPhases}
\mathrm{Symmetric\,\,phase (SYM):} && \bar\rho_0 =0, \quad \bar U'(0) > 0,\nonumber\\\nonumber
\mathrm{Symmetry\,\, broken\,\, phase (SSB):} &&\bar\rho_0  > 0, \quad U'(\bar\rho_0) = 0,\\\nonumber
\mathrm{Phase\,\, transition (PT):} && \bar\rho_0 = 0,\quad U'(0) = 0\\
\end{eqnarray}
where $\bar\rho_0 = \bar\phi_0^*\bar\phi_0$ denotes the minimum of $U(\rho)$. For high temperatures, the minimum of the
effective potential occurs at $\bar \phi =0$ and we deal with a normal gas phase. For low enough $T$, on the other hand,
we expect the minimum of $U(\sigma, \bar{\phi})$ to occur for $\bar{\phi}_0\neq 0$. The spontaneous breaking of the
$U(1)$ symmetry is signalled by a nonzero field expectation value and indicates the condensation phenomenon.
This has an important aspect of universality: one and the same criterion can be used for the whole parameter space,
both for the BCS-type condensation of Cooper pairs and for the BEC of microscopic molecules.

In the remainder of this section, we will evaluate the effective action in the mean field approximation (MFT). This scheme
is defined by only considering the effects generated by fermion fluctuations. We will include bosonic
fluctuations in later chapters. Beyond MFT, we first include the contribution to the density from dressed molecules.
This is connected to the effective bosonic
two-point function (connected part of the bosonic particle density). This effect is
included in different current approaches to the crossover problem in the limit of broad Feshbach resonances
\cite{CCStrinati,HHChenReview,Stoof05}. We will refer to it as extended Mean Field Theory. Furthermore, we include
in sect. \ref{sec:beyond} the modifications of the effective potential due to bosonic fluctuations, using suitable
Schwinger-Dyson equations.

In a realistic physical situation the validity of our model is restricted to momenta smaller than some microphysical
``ultraviolet cutoff'' $\Lambda$. In turn, $\Lambda$ is typically given by the range of the van der Waals interactions.
One may use $\Lambda \approx a_B/100$ with $a_B$ the Bohr radius.
For practical computations it is often convenient to consider the limit $\Lambda \to \infty$ such that no explicit
information about the ``cutoff physics'' is needed. This requires to express the couplings of the theory in terms of
suitably defined ``renormalized couplings'' that stay finite for $\Lambda\to \infty$. In the next subsection
we will discuss the additive renormalization of the detuning. The functions $Z_\phi, \bar A_\phi$,
in contrast, are UV finite. We will also subtract field-independent pieces linear in $\sigma$ and $\mu$ that obtain
in a naive MFT computation. This additive ``density renormalization'' will be traced back to the relation between the
functional integral and the operator formalism.

\subsection{Effective potential and additive renormalization}
\label{sec:renormalization}

In the mean field approximation, the effective potential reads, after carrying out the Matsubara summation and
omitting an irrelevant infinite constant,
\begin{eqnarray}\label{USigmaPhi}
U_\Lambda(\sigma,\bar{\phi}) &=& (\bar{\nu}_\Lambda -2\mu) \bar{\phi}^*\bar{\phi} + \Delta U_1\hF,\\\nonumber
\Delta U_1\hF &=&  - 2T\int_\Lambda\frac{d^3q}{(2\pi)^3}\ln\cosh\gamma_\phi
\end{eqnarray}
where
\begin{eqnarray}\label{DefGammaPhi}
\gamma_\phi&=&\frac{1}{2T}\Big(\Big(\frac{q^2}{2M} -\sigma\Big)^2 + r \Big)^{1/2} =
\big(\gamma^2 + \beta^2\big)^{1/2},\\\nonumber
\gamma     &=& \frac{\frac{q^2}{2M} -\sigma}{2T},\quad \beta = \frac{r^{1/2}}{2T},
\quad r = \hpb^2\bar{\phi}^*\bar{\phi}.
\end{eqnarray}
Here we have added an index $\Lambda$ in order to remind that this form still depends on an ultraviolet cutoff $\Lambda$.
We regularize the effective potential by limiting the integration over spacelike momenta by an upper bound $\Lambda$,
$q^2<\Lambda^2$.

Let us now discuss the additive renormalization needed to properly describe the physics encoded in this object. It
is needed in two instances: The first one concerns a zero-point shift of the two-point function and is
related to the quantization via the functional integral. Its removal does not involve a physical scale and can thus be
seen as a normalization of a certain observable, the particle number. The second one is related to a true ultraviolet
divergence which needs to be cured by an appropriate counterterm.
For the ultraviolet renormalization we can restrict to the simpler situation where there is no spontaneous symmetry
breaking, since this effect occurs only in the low energy sector and cannot affect the ultraviolet physics. For the
physical value of the field expectation value, this implies $\bar\phi =0$ and $\gamma_\phi= \gamma$.

The fermionic part of the particle number is naively obtained from the thermodynamic relation $\bar{n}_\Lambda = -
\partial U_\Lambda/\partial\sigma$ in a homogeneous setting.
This yields the explicit expression
\begin{eqnarray}\label{BarNFMFT}
\bar{n}_{F,\Lambda} = -\int_\Lambda\frac{d^3q}{(2\pi)^3}\tanh \gamma
\end{eqnarray}
and we observe that this number may get negative for large negative $\sigma/T$.
In order to clarify the precise relation between $\bar{n}_\Lambda$
and the particle density, we first consider the simpler situation of a single fermionic degree of freedom.
In this case the expectation value $\langle\hat\psi^\dagger\hat\psi\rangle$ can be related to the expectation values of products
of the usual annihilation and creation operators $a,a^\dagger$, which obey the anticommutation relation
$a^\dagger a+aa^\dagger=1$,
\begin{eqnarray}\label{24}
\langle\hat\psi^\dagger\hat\psi\rangle&=& \frac{1}{2}\langle \hat\psi^\dagger\hat\psi-\hat\psi\hat\psi^\dagger\rangle=
\frac{1}{2}\langle a^\dagger a-aa^\dagger\rangle\nonumber\\
&=&\langle a^\dagger a\rangle  -1/2=n-1/2.
\end{eqnarray}
Here the second equality holds since this combination of operators is covariant with respect to permutations of the
ordering. For a lattice model (as, for example, the Hubbard model) with $f$ degrees of freedom per site the fermion number
per site therefore reads $n=\bar{n}_\Lambda+\frac{f}{2}$. For electrons in a solid ($f=2$) one can associate
$\bar{n}_\Lambda$ with the difference of the electron density from half filling (where $n=1$), i.e. the average number of
electrons minus holes per site as compared to the half-filling density. For relativistic charged fermions
$\hat\psi^\dagger\hat\psi$ measures the difference between particle and antiparticle density and the additive constant drops out.
For nonrelativistic atoms, however, the relation between the atom density $n$ and $\bar{n}_\Lambda$ becomes \footnote{
Note that the constant shift $\hat{n}$ diverges if the ultraviolet cutoff for the momentum integration $(q^2<\Lambda^2)$
goes to infinity, $\hat{n}=\Lambda^3/(6\pi^2)$.}
\begin{eqnarray}\label{25}
&& \bar{n}_{F,\Lambda}(x)=\langle\hat\psi^\dagger(x)\hat\psi(x)\rangle \\\nonumber
&=&\frac{1}{2}\langle\int\limits_y\sum \limits_{i,j}
\left[\hat\psi_i^\dagger(x)\hat\psi_j(y)-\hat\psi_j(y)\hat\psi_i^\dagger(x)\right]\delta_{ij}\delta(x-y)\rangle\\\nonumber
&=&\frac{1}{2}\langle\int\limits_y\sum \limits_{i,j}\left[a_i^\dagger(x)a_j(y)-a_j(y)a_i^\dagger(x)
\right]\delta_{ij}\delta(x-y)\rangle\\\nonumber
&=&\frac{1}{2}\langle\int\limits_y\sum \limits_{i,j}\left[2a_i^\dagger(x)a_j(y)-\delta_{ij}\delta(x-y)
\right]\delta_{ij}\delta(x-y)\rangle\\\nonumber
&=&\langle a^\dagger(x)a(x)\rangle - \frac{f}{2}\delta(0)
= n_F(x) - \frac{f}{2}\int \frac{d^3q}{(2\pi)^3}\\\nonumber
&=& n_F(x)-\hat{n}.
\end{eqnarray}
The volume factor in momentum space, $\delta(0)$, diverges in the limit of infinite momentum cutoff. The physical
fermionic particle density $n_F(x)$ and the relative particle density $\bar{n}_{F,\Lambda}(x)= \langle\hat\psi^\dagger
(x)\hat\psi(x)\rangle$ are therefore related by an additive shift that depends on the momentum cutoff.

In consequence, one finds now a manifestly positive total fermionic particle number
\begin{equation}\label{26}
N_F=\int d^3x(\bar{n}_{F,\Lambda}+\hat{n})= V \int\frac{d^3q}{(2\pi)^3}
\big(\exp(2\gamma)+1\big)^{-1}.
\end{equation}
The momentum integral is now ultraviolet finite for $\Lambda\to \infty$. It becomes exponentially insensitive
to the ultraviolet cutoff $\Lambda$, such that we dropped the index. We recover the familiar Fermi distribution.
An analogous argument holds for the connected part of the  bosonic two-point function which will be implemented below.
Formally, this additive renormalization appears in the form of field independent terms linear in $\sigma$ and $\mu$
in the classical potential.

Let us now proceed to the second instance where UV renormalization is needed.
The microscopic action (\ref{PhiAction}) depends explicitly on two parameters $\bar{\nu}_\Lambda$, $\hpb$. A third
parameter is introduced implicitly by the ultraviolet cutoff $\Lambda$ for the momentum integration in the fluctuation
effects. (Besides this, the results will depend on the thermodynamic variables $T$ and $\sigma, \mu$.) Contact to
experiment is established by relating the microscopic parameters to observables of the concrete atomic system.
Here we choose these parameters to be the magnetic field dependent physical detuning $\bar\nu(B)$ and the Feshbach
coupling $\hpb$ \footnote{For vanishing background or open channel coupling, $\hpb$ is a free parameter. Else, a further
UV renormalization of the Feshbach coupling is necessary \cite{Diehl:2006}.}. The parameters $(\bar\nu, \hpb)$ can,
however, be replaced by an equivalent set $(a^{-1},\hpb)$ ($a$ the scattering length) in a second step as pointed out
below. Once the parameters $\bar{\nu}(B)$ and $\hpb$ are fixed by the properties of the molecules or
atom scattering in empty space we can proceed to compute the properties of the atom gas at nonzero temperature and density
without further free parameters.

In the vicinity of the Feshbach resonance at $B =B_0$ we may approximate $\bar{\nu}_\Lambda(B)$ by a linear behavior
(linear Zeeman effect)
\begin{eqnarray}
\frac{\partial\bar{\nu}_\Lambda}{\partial B} = \bar{\mu}_B.
\end{eqnarray}
Here $\bar{\mu}_B= \mu_+ + \mu_- - \mu_M$ reflects the difference between the sum of the magnetic moments of the two atomic
states ($\mu_+ + \mu_-$) and the molecule magnetic moment $\mu_M$. We relate the physical detuning $\bar{\nu}$ to
$\bar{\nu}_\Lambda$ by an additive, $B$-independent shift
\begin{eqnarray}\label{Watwees}
\bar{\nu} = \bar{\nu}_\Lambda - \frac{\hpb^2 M \Lambda}{2\pi^2}, \quad \frac{\partial\bar{\nu}_\Lambda}{\partial B}
= \frac{\partial\bar{\nu}}{\partial B}.
\end{eqnarray}
This is motivated by a consideration of the fermionic contribution to the ``boson mass term'',
\begin{eqnarray}
\bar m_\phi^2 = U'(\rho_0=0).
\end{eqnarray}
Indeed, the fluctuation contribution diverges for $\Lambda\to \infty$
\begin{eqnarray}\label{pedagogic}
\bar{m}_\phi^2 &=& \bar{\nu}_\Lambda - \frac{\hpb^2}{2}\hspace{-0.1cm} \int^\Lambda\hspace{-0.2cm} \frac{d^3q}{(2\pi)^3}\big(\frac{q^2}{2M} -
\sigma\big)^{-1} \tanh\frac{q^2/2M - \sigma}{2T}\nonumber\\\nonumber
 &=& \bar{\nu} - \frac{\hpb^2}{2}\hspace{-0.1cm} \int^\Lambda \hspace{-0.2cm} \frac{d^3q}{(2\pi)^3}\Big[\big(\frac{q^2}{2M}
 - \sigma\big)^{-1}
\tanh\frac{q^2/2M - \sigma}{2T} \\
&&\qquad\qquad\qquad- \frac{2M}{q^2}\Big] .
\end{eqnarray}
In the second equation the linear dependence of the fluctuation correction on the cutoff $\Lambda$ is absorbed
into the definition of the
physical detuning $\bar{\nu}$. The remaining integral in the second line of (\ref{pedagogic}) is very insensitive with
respect to the precise value of $\Lambda$, and we can formally send $\Lambda$ to infinity.

More generally, we choose $\bar \nu_\Lambda - \bar \nu$ such that the zero of $\bar \nu$ coincides with the
Feshbach resonance at $B_0$
\begin{eqnarray}
\bar{\nu} = \bar{\mu}_B(B - B_0).
\end{eqnarray}
In vacuum ($n=0, T=0$) the Feshbach resonance corresponds to a vanishing binding energy $\epsilon_M = 0$. This is realized
\cite{Diehl:2005an} for $\sigma=0$, $\bar{m}_\phi^2 = 0$. There is also no condensate in the vacuum, i.e. $\bar{\phi}_0=0$.
We note that there are no boson fluctuations contributing to the renormalization of the mass term in the physical vacuum
$n=T=0$ as can be seen from a diagrammatic argument \cite{Diehl:2006}.

The scattering length $a$ can be defined in terms of the scattering amplitude at zero momentum and zero energy \cite{Diehl:2005an}. At
the present stage we may consider a ``resonant scattering length'' $a_R$ related to $\bar{\nu}$ by
\begin{eqnarray}
\frac{M}{4\pi a_R} = -\frac{\bar{\nu}}{\hpb^2} = -\frac{\bar{\nu}_\Lambda}{\hpb^2} + \frac{M\Lambda}{2\pi^2}.
\end{eqnarray}
It accounts for the contribution of the molecule exchange to the atom scattering. 
For $B\neq B_0$ and low enough momenta and energies (low enough temperature) the molecule exchange can be described
by an effective pointlike interaction. We therefore also define the renormalized resonant four-fermion vertex
$\bar{\lambda}_R$,
\begin{equation}\label{CoupPointlike}
\bar{\lambda}_R = \frac{4\pi a_R}{M}.
\end{equation}
Comparison with eq. (\ref{BosonCond2}) yields
\begin{eqnarray}\label{UVRenorm}
\frac{1}{\bar{\lambda}_R}&=& -\frac{\bar{\nu}}{\hpb^2} =- \frac{\bar{\nu}_\Lambda}{\hpb^2} + \frac{M\Lambda}{2\pi^2}
= \frac{1}{\bar{\lambda}_\Lambda}
 + \frac{M\Lambda}{2\pi^2} .
\end{eqnarray}
This demonstrates that the renormalization of $\bar{\nu}$ (\ref{Watwees}) corresponds to the renormalization of the
effective atom interaction strength $\bar{\lambda}$.

It is instructive to consider in the mean field framework the explicit equation which determines the order
parameter $\bar \phi_0$ in the superfluid phase. A nonzero $\bar{\phi}_0$ obeys
\begin{eqnarray}\label{eq65}
\frac{\bar{\nu} - 2\mu}{\hpb^2}= \frac{1}{4T} \int\frac{d^3q}{(2\pi)^3}\big(\gamma_\phi^{-1}\tanh \gamma_\phi
- 4M T/q^2\big).
\end{eqnarray}
In an alternative purely fermionic formulation we may compute $\bar{\phi}_0$ in the local interaction approximation
(\ref{BosonCond2}) by solving the Schwinger-Dyson equation \cite{Dyson49,Schwinger51}. Eq. (\ref{eq65})
would correspond precisely to the BCS gap equation (lowest order Schwinger-Dyson equation) for the purely fermionic
formulation with local interaction \cite{AABiBa00,Jaeckel02}, provided we choose
\begin{eqnarray}\label{SD1}
 \frac{1}{\bar{\lambda}_\Lambda}=-\frac{\bar{\nu}_\Lambda - 2\mu}{\hpb^2} \quad \mathrm{or}\quad
 \frac{1}{\bar{\lambda}}=-\frac{\bar{\nu} - 2\mu}{\hpb^2} .
\end{eqnarray}
This suggests the definition of a $\mu$ or density dependent effective coupling $\bar\lambda (\mu)$
(cf. eq. (\ref{Mom4Fermion})) for the atom-molecule model. In the
many-body context and in thermodynamic equilibrium, where fermions and bosons share a common chemical potential
$\sigma=\mu$, the latter is determined by the density. In the physical vacuum, obtained by sending the density to
zero, $\mu$ describes the binding energy of a molecule, $\epm = 2\mu = 2\sigma$ and does not vanish on the BEC side of
the resonance \cite{Diehl:2005an}. We emphasize that the resonant scattering length $\bar \lambda_R$ in eq.
(\ref{CoupPointlike}) describes the scattering of fermions throughout the crossover and is directly related to the
observed scattering length for two atom scattering. On the other hand, $\bar\lambda (\mu)$ is a universal
combination characteristic for the ground state of the system \cite{Diehl:2006}. For broad Feshbach resonances
($\bar h_\phi\to \infty$) the two quantities coincide.

Expressing the effective potential $U$ (\ref{USigmaPhi}) in terms of $\bar{\nu}$ the momentum integral becomes ultraviolet
finite,
\begin{eqnarray}\label{finalActionEq}
U(\sigma,\bar{\phi}) &=& (\bar{\nu} - 2\mu)\bar{\phi}^*\bar{\phi} + U_1^{(F)}(\sigma,\bar{\phi}),  \nonumber\\
U_1^{(F)}(\sigma,\bar{\phi})&=& - 2T\int\frac{d^3q}{(2\pi)^3}\Big[\ln\big(\mathrm{e}^{\gamma_\phi - \gamma} +
               \mathrm{e}^{-\gamma_\phi-\gamma}\big)\nonumber\\
              &&\qquad\qquad\qquad - \frac{\hpb^2\bar{\phi}^*\bar{\phi} M}{2T q^2}\Big].
\end{eqnarray}
The remaining cutoff dependence is $\mathcal{O}(\Lambda^{-1})$, the precise value of $\Lambda$ therefore being unimportant.
For definiteness, the cutoff $\Lambda$ can be taken $\Lambda = (3\pi^2n_{gs})^{1/3}$ with $n_{gs}$ the density in the liquid
or solid ground state at $T=0$. This choice will be motivated in app. \ref{sec:Meta}.

At this point we have reached a simple but nevertheless remarkable result. When expressed in terms of measurable quantities,
namely scattering length $a_R$ (or $\bar{\lambda}_R$) and the open channel atom density $\nF$ the $\bar{\phi}$ dependence of the
effective potential becomes very insensitive to the microscopic physics, i.e. the value of the cutoff $\Lambda$. Without
much loss of accuracy we can take the limit $\Lambda \to \infty$ for the computation of $U$. The effective chemical
potential $\sigma$ depends on $\nF$ and $\bar{\phi}$ via $\partial U(\sigma,\bar{\phi})/\partial\sigma = -\nF$.

\subsection{Wave function renormalization}
\label{sec:WFR}
The wave function renormalization $Z_\phi$ is another important ingredient for the description of the crossover physics
in the Yukawa model. This can be seen from the fact that rescaling all couplings in the effective action with the
appropriate power of $Z_\phi$, we end up with an effective bosonic Bogoliubov theory in the BEC regime, as will be
discussed in detail in sect. \ref{sec:renconstZR}. Here we focus on the explicit computation of the wave function
renormalization.

As can be read off from eq. (\ref{GammaPosSpace}), the wave function renormalization $Z_\phi$ is related to the inverse
molecule propagator $\bar{\mathcal{P}}_\phi$. In app. \ref{app:WFR} we compute $\bar{\mathcal{P}}_\phi$ in the mean field
approximation,
\begin{eqnarray}\label{Zphi}
\bar{\mathcal{P}}_\phi(Q)&=& 2\pi \mathrm{i} n T + \frac{q^2}{4M}  \\\nonumber
   &&\hspace{-1.2cm}-\hpb^2\int\limits_{Q'} \frac{P_F(Q')-\sigma}{[P_F(Q')-\sigma][P_F(-Q')-\sigma] + \hpb^2\bar{\phi}^*\bar{\phi}}\\\nonumber
   &&\hspace{-1.2cm}\times\frac{P_F(-Q' + Q)-\sigma}{[P_F(Q' - Q)-\sigma][P_F(-Q' + Q)-\sigma] + \hpb^2\bar{\phi}^*\bar{\phi}}\\
   &=& \bar{\mathcal{P}}_\phi^*(-Q)\nonumber
\end{eqnarray}
where the kinetic part of the inverse fermion propagator reads
\begin{eqnarray}
P_F(Q) = \mathrm{i}\omega_B  + \frac{q^2}{2M}
\end{eqnarray}
with $Q = (\omega_B, \vec{q})$, and the frequency variable $\omega_B$ represents the discrete fermionic Matsubara
frequencies at finite temperature, $\omega_B = (2n + 1)\pi T$.

Here we interpret the wave function renormalization $\ZpR$ as a renormalization of the term in the effective action
with a timelike derivative. We may then evaluate the propagator correction (\ref{Zphi}) for analytically
continued frequencies $\omega_B \to \omega_B + \mathrm{i}\omega$ and set $\omega_B=0$. Now
$\Delta P_\phi (\omega, \vec q)$ becomes a continuous function of $\omega$. Defining
\begin{eqnarray}\label{WFRDefini}
\ZpR= 1 - \frac{\partial \Delta P_\phi}{\partial\omega}\Big|_{\omega=0}
\end{eqnarray}
one finds
\begin{eqnarray}\label{ZRFormula}
\ZpR \hspace{-0.1cm} &=& \hspace{-0.1cm} 1 + \frac{\hpn^2}{16\Tn^2}\int \frac{d^3\qn}{(2\pi)^3}
\gamma \gamma_\phi^{-3}\big[\tanh\gamma_\phi - \gamma_\phi\cosh^{-2}\gamma_\phi\big].\nonumber\\
\end{eqnarray}
Here we have rescaled the integration variable and the Yukawa coupling as $\tilde q = q/k_F$, $\hpn^2 = 4M^2
\bar h_\phi^2 /k_F$. In the symmetric phase the simplification $\gamma_\phi=\gamma$ applies. We note that
$\ZpR$ is closely related to the spectral function for the molecules.
If we only consider fermionic diagrams, we can give an equivalent definition of the wave function
renormalization using the mean field effective potential,
\begin{eqnarray}\label{ZphiinMFT}
Z_\phi^{\sigma} = 1 -\frac{1}{2} \frac{\partial^3 U}{\partial\sigma\partial\bar\phi^*\partial\bar\phi}.
\end{eqnarray}
The property $\ZpR^{\sigma} = \ZpR$ holds since the integral in eq. (\ref{Zphi}) depends on the combination
$\omega - 2 \sigma$ only.

\subsection{Gradient coefficient}
\label{GradCoeff}
In order to compute the gradient coefficient, we proceed in complete analogy to the wave function renormalization:
For the spacelike momenta, we define
\begin{eqnarray}\label{Zphi1}
\Apb(q) &=& \frac{\bar{\mathcal{P}}_\phi(0,q) - \bar{\mathcal{P}}_\phi(0,0)}{q^2},\\\label{Zphi2}
\Apb &=& \lim\limits_{q^2\to 0} \Apb(q).\nonumber
\end{eqnarray}
More explicit formulae are given in app. \ref{app:WFR} in eqs. (\ref{Aphi1},\ref{AphiSYM}).

\begin{figure}[t!]
\begin{minipage}{\linewidth}
\begin{center}
\setlength{\unitlength}{1mm}
\begin{picture}(85,55)
      \put (0,0){
     \makebox(80,49){
       \begin{picture}(80,49)
      \put(0,0){\epsfxsize80mm \epsffile{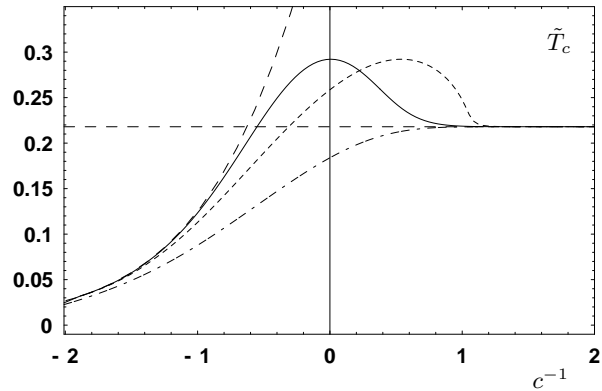}}
      \put(70,-2){$c^{-1}$}
      \put(72,43){$\Tn_c$ }
      \end{picture}
        }}
   \end{picture}
\end{center}
\vspace*{-1.25ex} \caption{Crossover phase diagram $\Tn_c (c^{-1})$. For large $\hpn$ we compare a calculation with
a gap equation modified by boson fluctuations (dashed) to the result obtained with the standard BCS gap equation (solid).
The universal narrow resonance limit $\hpn \to 0$ is indicated by the dashed-dotted line.
We further plot the result of the standard BCS approach (BCS gap equation, only fermionic contributions in the
density equation; long-dashed rising line) and the result for noninteracting bosons (long dashed horizontal line). The
solid line at resonance $\Tn_c(c^{-1}=0)=0.292$ coincides with the result obtained in \cite{Levin052} which omits boson
fluctuations. The dashed line ($\Tn_c(c^{-1}=0)=0.259$ is in good agreement with the value obtained in \cite{ZZZStrinati}
($\Tn_c = 0.266$), who work in the broad resonance limit from the outset.}
\label{CrossoverTcAll}
\end{minipage}
\end{figure}

\section{Relevant Parameters, Momentum and Energy Scales}
\label{sec:Relevant}
\subsection{Concentration}
For a mean field computation with nonzero density the effective renormalized four-fermion coupling is related by an equation
similar to eq. (\ref{BosonCond2}) to the resonant scattering length in vacuum.
We define \footnote{For $\sigma=\mu$ this combination appears in the term quadratic
in $\bar{\phi}$ in eq. (\ref{finalActionEq}).}
\begin{eqnarray}\label{CoupMu}
\frac{1}{\bar{\lambda}(\sigma)} = \frac{M}{4\pi a_R} + \frac{2\sigma}{\hpb^2}.
\end{eqnarray}
Therefore the \emph{effective} scattering length in an atom gas differs from the (vacuum) scattering length which is
measured by the scattering of individual atoms. This density effect is reflected by the dependence on the effective
chemical potential and we define
\begin{eqnarray}\label{ScatterDens}
\frac{1}{\bar{a}} = \frac{1}{a_R} + \frac{8\pi\sigma}{\hpb^2M}.
\end{eqnarray}
On the BCS side $\sigma$ is positive and $a_R$ is negative - the size of $|\bar{a}|$ is
therefore larger than $|a_R|$. A similar enhancement occurs in the BEC regime where $\sigma$ will turn out to be negative
and $a_R$ positive. Roughly speaking, on the BCS side the presence of a nonvanishing atom density favors the
formation of (virtual) molecules by reducing the ``cost'' of forming molecules with positive $\bar{\nu}$ (or positive binding
energy) to $\bar{\nu} - 2\sigma$. In the BEC regime the presence of molecules reduces the absolute size of the effective
binding energy. It should, however, be pointed out that the effect of a density dependence is small in case of broad
resonances $\hpn\to \infty$, as visible from eq. (\ref{CoupMu}).

The atom density $n$ defines a characteristic momentum scale by the Fermi momentum $k_F$, i.e.
\begin{eqnarray}
\label{TotalDensity}
n= \frac{k_F^3}{3\pi^2}.
\end{eqnarray}
The inverse of the Fermi momentum defines the most important characteristic length scale in our problem. Roughly speaking,
it corresponds to the average distance $d$ between two unbound atoms or molecules. We emphasize that our definition of
$k_F$ involves the total density $n$ and therefore includes unbound atoms, molecules and condensed atom pairs. In terms
of $k_F$ we can form a characteristic dimensionless ``concentration''
\begin{equation}\label{defconc}
c  = \bar{a} k_F.
\end{equation}
The concentration is a measure for the ratio between the in-medium scattering length $\bar{a}$ and average distance,
$c\propto \bar{a}/d$. As mentioned in the introduction, the concentration is the crucial parameter for the description
of the crossover between the BEC and BCS regimes. For a small concentration $|c|$ the gas is dilute in the sense that
scattering can be treated as a perturbation. In this range mean field theory is expected to work reasonably well. On the
other hand, for large $|c|$ the scattering length exceeds the average distance between two atoms and fluctuation effects
beyond mean field might play a crucial role.

An alternative definition of the concentration could use the measured (vacuum)
scattering length, $c = a(B) k_F$. This definition has the advantage of a simple relation to the magnetic field. The
choice between the two definitions is a matter of convenience. We adopt here the definition (\ref{defconc}) since this
reflects universality in an optimal way. For broad Feshbach resonances the two definitions coincide since $\bar a = a(B)$.
In presence of an additional ``background scattering length'', $a(B) = a_{bg} + a_R(B)$ we will include $a_{bg}$ in the
definition of $c$.

The concentration $c$ is the most important parameter for the description of the crossover (besides $T$ and $n$).
The inverse concentration $c^{-1}$ corresponds to a ``relevant parameter'' which vanishes at the location of the resonance.
Once described in terms of $c^{-1}$ the ultracold fermionic gases show a large universality. We demonstrate this
universality on the crossover phase diagram in fig. \ref{CrossoverTcAll} which plots the critical temperature
$\Tn_c = T_c/\epsilon_F$ for the transition to superfluidity as a function of $c^{-1}$. The narrow resonance limit
$\hpn \to 0$ is exact, while our results for broad resonances ($\hpn \to \infty$) still have substantial uncertainties, as
demonstrated by two approximations that will be explained later. For large $\hpn$ the actual value of $\hpn$ becomes
irrelevant and all curves coincide with the broad resonance limit. Intermediate $\hpn$ interpolate between the broad
and narrow resonance limits.

We have argued in sect.
\ref{sec:metastabledilutegas} that in the limit of a pointlike approximation for the effective fermionic interaction all
results should only depend on the effective scattering length. This only involves the ratio $\bar{\nu}/\hpb^2$ such that
for fixed $c$ the separate value of $\hpb$ should not matter. On the BCS side for small $|c|$ we therefore expect a universal
behavior independent of the value of the Yukawa coupling $\hpb$. This is clearly seen in fig. \ref{CrossoverTcAll}
where the critical line approaches the BCS result independently of $\hpb$. Furthermore, all results become independent
of the value of $\hpb$ in the broad resonance limit ($\hpn \to
\infty$). The concentration remains then as the only parameter (besides $T$ and $n$) for the description of the
crossover. These new universal aspects, adding to those that will be presented in \ref{sec:Univ}, are discussed in
\cite{Diehl:2005an}. In particular, the broad resonance limit $\hpn\to \infty$ corresponds to a pointlike microscopic
interaction.

For the broad resonance limit the first approximation (solid line) corresponds to extended MFT and neglects the
modifications of the effective potential induced by the molecule fluctuations. The second approximation (dashed line)
includes these fluctuation effects via the solution of a gap equation for the molecule propagator (sect. \ref{sec:beyond}).
The fast approach to the BEC value for $c^{-1} \to 0_+$ does not reflect the expected behavior $\Tn_c = \Tn_c^{BEC} +
\kappa c$ with a dimensionless constant $\kappa$. This shortcoming of our treatment is remedied by a functional
renormalization group study \cite{Diehl:2007th}. The
solid line at resonance $\Tn_c(c^{-1}=0)=0.292$ coincides with the result obtained in \cite{Levin052} which omits boson
fluctuations. The dashed line ($\Tn_c(c^{-1}=0)=0.259$ is in good agreement with the value obtained in \cite{ZZZStrinati}
($\Tn_c = 0.266$), who work in the broad resonance limit from the outset.
\begin{figure}[t!]
\begin{minipage}{\linewidth}
\begin{center}
\setlength{\unitlength}{1mm}
\begin{picture}(85,52)
      \put (0,0){
     \makebox(80,49){
     \begin{picture}(80,49)
      \put(0,0){\epsfxsize80mm \epsffile{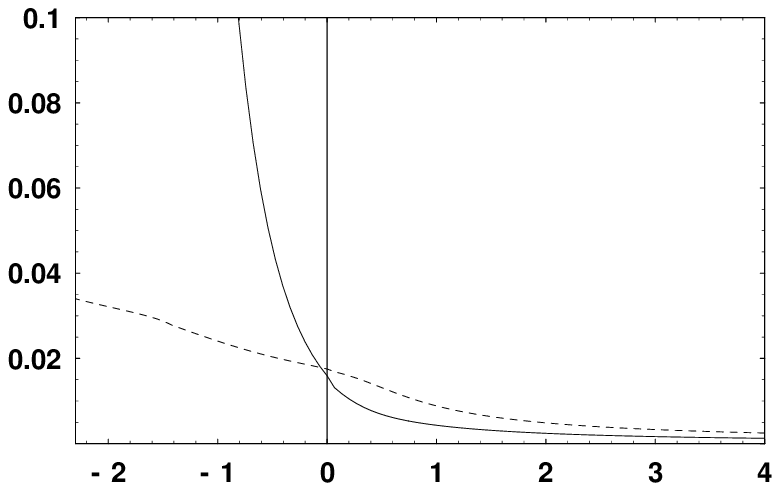}}
      \put(70,-2){$c^{-1}$}
      \put(12,42){$\tilde{A}_\phi/\hpn^2$ }
      \put(12,22){$Z_\phi/\hpn^2$ }
      \put(40,38){$T=0$ }
      \end{picture}
      }}
   \end{picture}
\end{center}
\vspace*{-1.25ex} \caption{Gradient coefficient $\Apn = 2M\Apb$ and wave function renormalization $\ZpR$ in dependence
on $c^{-1}$. We divide by $\hpn^2$ in order to get numbers $\mathcal{O}(1)$ and use $\hpn^2 = 3.72\cdot 10^5$
as appropriate for $\lit$ \cite{Diehl:2005an}. The ratio $\ApR=\Apn/\ZpR$ is displayed in fig. \ref{BoseCoeffs}.}
\label{UnrenormWFR}
\end{minipage}
\end{figure}

In fig. \ref{UnrenormWFR} we show the wave function renormalization $Z_\phi$ and the dimensionless gradient coefficient
$\tilde A_\phi = 2M \bar A_\phi$ as a function of $c^{-1}$ for $T=0$. We use the large value for the Feshbach
coupling in $\lit$, $\hpn^2 =3.72 \cdot 10^5$
for $k_F = 1 \mathrm{eV}$ (cf. eq. (\ref{DimlessDimful})). We note that for large $\hpn$, as appropriate for the Feshbach
resonances in $\lit$ or $\kal$, the wave function renormalization $\ZpR$ is large, the ratio $\ZpR/\hpn^2$ being an
$\mathcal{O}(1)$ quantity. We observe a strong increase of $\tilde A_\phi$ when
evolving to the BCS side of the resonance. This strongly suppresses the propagation of the effective
bosonic degrees of freedom, leading to a situation where they are completely irrelevant. This is an aspect of
universality, where the system ``looses memory'' of the bosonic degrees of freedom and a
purely fermionic BCS-type description becomes appropriate. These curves are essentially identical in the limit
$\hpn^2 \to \infty$.

Indeed, large values of $\hpn^2$ exhibit an ``enhanced universality'' \cite{Diehl:2005an}. In this case, all ``microscopic
quantities'' depend only on the concentration $c$. In the limit $\hpn \to \infty$ at fixed scattering
length there is a loss of memory concerning the details of the bosonic sector in the microscopic action (\ref{PhiAction}).
These aspects are worked out in more detail in \cite{Diehl:2005an}. In a renormalization
group treatment this universality will be reflected in strongly attractive partial infrared fixed points, similar to the
quark meson model in strong interactions \cite{Jungnickel95}. This universality property will be valid beyond mean field
theory.

\subsection{Dimensionless parameters}
The characteristic energy scales for $T$, $\sigma$ and the gap $\Delta = \hpb\bar{\phi}$ are set by the Fermi energy
$\epsilon_F= k_F^2/(2M)$. It is appropriate to define dimensionless quantities
\begin{eqnarray}\label{DimlessDimful}
\tilde{T} &=& 2MT/k_F^2= T/\epsilon_F, \quad \tilde{\sigma}= 2M\sigma/k_F^2,\nonumber\\\nonumber
 \qn&=& q/k_F, \quad \tilde{\Delta}=\Delta/\epsilon_F= 2M \hpb \bar{\phi}/k_F^2, \quad \rex = \tilde{\Delta}^*\tilde{\Delta},\\
 \hpn &=& 2M k_F^{-1/2}\hpb,\quad  \Apn = 2M\Apb.
\end{eqnarray}
Once all quantities are expressed in this way in units of the Fermi momentum or the Fermi energy the atom mass $M$ will no
longer be present in our problem. Indeed, in dimensionless units one has for the expressions (\ref{DefGammaPhi})
\begin{eqnarray}\label{GammaGammaPhiDimless}
\gamma&=&\frac{1}{2\tilde{T}}\big(\qn^2 -\tilde{\sigma}\big),\\\nonumber
\gamma_\phi&=&\frac{1}{2\tilde{T}}\big((\qn^2 -\tilde{\sigma})^2 +\tilde{r}\big)^{1/2}
\end{eqnarray}
such that the atom mass $M$ drops out in the computation (\ref{USigmaPhi},\ref{finalActionEq}) of the appropriately
rescaled effective potential $U$. All dimensionless quantities in $\gamma_\phi$ are typically of the
order one. For practical computations we may therefore choose units $k_F=1 \mathrm{eV}$.

The rescaled potential
\begin{eqnarray}
\label{SigEq}
\tilde{u} = k_F^{-3}\frac{\tilde{T}}{T}U=2M k_F^{-5} U
\end{eqnarray}
is composed of a classical contribution \footnote{In the formulation of sect. \ref{EffAtDens} $\hat{c}$ will be replaced by
$c$ since $\sigma$ and $\mu$ will be identified.} and a contribution from the fermion fluctuations ($\tilde{u}_1\hF$)
\begin{eqnarray}\label{MFTFermionPot}
\tilde{u} = -\frac{\rex}{8\pi \hat{c}} + \tilde{u}_1^{(F)},\\\nonumber
\frac{1}{8\pi \hat{c}} = \frac{1}{8\pi c} -\frac{\sigma - \mu}{\hpb^2 M k_F}.
\end{eqnarray}
Then the equation determining $\sigex(\tilde{r},\tilde{T})$ becomes
\begin{eqnarray}\label{SigEq2}
\frac{\partial\tilde{u}}{\partial\sigex}= -\frac{1}{3\pi^2}\frac{\nF}{n}
\end{eqnarray}
and is indeed independent of $M$.
It depends, however, on the ratio $\nF/n$ since we have defined in eq. (\ref{TotalDensity}) $k_F$ as a function of $n$
rather than $\nF$. In the mean field approximation the effective chemical potential $\sigex$ obeys, using
the dimensionless shorthands (\ref{GammaGammaPhiDimless}),
\begin{eqnarray}\label{anotherDens}
\int\limits_0^\infty d\qn\,\qn^2\big\{\frac{\gamma}{\gamma_\phi}\tanh\gamma_\phi - 1\big\}
= -\frac{2}{3}\frac{\nF}{n}.
\end{eqnarray}
In particular, for $\rex=0$ this determines $\sigex$ as a function of $\tilde{T}$ and $\nF/n$.

For given $\sigex$, $\Tn$ and $c$ one can compute the order parameter $\rex$ in the low temperature phase according to
\begin{eqnarray}\label{dimlessFieldEq}
 \frac{\partial\tilde{u}}{\partial\rex}=0 , \quad \frac{\partial\tilde{u}_1\hF}{\partial\rex}= \frac{1}{8\pi \hat{c}}.
\end{eqnarray}
The critical temperature for the phase transition corresponds to the value $\Tn_c$ where $\rex$ vanishes. This part of the
mean field computation is independent of the Yukawa coupling $\hpb$. In sect. \ref{sec:MolFrac} we will determine $\nF/n$
as a function of $\sigex, \Tn$ and $c$ such that eq. (\ref{SigEq2}) can be used to determine $\sigex$ as a function of $c$
and $\Tn$. As an alternative, we will modify in sect. \ref{EffAtDens} the definition of $U$ such that a modified
equation $\partial\tilde{u}/\partial\sigex = -1/3\pi^2$ becomes independent of $\nF/n$. Again, $\sigex$ can now be fixed as a function
of $c$ and $\tilde{T}$. We will see that the relation $\sigex (c,\Tn)$ depends on the choice of the Yukawa coupling $\hpb$.
The results will therefore depend on the additional dimensionless parameter $\hpn$ (\ref{DimlessDimful}).
Away from the narrow and broad resonance limits, the model is therefore characterized by two dimensionless quantities,
$c$ and $\hpn$.

\subsection{Universality}
\label{sec:Univ}
\begin{figure}[t!]
\begin{minipage}{\linewidth}
\begin{center}
\setlength{\unitlength}{1mm}
\begin{picture}(85,55)
     \put (0,0){
    \makebox(80,49){
    \begin{picture}(80,49)
      \put(0,0){\epsfxsize80mm \epsffile{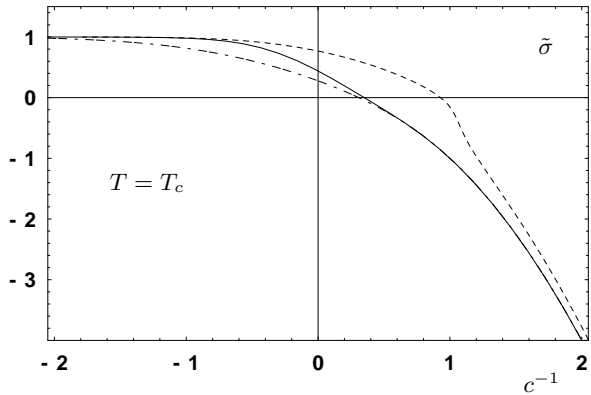}}
      \put(70,-2){$c^{-1}$}
      \put(72,43){$\sigex$ }
      \put(15,25){$T=T_c$ }
      \end{picture}
      }}
      \end{picture}
\end{center}
\vspace*{-1.25ex} \caption{Crossover at the critical temperature in the broad resonance limit: Effective dimensionless
chemical potential $\sigex$ at the critical temperature as a function of the inverse concentration $c^{-1}$. We compare the
results for two versions of the gap equation as in fig. \ref{CrossoverTcAll} (solid and dashed line). Additionally, the
result for the narrow resonance limit is indicated (dashed-dotted).}
\label{CrossoverPhaseDiagramSigmaAll}
\end{minipage}
\end{figure}

All observables can be expressed in terms of $c$, $\Tn$ and $\tilde{h}_\phi$. For example, using the
definition
\begin{eqnarray}
\tilde{\nu} = \frac{2M}{k_F^2}\bar{\nu}, 
\end{eqnarray}
one finds the relation
\begin{eqnarray}\label{NuMuC}
\nun - 2\tilde{\sigma} = -\frac{\tilde{h}_\phi^2}{8\pi c}.
\end{eqnarray}
In particular, the
phase diagram $\Tn_c(c)$ depends only on $\hpn$ as shown in fig. \ref{CrossoverTcAll} for the case of narrow and broad
resonances. For $\Tn=\Tn_c$ we find that the relation $\sigex (c)$ depends only mildly on the value of $\hpn$ as shown in
fig. \ref{CrossoverPhaseDiagramSigmaAll} where we compare narrow and broad resonance limits again.
Furthermore, for small $|c|$ the universal curves $T_c (c)$ and $\sigex (c,\Tn)$ become independent of $\hpn$: Both in the
BEC and BCS regime, an enhanced universality sets in, making $\hpn$ irrelevant for all its possible values! These issues are
investigated more systematically in \cite{Diehl:2005an}.

In summary we have now an effective low energy formulation where neither $M$ nor $\Lambda$ enter anymore. Everything is
expressed in terms of $k_F$ and three dimensionless parameters, namely $c,\hpn$ and $\Tn$. This scaling
property is an important aspect of universality. In \cite{Diehl:2005an} we discuss the relation of the parameters $c$ and $\hpn$ with
the physical observables for an experimental
setting, namely the magnetic field $B$ and the binding energy in vacuum. In the present paper we treat $\hpn$ as a
free parameter. The values of $\hpn$ for
$\lit$ and $\kal$ turn out to be large such that the broad resonance limit applies to these systems.

We display in tab. \ref{Scales} the values of the dimensionless scale rations $M/k_F$, $\Lambda/k_F$ and $\epsilon_F/k_F$.
The first two give an idea how well the detailed microphysics decouples for realistic ultracold fermionic gases.
We use a density $n = 4.4\cdot 10^{12}\mathrm{cm}^{-3}$, $k_F=1\mathrm{eV}\hat{=}(3.7290\cdot 10^3 a_B)^{-1}$.

\begin{table}[h] \caption{\label{Scales} Typical values for the dimensionless scale ratios for $^6$Li and $^{40}\mathrm{K}$
($k_F=1\mathrm{eV}\hat{=}(3.7290\cdot 10^3 a_B)^{-1}$).}
  \begin{ruledtabular}
    \begin{tabular}{cccccccc}
       & $M/k_F$ & $\Lambda/k_F$ &  $\epsilon_F/k_F$ \\\hline 
     $^6\mathrm{Li}$& $5.65\cdot 10^{9}$ & $1.6\cdot 10^3$ & $8.9\cdot 10^{-11}$   \\\hline 
     $^{40}\mathrm{K}$& $40.0\cdot 10^{9}$ & $1.2\cdot 10^3$ & $1.2\cdot 10^{-11}$ \\  
    \end{tabular}
  \end{ruledtabular}
\end{table}

\section{Molecule and Condensate Fraction}
\label{sec:MolFrac}

At this point the functional integral setting for the ultracold fermionic atom gases is fully specified. The parameters
$a_R$ and $\bar h_\phi^2$ can  be extracted from a computation of two-atom scattering in the vacuum, taking the limit
$T\to 0$, $n\to 0$ \cite{Diehl:2005an}. Rescaling with appropriate powers of $k_F$ yields the parameters $c$ and $\hpn^2$
for the many body system. The relation between $\sigma$ and $n$ is determined by eq. (\ref{TheDensEq}) and we identify $\mu =
\sigma$ at the end of the computation. An approximate solution of the functional integral for the effective action
$\Gamma$ gives access to thermodynamic quantities and correlation functions.

The relation (\ref{TheDensEq}) between $\sigma$ and $n$ seems to be rather formal at this stage. In this section
we will develop the physical interpretation of this formula. In this context the important distinction between
microscopic and dressed molecules will appear. In a quantum mechanical computation for the physics of a Feshbach resonance
the concept of dressed molecules arises from mixing effects between the open and closed channels. This is an important
ingredient for the understanding of the crossover. Our functional integral formalism has to reproduce this channel
mixing in vacuum and to extend it to the many body situation. In the functional
integral formulation the quantum mechanical ``mixing effects'' are closely related to the wave function renormalization
$Z_\phi$. The concept of dressed molecules and their contribution to the density is directly related to the
interpretation of eq. (\ref{TheDensEq}). We stress however, that the functional integral evaluation of $n$ for given
$\sigma = \mu$ can be done completely independently of this interpretation.

\subsection{Exact expression for the bare molecule density}
\begin{figure}
\begin{minipage}{\linewidth}
\begin{center}
\setlength{\unitlength}{1mm}
\begin{picture}(85,52)
      \put (0,0){
     \makebox(80,49){
     \begin{picture}(80,49)
      \put(0,0){\epsfxsize80mm \epsffile{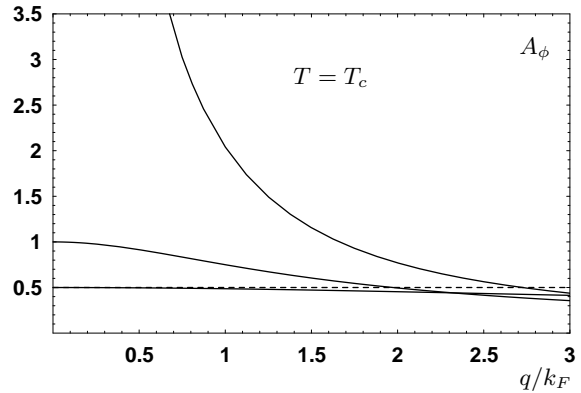}}
      \put(70,-2){$q/k_F$}
      \put(70,42){$\ApR$ }
      \put(40,38){$T=T_c$ }
      \end{picture}
      }}
   \end{picture}
\end{center}
\vspace*{-1.25ex} \caption{Momentum dependence of the gradient coefficient in the broad resonance regime
$\hpn^2\gg 1$. We plot $\ApR= 2M\Apb/\ZpR$ for $T=T_c$
in the different regimes, BCS ($c^{-1}=-1.5, \Tn_c = 0.057$, topmost), crossover  ($c^{-1}=0, \Tn_c = 0.29$) and BEC
$c^{-1}=1.5, \Tn_c=0.218$). The dashed line is the value for elementary pointlike bosons, $\ApR=1/2$.}
\label{Zqtot}
\end{minipage}
\end{figure}

The total density of atoms is composed of three components according to eq. (\ref{TotDens}),
\begin{eqnarray}\label{DensComps}
n=\nF + 2\nM +\nC.
\end{eqnarray}
Here we have split up the contribution $\nB$ from (\ref{TotDens}) in a contribution from uncondensed bare molecules
(connected two-point function)
\begin{eqnarray}\label{MolDens}
\nM &=& \langle\hat{\phi}^*\hat{\phi}\rangle - \langle\hat{\phi}^*\rangle\langle\hat{\phi}\rangle - \hat{n}_B\\\nonumber
    &=& \langle\hat{\phi}^*\hat{\phi}\rangle_c - \hat{n}_B
\end{eqnarray}
and one from the condensate which only occurs below the critical temperature $T_c$
\begin{eqnarray}
\nC= 2\langle\hat{\phi}^*\rangle\langle\hat{\phi}\rangle = \frac{k_F^4}{2\hpb^2 M^2}\,
\rex = 2 k_F^3\frac{\rex}{\tilde{h}_\phi^2}.
\end{eqnarray}
The ratio of atoms in the condensate arising from bare molecules is defined as
\begin{eqnarray}\label{CondFrac}
\OmC = \frac{\nC}{n} = \frac{6\pi^2}{\tilde{h}_\phi^2} \,\tilde{r}
\end{eqnarray}
and involves the Yukawa coupling $\tilde{h}_\phi=2M \hpb/k_F^{1/2}$.

The density of uncondensed bare molecules $\nM$ (\ref{MolDens}) can be written in terms of the bare molecule propagator or
connected two-point function
\begin{eqnarray}\label{ExactDens}
\nM(x) = T \int d\tau \bar{G}_\phi(x,\tau;x,\tau) - \hat{n}_B.
\end{eqnarray}
This formula is exact \footnote{Eq. (\ref{ExactDens}) is valid for the normal phase $T\geq T_c$. For the superfluid phase
$\bar{G}_\phi$ becomes a $2\times 2$ matrix and the corresponding generalization of eq. (\ref{ExactDens}) is discussed in
sect. \ref{sec:MolDensSFL}.}. It involves an additive renormalization $\hat n_B$ of a similar origin as for the
fermionic density, cf. sect. \ref{sec:renormalization}, but with opposite sign due to Bose-statistics.
In the homogeneous limit we can write eq. (\ref{ExactDens}) as a sum over integer Matsubara frequencies $2\pi n T$
\begin{eqnarray}\label{MolProExact}
\nM = T \sum\limits_n\int \frac{d^3q}{(2\pi)^3}\bar{G}_\phi(q,n) - \hat{n}_B.
\end{eqnarray}
The ratio of bare molecules to open channel atoms, $\nM/\nF$, is important for the understanding of the atom gas.
Technically, it enters the field equation for the effective chemical potential (\ref{SigEq2}) which involves $\nF/n$.
In the regions in parameter space where $\bar{n}_M/n$ is not very small a reliable computation of $\nM$ is crucial
for a quantitative understanding of the phase diagram. Such situations are e.g. realized in the BEC regime for narrow
and intermediate resonances. However, for the crossover region for a broad Feshbach resonance
(as for $\lit$ and $\kal$) it turns out that $\bar{n}_M/n$ is small such that an approximation $\bar{n}_F =n$ yields
already a reasonable result.

Nevertheless, the mean field approximation (\ref{anotherDens}) does not remain valid in all regions of the phase diagram.
This is due to an additional $\sigma$-dependence of $u$ arising from the boson fluctuations which are neglected in MFT.
We will argue below that this contribution from the bosonic fluctuations can be interpreted as the density of
dressed molecules. Thus, again an estimate of the molecule density will be mandatory for the understanding of the
crossover physics. A simple estimate of the dressed molecule density is the minimal ingredient beyond mean
field theory needed for a qualitatively correct description. We will refer to this as ``extended mean field
theory''.

As a first step the evaluation of the molecule density we may evaluate the classical approximation where the Yukawa
interaction between open and closed channel atoms is
neglected. This corresponds to the limit $\hpb\to 0$ (for fixed $\bar{a}$). Then $G_\phi^{(cl)}(q,n)$ is the free
propagator \footnote{Note that we use $\bar{\nu}$ instead of $\bar{\nu}_\Lambda$ - this includes already the dominant fluctuation
effects as motivated by the following paragraphs. Strictly speaking, the classical propagator features $\bar{\nu}_\Lambda$.}
\begin{eqnarray}\label{BosPropClass}
G_\phi^{(cl)} (q,n) = \Big(2\pi \mathrm{i}n T + \frac{q^2}{4M} + \bar{\nu} -2\mu \Big)^{-1}.
\end{eqnarray}
Performing the Matsubara sum and inserting $\hat{n}_B = \hat{n}/2= 1/2\int d^3q/(2\pi)^3$ one obtains the familiar
expression involving the occupation numbers for bosons
\begin{eqnarray}\label{BosNumber}
\nM&=& \int \frac{d^3q}{(2\pi)^3}\Big[\exp\Big(\frac{\bar{P}_\phi^{(cl)}(q)}{T}\Big) - 1\Big]^{-1},\\\nonumber
\bar{P}_\phi^{(cl)}(q)&=&\frac{q^2}{4M} + \bar{\nu} -2\mu.
\end{eqnarray}
We note the role of $\hat{n}_B$ for the removal of pieces that do not vanish for $\Lambda\to \infty$.

\subsection{Fluctuation effects}
The fluctuation effects will change the form of $\bar{G}_\phi(q,n)$. Details of the computation of $\bar{G}_\phi =
\MatPpb^{-1}$ are presented in app. \ref{app:WFR}. In the symmetric phase we may account for the fluctuations by
using
\begin{eqnarray}\label{Pphifull}
\MatPpb(Q)&=& ( 2\pi \mathrm{i} n T \ZpR(\sigma,T) + \bar{P}_\phi(q))\delta_{ab},\nonumber\\
\bar{P}_\phi(q)&=& \Apb(\sigma,T,q^2)q^2 +\bar{m}_{\phi}^2(\sigma,T).
\end{eqnarray}
We include here the fluctuations of the fermions (open channel atoms) for the computation of $\Apb$ and $\ZpR$.
The mass term $\bar{m}_\phi^2$ will be evaluated by a gap equation which also includes the bosonic molecule fluctuations.
In the superfluid phase ($\bar{\phi} \neq 0$) the diagonalization of the propagator has to be performed carefully as
discussed below in subsects. G,H.

The gradient coefficient $\Apb(\sigma,T,q^2)$ is defined  by eq. (\ref{Zphi1}) and depends on the Yukawa
coupling $\hpb$. For large $q^2$ it comes close to the classical value $1/4M$. We plot the renormalized gradient coefficient
$\ApR(q^2)= 2M\Apb(q^2)/\ZpR$
for different $c$ corresponding to the BCS, BEC and crossover regimes in fig. \ref{Zqtot}. It is
obvious that this gradient coefficient plays a major role. Large $\ApR$ leads to an additional \emph{suppression} of
the occupation number for modes with high $q^2$, as anticipated in sect. \ref{GradCoeff}.
\begin{figure}[t!]
\begin{minipage}{\linewidth}
\begin{center}
\setlength{\unitlength}{1mm}
\begin{picture}(85,155)
      \put (0,0){
     \makebox(80,49){
     \begin{picture}(80,49)
      \put(0,0){\epsfxsize80mm \epsffile{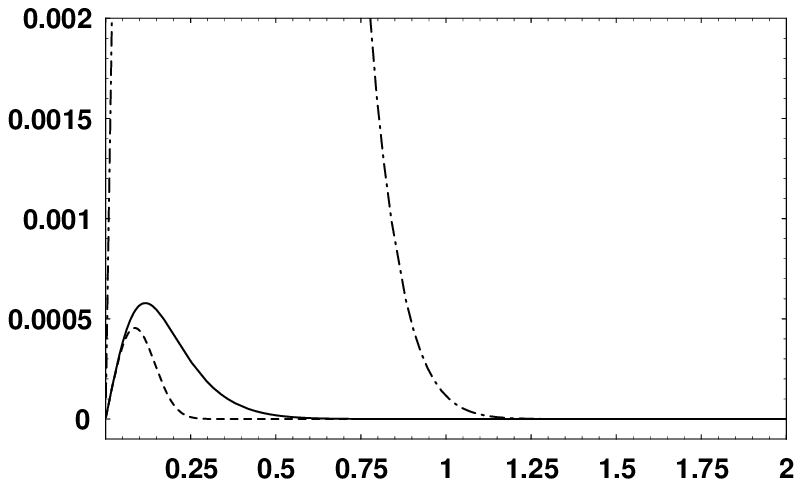}}
      \put(70,-2){$\qn$}
       \put(60,27){BCS}
      \put(55,22){$c^{-1}=-1.5$}
      \put(-3,1){(c)}
      \put(60,41){$\qn^3N_M(\qn)$ }
      \end{picture}
      }}
      \put (0,52){
    \makebox(80,49){
    \begin{picture}(80,49)
      \put(0,0){\epsfxsize80mm \epsffile{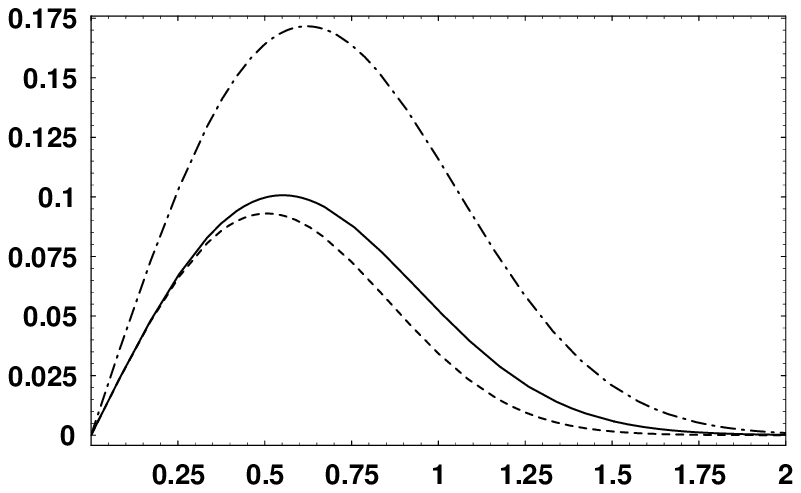}}
      \put(70,-2){$\qn$}
      \put(60,41){$\qn^3N_M(\qn)$ }
       \put(57,30){crossover}
      \put(58,25){$c^{-1}=0$}
      \put(-3,1){(b)}
      \end{picture}
      }}
      \put (0,104){
    \makebox(80,49){
    \begin{picture}(80,49)
      \put(0,0){\epsfxsize80mm \epsffile{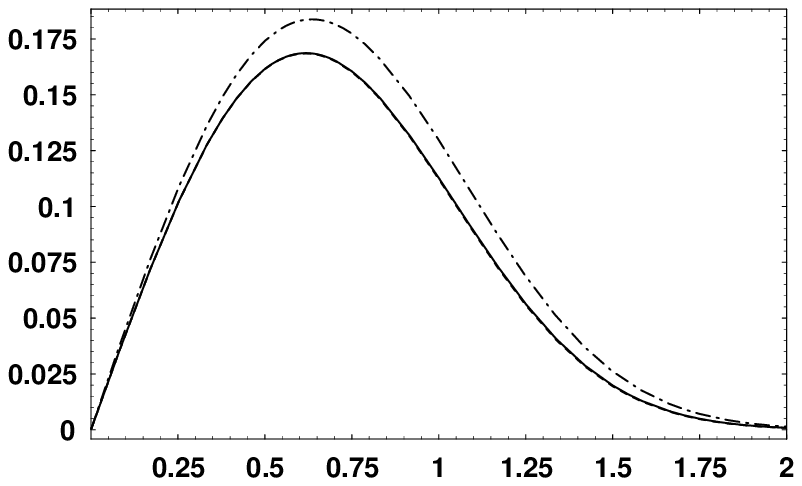}}
      \put(70,-2){$\qn$}
      \put(60,41){$\qn^3N_M(\qn)$ }
      \put(59,30){BEC}
      \put(55,25){$c^{-1}=1.5$}
      \put(-3,1){(a)}
      \end{picture}
      }}
   \end{picture}
\end{center}
\vspace*{-1.25ex} \caption{Weighted Bose distribution $N_M^{T=T_c}(\qn)=(\exp(\ApR \qn^2/\Tn) - 1)^{-1}$ as a
function of dimensionless momentum $\tilde{q} = q/k_F$, for $T=T_c$. We show the three regimes:
(a) BEC ($c^{-1}=1.5, \Tn_c=0.218$), (b) Crossover ($c^{-1}=0, \Tn_c = 0.292$) and (c) BCS ($c^{-1}=-1.5, \Tn_c = 0.057$).
We compare our best estimate using $\ApR(\qn)=\Apn(\qn)/\ZpR$ (solid curve) with the approximation $\ApR
=\ApR(\qn =0)$ (dashed curve) that we employ for the numerical estimates in this paper. Also indicated
is the result for the classical gradient coefficient $\ApR =1/2$ (dashed-dotted) that strongly
overestimates the molecule number in the crossover and BCS regimes. }
\label{nqtot}
\end{minipage}
\end{figure}

In the symmetric phase ($\bar{\phi}_0=0$) the ``mass term'' is given by
\begin{eqnarray}
\bar{m}_{\phi}^2 = \frac{\partial^2 U}{\partial\bar{\phi}^*\partial\bar{\phi}}\Big|_{\bar{\phi}=0}.
\end{eqnarray}
We use here a language familiar in quantum field theory since the molecule wave $\bar{\phi}$ behaves like a massive field for
$\bar{m}_\phi^2 >0$. (The propagator $G_\phi$ has a ``gap''.) For $\bar{m}_\phi^2 >0$ the symmetric solution of the
field equation, $\bar{\phi}_0 =0$, is stable whereas it becomes unstable for negative $\bar{m}_\phi^2$. For a second order phase
transition the critical temperature therefore corresponds precisely to a vanishing mass term, $\bar{m}_\phi^2(T=T_c) =0$.
The mass term reads in the mean field approximation
\begin{eqnarray}\label{MFTMass}
\bar{m}_{\phi}^{(F) 2}&=& \frac{\partial^2 U_{MFT}}{\partial \bar{\phi}^*\partial\bar{\phi}}\Big|_{\bar{\phi}=0} =
\bar{\nu} -2\mu+ \frac{\partial^2 U_1^{(F)}}{\partial\bar{\phi}^*\partial\bar{\phi}}\Big|_{\bar{\phi}=0}.\nonumber\\
\end{eqnarray}
For $T=0$, $\sigma =0$ the definition of $\bar{\nu}$ (\ref{Watwees}) implies $\bar{m}_\phi^{(F) \,2}=\bar{\nu} -2\mu$.
We note that the fermion fluctuations lower $\bar{m}_{\phi}^2$ as compared to the microscopic term $\bar{m}_{\phi,
\Lambda}^2=\bar{\nu}_\Lambda -2\mu$, cf. eq. (\ref{USigmaPhi}). This effect \emph{enhances} the occupation number for
molecules - using $\bar{m}_{\phi, \Lambda}^2$ instead of $\bar{m}_\phi^2$ would yield a much too small density of
molecules! At the critical temperature the mass term $\bar{m}_\phi^2$ vanishes (see below). For $T=T_c$ the fluctuation
effects therefore concern the size and shape of $\Apb(q)$.

As we have seen in our mean field computation in sect. \ref{sec:EffActMFT} the quantities $Z_\phi, \bar A_\phi$ and $\bar m_\phi^2$
depend strongly on $\sigma$. We may imagine to integrate first the fermion fluctuations in the functional integral
(\ref{GammaFuncInt}) (with $\eta^\dagger =\eta =0$). The result is an intermediate ``mean field action'' for the remaining
functional integral over bosonic fluctuations. The quadratic part $\sim \phi^*(Q) \phi(Q)$ in this action will be
of the type of eq. (\ref{Pphifull}). Performing now the boson fluctuations will induce an additional
contribution to $\partial U /\partial \sigma = - \bar n_F$. This will be interpreted below as the density of dressed
molecules.

\subsection{Dressed Molecules}

Dressed molecules are quasi-particles with atom number two. They are described by renormalized scalar fields $\phi$
with a standard non-relativistic $\tau$-derivative in the effective action. This allows for a standard association of the
number density of quasi-particles with the correlation function for renormalized fields. With the effective action
(\ref{GammaPosSpace}) the relation between the fields for dressed and bare molecules reads
\begin{eqnarray}
\phi_R = Z_\phi^{1/2} \bar \phi.
\end{eqnarray}
Correspondingly, the dressed molecule density $n_M$ becomes
\begin{eqnarray}
n_M = Z_\phi \bar n_M, \quad \Omega_M = \frac{2 n_M}{n}
\end{eqnarray}
and we find for large $Z_\phi$ a very substantial enhancement as compared to the bare molecule density $\bar n_M$.

We may define a renormalized gradient coefficient $\ApR$ and mass term $m_\phi^2$ by
\begin{eqnarray}
\bar{A}_{\phi,R} = \frac{\Apb}{\ZpR}, \quad \bar{m}^2_{\phi,R} = \frac{\bar{m}_{\phi}^2}{\ZpR},
\end{eqnarray}
or, in dimensionless units,
\begin{eqnarray}\label{DimlessRenorm}
\ApR = \frac{\Apn}{\ZpR}, \quad \mpR = \frac{\tilde{m}_{\phi}^2}{\ZpR} = \frac{2M}{k_F^2\ZpR} \bar{m}_\phi^2.
\end{eqnarray}
Then the quadratic part in the effective action for the bosons can be written in terms of $\phi$ and $m_\phi^2, A_\phi$,
without explicit reference to $Z_\phi$.
(Similar rescalings can be made for any quantity entering our calculations. For a complete list of dimensionful,
dimensionless and dimensionless renormalized quantities, cf. app. \ref{app:numerics}.) Using eq. (\ref{Pphifull}) in eq.
(\ref{MolProExact}) the wave function $\ZpR$ can be factored out in $\bar{P}_\phi$ such that
\begin{eqnarray}\label{RenormBosDens}
\nRM &=& \ZpR \nM = \int\frac{d^3q}{(2\pi)^3} \big[\exp\big(\frac{\bar{P}_{\phi,R}(q)}{T}\big) - 1\big]^{-1}.\nonumber\\
\end{eqnarray}
This is a standard bosonic particle number without the appearance of $Z_\phi$, now expressed in terms of
the effective renormalized inverse boson propagator (dimensionful and dimensionless version displayed)
\begin{eqnarray}\label{MFTCorrProp}
\bar{P}_{\phi,R}(q) &=&  \frac{\bar{P}_{\phi}(q)}{\ZpR} = \bar{A}_{\phi,R} q^2 + \bar{m}_{\phi,R}^2,\nonumber\\
P_\phi(q) &=&  \frac{2M}{k_F^2}\frac{\bar{P}_{\phi}(q)}{\ZpR} = \ApR \qn^2 + \mpR.
\end{eqnarray}
The ``dressed molecules'' \cite{XXStoofBos,Stoof05} include both bare
and fluctuation induced, effective molecules (cf. eq. (\ref{WFRDefini})). We will see how the dressed molecule density
$n_M$ emerges naturally in the equation of state for the particle density below.

We plot in fig. \ref{nqtot} the mode occupation number for the dressed molecules ($\qn = |\vec{q}\,|/k_F$)
\begin{eqnarray}\label{116}
N_M(\qn) &=& (\exp (\bar{P}_{\phi,R}(|\vec{q}\,|)/T)-1)^{-1}\\
&=& (\exp (P_\phi(\qn)/\Tn)-1)^{-1}\nonumber
\end{eqnarray}
(weighted by a volume factor $\qn^3$). There we compare the ``classical case'' $\bar{P}_\phi = q^2/4M$ with the result
including the fluctuation corrections. The normalization in the figure reflects directly the relative contribution
to $\nRM$
\begin{eqnarray}\label{117}
\nRM = \frac{k_F^3}{2\pi^2} \int d (\ln \qn ) \,\,\qn^3N_M(\qn).
\end{eqnarray}
We observe a large fluctuation effect in the BCS regime (beyond the renormalization of $\bar{m}_\phi^2$). In this regime,
however, the overall role of the molecules is
subdominant. On the other hand, in the BEC regime the molecule distribution is rather insensitive to the details of the
treatment of the fluctuations. The most important uncertainty from the ``molecule'' fluctuations therefore concerns the
crossover regime. In the BEC and crossover regime the replacement of $\Apb(q^2)$ by $\Apb (q^2=0)$ results only in a
moderate error. For simplicity we neglect the momentum dependence of $\Apb(q^2)$ for the numerical results in this work.
\begin{figure}[t!]
\begin{minipage}{\linewidth}
\begin{center}
\setlength{\unitlength}{1mm}
\begin{picture}(85,105)
      \put (0,55){
    \makebox(80,49){
   \begin{picture}(80,49)
      \put(0,0){\epsfxsize80mm \epsffile{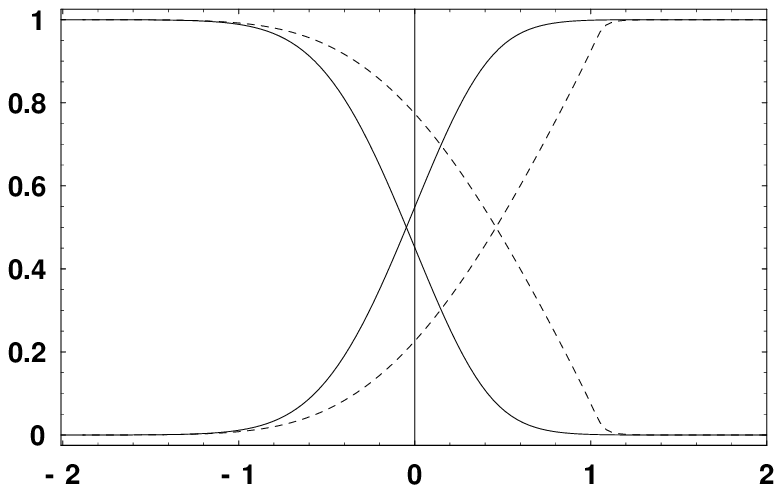}}
      \put(70,-2){$c^{-1}$}
      \put(-3,1){(a)}
      \put(10,43){$\OmRF$ }
      \put(69,43){$\OmRM$ }
      \put(60,23){$T=T_c$ }
      \end{picture}
      }}
   \put (0,0){
    \makebox(80,49){
   \begin{picture}(80,49)
      \put(0,0){\epsfxsize80mm \epsffile{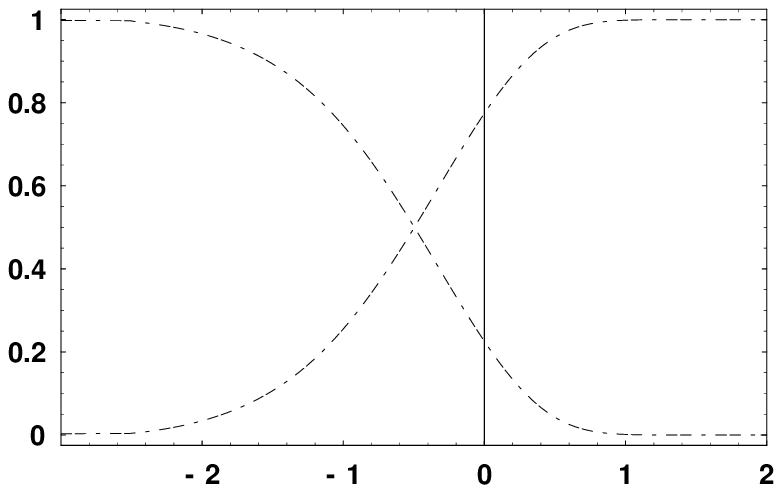}}
      \put(70,-2){$c^{-1}$}
      \put(-3,1){(b)}
      \put(10,41){$\bar{\Omega}_F$ }
      \put(69,41){$\bar{\Omega}_M$ }
      \put(60,23){$T=T_c$ }
      \end{picture}
      }}
   \end{picture}
\end{center}
\vspace*{-1.25ex} \caption{Crossover at the critical temperature: Contributions to the total particle number, showing the
crossover from fermion to boson dominated physics. (a) Fractions of ``dressed'' densities in the large $\hpn$ limit. We
compare the results for two versions of the gap equation as in fig. \ref{CrossoverTcAll}. (b) Fractions of ``bare''
densities in the exact narrow resonance limit $\hpn\to 0$. Though the pictures are similar, the physical interpretation
of the two plots differs as described in the text. }
\label{CrossoverDensAll}
\end{minipage}
\end{figure}

\subsection{Contribution of molecule fluctuations to the effective potential}
\label{sec:ContribMolFluct}

The computation of $\nRM$ evaluates a one loop integral which involves the molecule fluctuations. (Graphically, eqs.
(\ref{MolProExact}), (\ref{BosNumber}) correspond to a closed loop for the molecule fluctuations with an insertion of a
$\mu$ - derivative.) A self-consistent approximation should therefore also include the effects of the molecule fluctuations
in the computation of $U$ and therefore $\bar{\phi}_0$ or $T_c$. Our functional integral approach makes this necessity
particularly apparent: The computation of the partition function $Z$ involves the fluctuations of open channel atoms and
molecules in a completely symmetric way (the variables $\hat\psi$ and $\hat{\phi}$ in eq. (\ref{PhiAction})). There is no
reason why the fluctuations of the fermionic atoms should be included and not the ones for the bosonic molecules.
In particular, the critical region very close to $T_c$ will be dominated by the boson fluctuations.

For the effective potential the incorporation of the molecule fluctuations is achieved by adding to $U$ the one loop
contribution $U_1\hB$ from the fluctuations of $\hat{\phi}$
\begin{eqnarray}
\hspace{-0.1cm}U = U_{MFT} + U_1^{(B)} \hspace{-0.06cm}= (\bar{\nu} - 2 \mu)\bar{\phi}^*\bar{\phi} + U_1^{(F)}\hspace{-0.1cm}+U_1^{(B)}.
\end{eqnarray}
We can construct the bosonic contribution as the leading order correction to the mean field result which omits
boson fluctuations completely. For this purpose, we note that the fermion field appears only at quadratic order in the
classical action (\ref{PhiAction}). We can therefore integrate them out, which turns eq. (\ref{GammaFuncInt}) into a
purely bosonic functional integral,
\begin{eqnarray}\label{PurelyBosonic}
\Gamma[\psi=0,\bar\phi] =
 - \ln \int \mathcal D \delta \phi \exp\big( -  S_{MFT}[ \bar\phi + \delta\phi]\\\nonumber
 \qquad\qquad+  j^*\delta\phi + \mathrm{h.c.} \big)
\end{eqnarray}
with an intermediate action $ S_{MFT}$ depending on the field $\hat \phi = \bar\phi + \delta\phi$. This is
given by the exact expression
\begin{eqnarray}\label{IntermediateAct}
 S_{MFT} [\hat{\phi}] &=& S_\phi^{(cl)}[\hat{\phi}] - \frac{1}{2}\ln\det S  ^{(\psi\psi)}[\hat{\phi}]\\\nonumber
&=& S_\phi^{(cl)}[\hat{\phi}] - \frac{1}{2}\Tr \ln S  ^{(\psi\psi)}[\hat{\phi}]
\end{eqnarray}
where $S^{(\psi\psi)}$ denotes the second variation w.r.t. the fermion fields. In the ``classical approximation''
one has $\Gamma[\bar\phi] = S_{MFT}[\bar\phi]$, while the one loop approximation corresponds to an expansion of
$S_{MFT}[\bar\phi + \delta\phi]$ to second order in $\delta\phi$.

The mean field effective potential (\ref{USigmaPhi}) is obtained from eq. (\ref{PurelyBosonic}) in the classical
approximation. The next order contribution takes the Gaussian approximation for the fluctuations of the
molecule field into account. In principle, this requires the evaluation of highly nonlocal objects -- the one-loop
fermion fluctuations  encoded in the $\Tr\ln$-term in eq. (\ref{IntermediateAct}) feature a complex frequency and
momentum dependence. However, since we are interested in the observable low energy properties of the system, we may
apply a derivative expansion which only keeps the leading order terms in the frequency and momentum dependence.
This precisely generates the wave function renormalization
$Z_\phi$ (\ref{WFRDefini},\ref{ZRFormula}) and the gradient coefficient (\ref{Zphi1},\ref{Aphi1}). In this
approximation, we find (up to an irrelevant infinite constant and evaluated in the symmetric phase) the one loop result
\begin{eqnarray}\label{BosPotNew}
U_1^{(B)} = T \int\frac{d^3q}{(2\pi)^3} \ln \big|1- \mathrm{e}^{-\bar{P}_{\phi,R}/T}\big|,
\end{eqnarray}
where the spacelike part of the boson propagator is precisely given by eq. (\ref{MFTCorrProp}) in the symmetric phase
(for the result in the symmetry broken phase, cf. eq. (\ref{MFTCorrBroken})). This one loop formula has shortcomings
and we will improve on it by solving appropriate gap equations in sect. \ref{sec:beyond}. Nevertheless, it
already contains the essential information for the different contributions to the density.

Let us compare the effect of the $\mu$- and $\sigma$-derivatives on the bosonic part of the effective potential, eq.
(\ref{BosPotNew}). With the ``classical'' inverse boson propagator $\bar{P}_\phi^{(cl)} = q^2/4M + \bar{\nu} - 2\mu$
one has $\partial\bar{P}_\phi/\partial \mu = -2$ and we note the simple relation
\begin{eqnarray}\label{simplerel}
\frac{\partial U_1^{(B)}}{\partial \mu } = - 2 \nM.
\end{eqnarray}
On the other hand, the fermion loop corrections induce a $\sigma$ - dependence (at fixed $\mu$) of $U_1\hB$, which
contributes to $\nF$. This contribution can be interpreted as the number of open channel atoms that are bound in the
dressed molecules
\begin{eqnarray}
2n_{FM} = -\frac{\partial U_1\hB}{\partial \sigma}\big|_\mu.
\end{eqnarray}
The total number of dressed molecules is then given by
\begin{eqnarray}
\nRM = \nM + n_{FM} = \ZpR \nM.
\end{eqnarray}

This could be taken as a possible alternative definition of $\ZpR$
\begin{eqnarray}
\ZpR -1 = \frac{n_{FM}}{\nM} = \frac{\partial U_1\hB/\partial \sigma}{\partial U_1\hB/\partial \mu}.
\end{eqnarray}
In the limit where the $\sigma$ - dependence of $\ZpR$ and $\Apb$ can be neglected and $\partial \bar{P}_\phi/\partial \mu =-2$
one finds from
\begin{eqnarray}
\bar{P}_\phi = \Apb q^2 + \bar{m}_\phi^2
\end{eqnarray}
(cf. eq. (\ref{Pphifull})) the relation
\begin{eqnarray}\label{ZRdef}
\ZpR - 1 = \frac{\partial \bar{P}_\phi/\partial \sigma}{\partial \bar{P}_\phi/\partial \mu}  =-\frac{1}{2}\frac{\partial\bar{m}_\phi^2}
{\partial\sigma}\big|_\mu = -\frac{1}{2}\frac{\partial^3 U}{\partial \sigma\partial\bar{\phi}^*\partial\bar{\phi}}\big|_\mu.\nonumber\\
\end{eqnarray}
We recover the MFT result in eq. (\ref{ZphiinMFT}). The combined effect of the derivatives with respect to $\sigma$
and $\mu$ yields directly the number of dressed molecules
\begin{eqnarray}\label{DressedMol}
\nRM = \ZpR \nM = -\frac{1}{2}\Big(\frac{\partial U_1\hB}{\partial \mu}\big|_\sigma +
\frac{\partial U_1\hB}{\partial \sigma}\big|_\mu\Big).
\end{eqnarray}
The quantitative results shown in the figures are obtained by including the $\sigma$ - and $\bar{\phi}$ - dependence of $P_\phi$.
They will be discussed in more detail in the next sections.

\subsection{Open and closed channel atoms}
\label{sec:OpenandClosed}

Some characteristic properties of the ultracold gas involve the fractions of open channel atoms, uncondensed bare
molecules and condensed bare molecules
\begin{eqnarray}
\OmF &=& \frac{\nF}{n}, \quad \OmM = \frac{2\nM}{n},\quad\OmC = \frac{\nC}{n},\nonumber\\
&&\OmF + \OmM + \OmC = 1.
\end{eqnarray}
For example, the sum $\OmB = \OmM + \OmC$ measures the fraction of closed channel atoms, as observed in \cite{Partridge05}.
The formal use of two different effective chemical potentials in the action $S_B$ (\ref{PhiAction}), i.e. $\sigma$ for the
fermions and $\mu$ for the bosons, allows the simple association
\begin{eqnarray}
-3\pi^2 \bar{\Omega}_F = \frac{\partial U}{\partial \sigma}\Big|_\mu , \quad -3\pi^2 \bar{\Omega}_B =
\frac{\partial U}{\partial \mu}\Big|_\sigma,
\end{eqnarray}
and we recall that $\bar{\Omega}_F$ also receives a contribution from $U_1\hB$. We have computed $\bar \Omega_B$ in
\cite{Diehl:2005an} and the results agree well with \cite{Partridge05} over several orders of magnitude.

In the symmetric phase for $T\geq T_c$ one has $\OmF + \OmM =1$. For small $\hpn\lesssim 1$ the BEC-BCS crossover is the
crossover from
small to large $\OmF$. In fig. \ref{CrossoverDensAll} (b) we plot $\OmF$ and $\OmM$ as a function
of the inverse concentration $c^{-1}$ for $T=T_c$. The modifications of $\ZpR$, $\Apb$ and $\bar{m}_\phi^2$ in the inverse
propagator $\bar{P}_\phi$ (\ref{Pphifull}) depend on the Yukawa coupling $\hpn$. However, this influences the precise shape
of the crossover between the BEC and BCS regime only for moderate $\hpn\gtrsim 1$. For smaller $\hpn$, the density
fractions are insensitive to the fluctuation modifications.

For the large values of $\hpn$ encountered for the broad Feshbach resonances in $\lit$ and $\kal$ the contributions from
the closed channel molecules $\OmM, \OmC$ become very small (cf. fig. \ref{FractionsT0R} (b)). The dressed
molecules differ substantially from the bare molecules (large $\ZpR$) and the crossover physics is better described in terms
of dressed molecules. We display in fig. \ref{CrossoverDensAll} (a) the fraction of dressed unbound atoms
$\Omega_F$ and dressed
molecules $\Omega_M$ for large values of $\hpn$ and $T= T_c$. The fractions are not sensitive to the
precise value of $\hpn$ in the broad resonance limit $\hpn\to \infty$, similar to the behavior found for small $\hpn$.

\subsection{Condensate fraction}
\label{sec:CondFrac}

\begin{figure}[t!]
\begin{minipage}{\linewidth}
\begin{center}
\setlength{\unitlength}{1mm}
\begin{picture}(85,110)
      \put (0,54){
     \makebox(80,49){
        \begin{picture}(80,49)
      \put(0,0){\epsfxsize80mm \epsffile{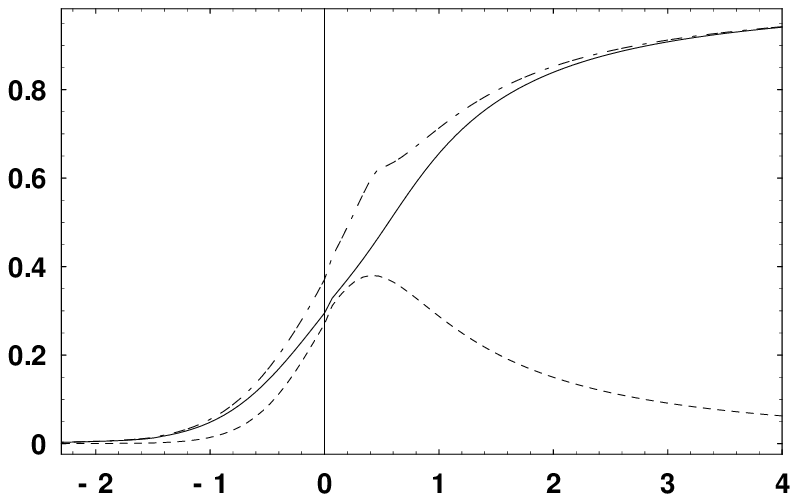}}
      \put(66,-2){$c^{-1}$}
      \put(-3,1){(a)}
      \put(10,30){BCS}
      \put(12,20){$T=0$}
       \put(65,12){$\OmRM$ }
       \put(43,28){$\OmRC$ }
       \put(25,23){$\Omega'_C$ }
       \put(69,30){BEC}
      \end{picture}
      }}
        \put (0,0){
     \makebox(80,49){
     \begin{picture}(80,49)
      \put(0,0){\epsfxsize80mm \epsffile{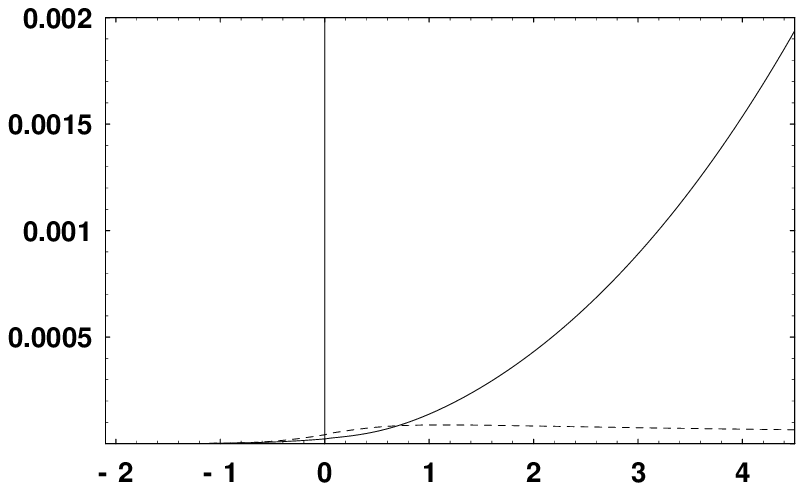}}
      \put(70,-3){$c^{-1}$}
      \put(-3,1){(b)}
      \put(68,39){$\OmC$ }
      \put(66,7){$\OmM$ }
      \put(15,15){$T =0$ }
     \end{picture}
      }}
\end{picture}
\end{center}
\vspace*{-1.25ex} \caption{(a) Contributions to the total particle density  for $T=0$ in the
large $\hpn$ limit: The fraction of dressed molecules $\Omega_M$ (dashed line) is largest in the crossover regime.
The condensate fraction $\Omega_C$ (solid line) grows to one in the BEC regime. The solid line corresponds
to $\Omega_C = \ZpR \bar{\Omega}_C$ whereas the dashed-dotted line uses $\ZpR'$ (eq. (\ref{ZRaltern})) instead of $\ZpR$.
(b) Fractions of the bare or closed channel molecules. In contrast to the dressed molecules, they are $\mathcal{O}
(\hpn^{-2})$. The dominant contribution arises from the condensed bare molecules $\OmC$ (solid line). The
contribution from noncondensed bare molecules $\OmM$ at $T=0$ remains very small. }
\label{FractionsT0R}
\end{minipage}
\end{figure}

The total number of atoms in the condensate depends on the expectation value of the \emph{renormalized} field
$\phi_R = \langle \hat\phi_R\rangle = Z_\phi^{1/2} \bar \phi$. We will mainly use the dimensionless field
\begin{eqnarray}
\phi = k_F^{-3/2} \langle\hat{\phi}_R \rangle = k_F^{-3/2}\ZpR^{1/2}\bar{\phi}.
\end{eqnarray}
With $\rho = \phi^*\phi$, $\rho_0 = \phi_0^*\phi_0$ the dressed condensate fraction can be defined as
\begin{eqnarray}\label{OmCRDef}
\OmRC =\frac{2\langle\hat{\phi}^*_R \rangle\langle\hat{\phi}_R \rangle}{n} = 2 k_F^3 \phi_0^*\phi_0 /n
= 6 \pi^2 \rho_0
= \ZpR\OmC.\nonumber\\
\end{eqnarray}
In fig. \ref{FractionsT0R} (a) we plot the condensate fraction $\Omega_C$ as a function of $c^{-1}$ for $T=0$. We
also show the fraction of closed channel atoms \footnote{The fig. for $\bar{\Omega}_C$ includes renormalization effects
for $\hpn$ discussed in \cite{Diehl:2005an}.} in the condensate, $\bar{\Omega}_C$, in fig. \ref{FractionsT0R} (b).
Both correspond to the broad resonance limit and we choose the Yukawa coupling appropriate for $\lit$, $\hpn=610$.

For large $\ZpR$ one has $\OmRC \gg \OmC$ - indeed, the probability $\ZpR^{-1}$ that a condensed di-atom state
contains a bare molecule is small. In this case the major part of the condensate is due to open channel atoms, i.e. $\nRC$
is dominated by a contribution from $\nF$
\begin{eqnarray}
n_{FC} &=& \nRC - \nC= (\ZpR-1) \nC \\\nonumber
&=& \frac{\OmC}{\OmC +\OmM}\big( 1- \frac{ \OmRF}{\OmF}\big)\nF.
\end{eqnarray}
Here the second line defines implicitly $\Omega_F$ by the requirement
\begin{eqnarray}
\Omega_F + \Omega_M + \Omega_C = 1.
\end{eqnarray}
(From $\nRC + \nRM \leq n$, $\OmC + \OmM \leq \ZpR^{-1} \ll 1$ one concludes $\OmF \approx 1$, and for low
enough $T$ one further expects $\OmM < \OmC$.) In the BCS limit this result is not surprising since we could have
chosen a formulation without explicit molecule fields such that all atoms are described by $\nF$. (If the chemical
potential multiplies only $\hat\psi^\dagger\hat\psi$ the total condensate fraction must be $\OmRC = 1- \OmRF$
\footnote{The difference between
$\OmRC$ and $1- \OmRF$ can be traced back to the appearance of the chemical potential $\mu$ in the effective
four-fermion interaction (\ref{Mom4Fermion}).}.)

The total number of open channel atoms can therefore be found in three channels, $\nF = \nRF+ n_{FC}+ 2n_{FM}$. Here
$\nRF$ denotes the unbound dressed fermionic atoms, $2n_{FM}$ the open channel atoms contained in dressed molecules
and $n_{FC}$ the ones in the condensate. As long as the fermionic and bosonic contributions to $U$ can be separated we
have the identities
\begin{eqnarray}
n_{F,0}  &=& \nRF + n_{FC} = -\frac{\partial U_1\hF}{\partial \sigma},\\\nonumber
\nRM &=& -\frac{1}{2}\Big(\frac{\partial U_1\hB}{\partial \mu}\big|_\sigma +
\frac{\partial U_1\hB}{\partial \sigma}\big|_\mu\Big),
\end{eqnarray}
and
\begin{eqnarray}\label{RenDensConstr}
n &=& n_{F,0} + 2 \nRM + \nC\\\nonumber
  &=& \nRF + 2 \nRM + \nRC\\\nonumber
  &=& \nF + 2 \nM + \nC.
\end{eqnarray}

The definition of a condensate fraction in terms of the superfluid order parameter $\rho_0$ is rather simple and
appealing. Nevertheless, this may not correspond precisely to the condensate fraction as defined by a given
experimental setup. The ambiguity is even larger when we come to the concepts of uncondensed fermionic atoms and molecules.
The distinction becomes somewhat arbitrary if we include higher loops for the computation of it, where bosonic and
fermionic fluctuations are mixed. As an example for the ambiguities in the definition one
may try to extract the number of open channel atoms in the condensate directly from the $\phi$ - dependence of $\nF$.
We can decompose the fermion contribution into a part for vanishing condensate and a ``condensate contribution'' due to
$\phi \neq 0$
\begin{eqnarray}
U_1\hF = U_1\hF (\bar{\phi} = 0) + \Delta U_1\hF .
\end{eqnarray}
The association
\begin{eqnarray}
n_{FC} = - \frac{\partial\Delta U_1\hF}{\partial \sigma}
\end{eqnarray}
yields an alternative definition of $\ZpR$,
\begin{eqnarray}\label{ZRaltern}
\ZpR' - 1 = \frac{n_{FC}}{\nC} = -\frac{1}{2\bar{\phi}^*\bar{\phi}} \frac{\partial\Delta U_1\hF}{\partial \sigma}.
\end{eqnarray}
With this definition the number of unbound atoms $\nRF$ becomes
\begin{eqnarray}
\nRF = \nF - n_{FC} - n_{FM} = -\frac{\partial U_1\hF(\bar{\phi} = 0)}{\partial \sigma}
\end{eqnarray}
which amounts to the standard number density in a Fermi gas. We note that $\ZpR'$ (\ref{ZRaltern}) coincides with $\ZpR$
(\ref{ZRdef}) in the limit where $\partial^3 U/\partial\sigma\partial\bar{\phi}^*\partial\bar{\phi}$ is dominated by the
term quadratic in $\bar{\phi}$ in $U_1\hF$. From fig. \ref{FractionsT0R} (a) we see that this is the case in the BEC and BCS
regimes.

\subsection{Excitations in the superfluid phase}
\label{sec:Excitations}

The bosonic excitations in the superfluid phase are analogous to a purely bosonic theory as for superfluidity in
$^4\mathrm{He}$ \cite{Gollisch02}. Due to the nonvanishing order parameter the matrix for the inverse renormalized propagator
$\mathcal{P}_\phi$ contains now ``off-diagonal'' entries form terms $\sim \rho_0 \phi\phi$ or $\rho_0\phi^*\phi^*$
where $\rho_0 = k_F^{-3}\ZpR\langle \hat{\phi} \rangle^* \langle \hat{\phi}\rangle$. It is convenient to use a basis
of real fields $\phi_1,\phi_2$
\begin{eqnarray}
\phi = \frac{1}{\sqrt{2}}(\phi_1 + \mathrm{i}\phi_2) , \quad \rho = \frac{1}{2} (\phi_1^2 + \phi_2^2).
\end{eqnarray}
In this basis $\mathcal{P}_\phi$ and $G_\phi =\mathcal{P}_\phi^{-1}$ are $2\times 2$ matrices and the exact eq.
(\ref{MolProExact}) for the bare molecule density is replaced by
\begin{eqnarray}
\nM (x) = \frac{T}{2}\mathrm{tr} \int d\tau \bar{G}_\phi(x,\tau; x,\tau ) - \hat{n}_B.
\end{eqnarray}
We expand in the superfluid phase around the minimum of the potential at
$\rhob = \rhob_0$
\begin{eqnarray}
\tilde{u} = \frac{\lpR}{2}(\rhoR - \rhoR_0)^2 + ...
\end{eqnarray}
such that the mass matrix
\begin{eqnarray}
\big(\mpR\big)_{ab} = \frac{\partial^2 \tilde{u}}{\partial \phi_a\partial\phi_b}\Big|_{\rhoR=\rhoR_0}
\end{eqnarray}
becomes
\begin{eqnarray}
\mpR  =\left( \begin{array}{cc}
  {2\lpR \rhoR_0} & {0} \\
  {0} & {0}
\end{array}\right).
\end{eqnarray}
Without loss of generality we have taken here the order parameter in the $\phi_1$ direction, $\phi_{1,0} = \sqrt{2\rhoR_0}$,
$\phi_{2,0} =0$, and we recognize the flat direction in the potential (vanishing eigenvalue of $\mpR$) in the
``Goldstone direction'' $\phi_2$. In contrast, the ``radial mode'' $\phi_1$ has a nonvanishing mass term
$2\lpR\rhoR_0$.

In the basis $(\phi_1,\phi_2)$ the term containing the $\tau$ - derivative is off-diagonal (neglecting total derivatives)
\begin{eqnarray}
\int d\tau \phi^*\partial_\tau \phi = \mathrm{i} \int d\tau \phi_1\partial_\tau \phi_2
= - \mathrm{i} \int d\tau \phi_2\partial_\tau \phi_1.
\end{eqnarray}
In momentum space one therefore finds for the renormalized inverse propagator $\mathcal{P}_\phi =
2M/(\ZpR k_F^2) \MatPpb$\footnote{Note the structure $\Gamma_2
\approx 1/2 \int_q \phi^T(-Q) \mathcal{P}_\phi(Q) \phi (Q)$.}
\begin{eqnarray}\label{MFTCorrBroken}
\mathcal{P}_\phi  = \left(
\begin{array}{cc}
  {\ApR \qn^2 + 2\lpR \rhoR_0,} & {-\tilde{\omega}} \\
  {\tilde{\omega} \qquad,} & {\ApR \qn^2}
\end{array}\right)
\end{eqnarray}
where we use eq. (\ref{DimlessRenorm}) and the renormalized order parameter and quartic coupling
\begin{eqnarray}
\rhoR_0 = \ZpR \tilde{\rho}_0 , \quad  \lpR = \tilde{\lambda}_\phi/\ZpR^2.
\end{eqnarray}
(For a list of the relations between dimensionless and dimensionless renormalized parameters cf. app. \ref{app:numerics}.)
This has an important consequence: The propagating excitations correspond to frequencies $\omega$ which obtain from the
Matsubara frequencies $\omega_B$ by analytic continuation $\omega_B \to \mathrm{i}
\omega$. This corresponds to the analytic continuation from Euclidean time $\tau$ to real or ``Minkowski'' time
$t = -\mathrm{i} \tau$. Now $G_\phi$ has a pole or $\mathcal{P}_\phi$ a zero eigenvalue. The eigenvalues $\lambda$ of
$\mathcal{P}_\phi$ therefore obey
\begin{eqnarray}
(\ApR \qn^2 + 2\lambda_\phi \rhoR_0 - \lambda)(\ApR \qn^2- \lambda) - \tilde{\omega}^2 =0.
\end{eqnarray}
Vanishing eigenvalues $\lambda$ therefore lead to the dispersion relation
\begin{eqnarray}
\tilde{\omega}^2 = \ApR \qn^2 ( 2\lambda_\phi \rhoR_0 + \ApR \qn^2 ).
\end{eqnarray}
For small $\qn^2$ this yields the linear dispersion relation characteristic for superfluidity
\begin{eqnarray}
\tilde{\omega} = \sqrt{2\ApR\lambda_\phi\rhoR_0}\sqrt{\qn^2},
\end{eqnarray}
from which we can read off the speed of sound
\begin{eqnarray}
v_s =k_F/(2M) \, \sqrt{2\ApR\lambda_\phi\rhoR_0}.
\end{eqnarray}

\subsection{Molecule density in the superfluid phase}
\label{sec:MolDensSFL}

The density of dressed molecules in the superfluid phase obeys
\begin{eqnarray}
\nRM &=& \ZpR \nM = k_F^3\frac{\Tn}{2} \mathrm{tr} \sum\limits_n \int\frac{d^3\qn}{(2\pi)^3} \mathcal{P}_\phi^{-1}(\qn,
\tilde{\omega}_n)
- \hat{n}_B\nonumber\\
&=& k_F^3 \Tn\sum\limits_n \int\frac{d^3\qn}{(2\pi)^3}\frac{\ApR \qn^2 + \lambda_\phi\rhoR_0}{\tilde{\omega}_n^2 +
\ApR \qn^2(\ApR \qn^2 + 2\lambda_\phi\rhoR_0)} \nonumber\\
&& - \hat{n}_B,
\end{eqnarray}
where the Matsubara summation can be performed analytically
\begin{eqnarray}\label{SuperFlDens1}
\nRM &=&\frac{k_F^3}{2} \int\frac{d^3\qn}{(2\pi)^3}\Big\{\frac{\ApR \qn^2 + \lambda_\phi\rhoR_0}{
\sqrt{\ApR \qn^2(\ApR \qn^2 + 2\lambda_\phi\rhoR_0)}}\nonumber\\
&&\times\coth\frac{\sqrt{\ApR \qn^2(\ApR \qn^2 + 2\lambda_\phi\rhoR_0)}}{2\Tn} - 1\Big\}.
\end{eqnarray}
Due to the subtraction of $\hat{n}_B$ (the term $-1$ in the curled bracket) the momentum integral is UV finite.

We have used dimensionless units and may introduce
\begin{eqnarray}\label{DefAlpha}
\alpha &=&\left\{
\begin{array}{c}
  {(\ApR \qn^2 + \mpR)/(2 \Tn) \quad \text{symmetric phase}}  \\
  {\ApR \qn^2/(2 \Tn) } \qquad \,\qquad\text{superfluid phase}
\end{array}\right. \nonumber\\
\kappa &=&\left\{
\begin{array}{c}
  {0 \qquad\qquad\qquad \,\,\,\qquad \text{symmetric phase}}  \\
  {\lambda_\phi\rho_0/(2 \Tn) \,\qquad\quad \quad\text{superfluid phase}}
\end{array}\right.\label{DefKappa}
\end{eqnarray}
where (for SSB)
\begin{eqnarray}\label{DefAlphaPhi}
\alpha_\phi &=&\sqrt{\ApR \qn^2(\ApR \qn^2 + 2\lpR\rhooR)}/(2\Tn)\\\nonumber
&=& \sqrt{\alpha^2 +  2\kappa\alpha}
\end{eqnarray}
is the bosonic analog to eq. (\ref{DefGammaPhi}). This allows us to write the dressed molecule density in both
phases as
\begin{eqnarray}
\nRM^{(SYM)} &=& \int\frac{d^3q}{(2\pi)^3}\Big(\exp 2\alpha - 1\Big)^{-1}\label{SymmDens},\nonumber\\
\nRM^{(SSB)} &=& \frac{1}{2} \int\frac{d^3q}{(2\pi)^3}\Big(\frac{\alpha + \kappa}{\alpha_\phi}\coth\alpha_\phi - 1\Big).
\label{SuperFlDens}
\end{eqnarray}
At the phase boundary the two definitions coincide since $\alpha_\phi = \alpha$ ($m_\phi^2 = 0, \rho_0 = 0$ and $\kappa=0$).
\begin{figure}[t!]
\begin{minipage}{\linewidth}
\begin{center}
\setlength{\unitlength}{1mm}
\begin{picture}(85,108)
      \put (0,0){
     \makebox(80,49){
     \begin{picture}(80,49)
      \put(0,0){\epsfxsize80mm \epsffile{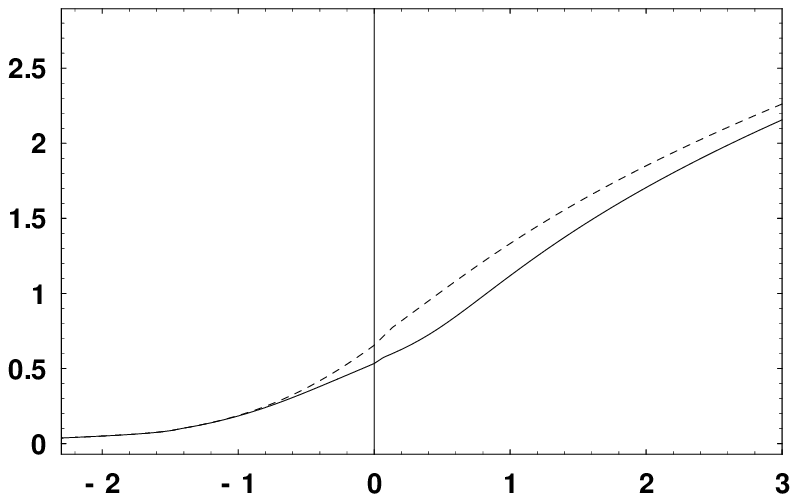}}
      \put(72,-2){$c^{-1}$}
      \put(72,42){$\tilde{\Delta}$ }
      \put(38.5,11.5){$\nwarrow$ }
      \put(36.8,12.9){{\large $\star$} }
      \put(41,8){QMC }
      \put(-3,1){(b)}
      \put(20,35){$T=0$}
      \end{picture}
      }}
      \put (0,54){
    \makebox(80,49){
    \begin{picture}(80,49)
      \put(0,0){\epsfxsize80mm \epsffile{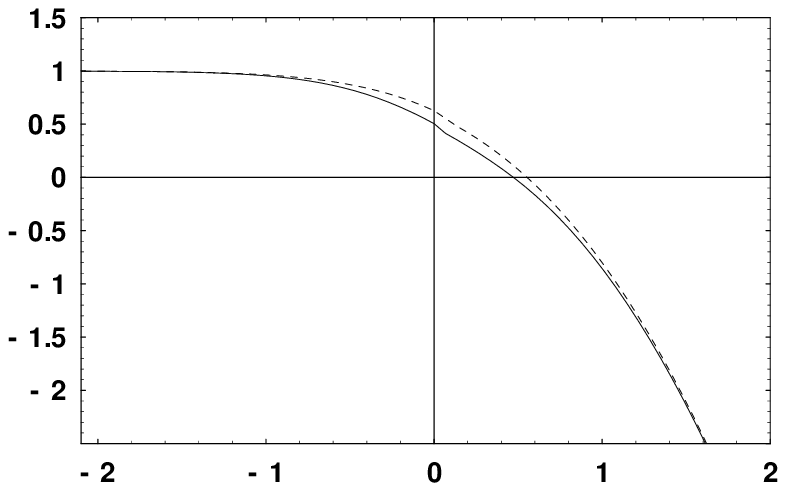}}
      \put(72,-3){$c^{-1}$}
      \put(72,42){$\sigex$ }
      \put(40,33){$\nearrow$ }
      \put(43,35.2){{\large $\star$} }
      \put(32,32){QMC }
      \put(20,15){$T=0$}
      \put(-3,1){(a)}
      \end{picture}
      }}
\end{picture}
\end{center}
\vspace*{-1.25ex} \caption{Solutions of the coupled gap and density equations at $T=0$, (a) effective chemical potential, (b) gap
parameter $\tilde{\Delta} = \sqrt{\rex}$. We also show (dashed)the mean field result (obtained by setting the bosonic
contribution $\nRM =0$ in the density equation). Our result can be compared to QMC calculations \cite{Carlson03} performed
at $c^{-1} =0$ which find $\sigex=0.44, \tilde{\Delta} =0.54$. Our solution yields $\sigex=0.50, \tilde{\Delta} =0.53$,
and improves as compared to the MFT result $\sigex=0.63, \tilde{\Delta} =0.65$. 
}
\label{CrossoverSnGap}
\end{minipage}
\end{figure}

The equations (\ref{SymmDens}) for the density of dressed molecules involve the renormalized coupling
$\lambda_\phi$. This describes the full vertex and therefore $\lambda_\phi$ is a momentum dependent function. This
also holds for $\ApR$. We will neglect the momentum dependence of $\ApR$, as motivated by fig. \ref{nqtot}
for $T= T_c$. For $\lambda_\phi$ this issue is more involved since $\lambda_\phi(q)$ vanishes for $q\to 0$
($T\leq T_c$) due to the molecule fluctuations, as shown in app. \ref{app:SDE}. We observe that for small $T$ the momentum
integration in eq. (\ref{SuperFlDens1}) is dominated by the range $\qn^2 \approx \lambda_\phi\rho_0/\ApR$. The
infrared suppression of $\lambda_\phi(q\to 0)$ is therefore not effective. For the density equation we will simply
omit the contribution of the molecule fluctuations to $\lambda_\phi$ and approximate $\lambda_\phi
= \lambda_\phi\hF$ with $\lambda_\phi\hF$ evaluated at $q^2 =0$.

The inclusion of the molecule density $n_M$ is important for quantitative accuracy even at $T=0$. This is demonstrated in
fig. \ref{CrossoverSnGap} where we show the crossover for the effective chemical potential $\sigex$ and the gap
$\tilde{\Delta}$ as a function of $c^{-1}$. The agreement with quantum Monte Carlo simulations \cite{Carlson03} is
substantially improved as compared to mean field theory, and also compares reasonably well with other analytical approaches
(Strinati \emph{et al} \cite{StrinatiPieri04}, $\tilde \Delta = 0.53, \sigex = 0.445$).

\section{Effective Field for Atom Density}
\label{EffAtDens}

Before proceeding in sect. \ref{sec:beyond} to a description of our computation of the effects from the molecule
fluctuations on the effective potential
we present in this section an improvement of our formalism which treats the fermionic and bosonic fluctuations in an
even more symmetric way. Indeed, in the thermodynamic equilibrium the molecule propagator should involve the same effective chemical
potential $\sigma$ as the propagator of the unbound atoms. (So far it involves $\mu$ instead of $\sigma$.)

If one is interested in the separate contributions from open and closed channel atoms one may formally consider different
chemical potentials $\sigma$ and $\mu$ multiplying $\hat\psi^\dagger\hat\psi$ and $\hat\phi^*\hat\phi$ in the action.
Then $\nF$ and $\nB$ can be associated to the variation of the effective action with respect to $\sigma$ and $\mu$.
At the end of the computations one has to identify $\sigma = \mu$). For many purposes, however,
one only needs the total number of atoms $n$. It seems then advantageous to modify 
our formulation such that the field $\sigma$ is associated to $n$ instead of $\nF$. In this section, we treat $\sigma$
as a classical field such that its role is reduced to a source term $\sigma =J$. In app. \ref{sec:partial} we will
consider the more general case where $\sigma$ is treated as a fluctuating field.

Identifying $\sigma = \mu$ in the bare action (\ref{PhiAction}), the $\sigma$-derivative of the effective potential
generates the total particle number density,
\begin{eqnarray}
-\frac{\partial U}{\partial\sigma} = n = \bar n_F + 2 \bar n_M + 2 \bar\phi^*\bar\phi.
\end{eqnarray}
In the presence of interactions it is hard
to evaluate the full bare correlation functions in the above form explicitly -- it premises the solution of the full
quantum field theory. As we have discussed in sects.
\ref{sec:ContribMolFluct}, \ref{sec:OpenandClosed} it is more practicable to decompose $n$ according to
\begin{eqnarray}\label{RenDensConstr}
n &=& -\frac{\partial(U_{MFT} + U_1\hB)}{\partial\sigma} = \nC + n_{F,0} + 2 n_M \\\nonumber
\end{eqnarray}
which is a ``mixed'' representation involving both bare and renormalized quantities. In the next section we
will see that this equation reduces to an equation of state for ``fundamental'' bosons in the BEC regime, which
involves dressed quantities only (eq. (\ref{BogoliubovDens})).

Beyond the classical bosonic propagator we have implemented \footnote{A more rigorous derivation of the equation of state
via the Noether construction for the conserved charge associated to the global $U(1)$-symmetry reveals that this choice
is uniquely fixed by requiring consistency within our truncation (linear frequency dependence) \cite{Diehl:2006}.} the
$\sigma$ - dependence of $U_1\hB$ by the $\sigma$ - dependence of $\bar{m}_\phi^2$. This yields
\begin{eqnarray}
\ZpR = -\frac{1}{2} \frac{\partial \bar{m}_\phi^2}{\partial\sigma} = -\frac{1}{2} \frac{\partial^3 U}
{\partial\sigma\partial\bar{\phi}^*\partial\bar{\phi}}
\end{eqnarray}
where we note that the contribution from the bare molecules is already included in the classical $\sigma$ - dependence of
$\bar{m}_\phi^2$, leading to the replacement $\ZpR \to \ZpR - 1$ in eq. (\ref{ZRdef}). In practice, we use the approximation
\begin{eqnarray}\label{ZRApprox}
\ZpR = 1 -\frac{1}{2} \frac{\partial^3 U_1\hF}{\partial\sigma\partial\bar{\phi}^*\partial\bar{\phi}}.
\end{eqnarray}
This has a simple interpretation: due to the fermionic fluctuations the classical ``cubic coupling'' $- 2 \sigma\bar{\phi}^*\bar{\phi}$
is replaced by a renormalized coupling $- 2 \ZpR \sigma\bar{\phi}^*\bar{\phi}$. We can compute $\nRM$ directly from eqs.
(\ref{117}),(\ref{116}),(\ref{RenormBosDens}),(\ref{Pphifull}).

\section{BEC Limit}
\label{sec:renconstZR}

In sect. \ref{sec:Excitations} we have defined a multiplicative renormalization scheme for the bosons by rescaling all
couplings and fields with an appropriate power of $Z_\phi$ such that the timelike derivative in the effective action
has a unit coefficient. This leads to the notion of dressed molecules in our formalism. In this section we discuss the
implications of this prescription in the BEC regime. Here we will see that the dressed molecules behave just as
``elementary'' bosons.

\begin{figure}[t!]
\begin{minipage}{\linewidth}
\begin{center}
\setlength{\unitlength}{1mm}
\begin{picture}(85,55)
      \put (0,0){
    \makebox(80,49){
   \begin{picture}(80,49)
      \put(0,0){\epsfxsize80mm \epsffile{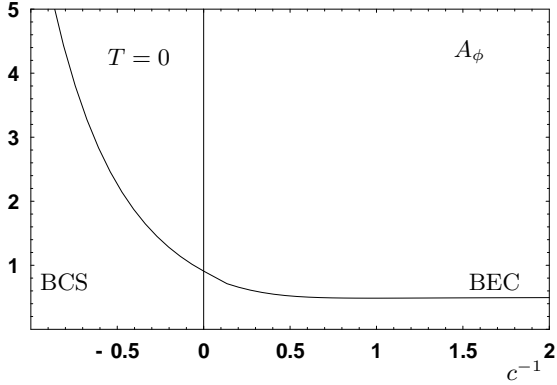}}
      \put(70,-2){$c^{-1}$}
      \put(65,10){BEC}
      \put(17,40){$T=0$}
      \put(8,10){BCS}
      \put(63,41){$\ApR$ }
      \end{picture}
      }}
   \end{picture}
\end{center}
\vspace*{-1.25ex} \caption{Gradient coefficient $\ApR = \Apn/\ZpR$ for $T=0$, as a function of the
inverse concentration. In the BEC limit, $\ApR$ takes the classical value $1/2$ as appropriate for elementary bosons
of mass $2M$.}
\label{BoseCoeffs}
\end{minipage}
\end{figure}
The propagator for the dressed molecules has still a nontrivial renormalization factor for its dependence on the
spacelike momenta (\ref{Pphifull}). The ratio $\ApR = \Apn/\ZpR$ is shown as a function of $c$ in fig. \ref{BoseCoeffs}.
It is instructive to evaluate $\ApR$ in the BEC limit where the propagator should describe the propagation of dressed
molecules. The integrals in $\Apn$ and $\ZpR$ can be evaluated analytically (cf. app. \ref{AnalBEC}, eqs. (\ref{ZphiBEC},
\ref{ZRBEC})) irrespective to the thermodynamic phase of the system. The universal result does not depend on $\hpn$,
\begin{eqnarray}
\ApR \to \frac{1}{2}.
\end{eqnarray}
This makes it particularly clear that for \emph{both} the limit of small $\hpn \to 0$ \emph{and} the limit of large $\hpn$
we recover composite, but pointlike  bosons with wave function $\ApR = 1/2$, corresponding to a kinetic energy
$p^2/4M$. However, in the first case of small $\hpn$ we deal with microscopic closed channel molecules,
while in the second case of large $\hpn$ the emerging bosons are open channel pairs, however behaving just as if they
were pointlike bosons in the BEC limit. For large $\hpn$ we could equally well drop the classical piece in $\Apb$ and
$\ZpR$. This corresponds to a purely pointlike interaction for the fermions without any explicit reference to molecules at
all, cf. eq. (\ref{Mom4Fermion}), and the discussion in \cite{Diehl:2005an}. Nevertheless, bound atom pairs emerge that
behave just as pointlike particles.

A similar aspect of universality is found for the four-boson coupling $\lambda_\phi = \tilde\lambda_\phi/Z_\phi^2$.
In the approximation where we neglect bosonic contributions to the four-boson coupling and to the gap equation for
the mass (BCS gap equation), we find $\lambda_\phi = \lambda_\phi\hF = 8\pi/\sqrt{-\sigex}$ such that the bosonic
scattering length becomes $a_{M} = 2a_R$. This is the Born approximation for the scattering of the composite bosons.
However, more accurate treatments on this issue have shown $a_M/a_R =0.6$. This result is obtained from the solution of the
four-body Sch\"odinger equation \cite{Petrov04} and form a numerically demanding diagrammatic approach \cite{Kagan05},
and has also been confirmed in QMC simulations \cite{Giorgini04}. A resummation of the effective boson interaction vertices
in vacuum as done in \cite{AAAAStrinati} yields $a_M/a_R =0.75(4)$. Our present Schwinger-Dyson approach
does not correctly account for the real situation since the contribution from molecule fluctuations in the vacuum is
missing. In the frame of functional renormalization group equations, this deficiency has been remedied. Simple truncations
yield $a_M/a_R = 0.71 - 0.92$ \cite{Diehl:2007th}.

It is instructive to investigate eq. (\ref{SymmDens}) in the limit $c^{-1} \to \infty$ (BEC limit) where
we have $\ApR \to 1/2, a_M=2a_R$. The density in the superfluid phase $\nRM$ (\ref{SymmDens})
then coincides precisely with the Bogoliubov formula for bosons of mass $2M$ and the above scattering length,
\begin{eqnarray}
\nRM &=& \frac{1}{2} \int\frac{d^3q}{(2\pi)^3}\Big(|v_q|^2  + \frac{|u_q|^2 +  |v_{-q}|^2}{\exp 2\alpha_\phi -1}
\Big)\label{BogoDens}
\end{eqnarray}
where we used the relations connecting the Bogoliubov transformation coefficients and our expressions (\ref{DefAlpha},
\ref{DefAlphaPhi}),
\begin{eqnarray}
|v_q|^2 &=& \frac{1}{2}\big(\frac{\alpha + \kappa}{\alpha_\phi} -1\big),\quad
|u_q|^2 +  |v_{-q}|^2  =\frac{\alpha + \kappa}{\alpha_\phi}.\nonumber\\
\end{eqnarray}
We emphasize again that we can perform first the limit $\hpn \to \infty$ where we recover a purely fermionic model with pointlike
interaction and no explicit molecule degrees of freedom (cf. \cite{Diehl:2005an}). Subsequently we may consider
large $c^{-1}$ where the approximations leading to eq. (\ref{BogoDens}) become valid. This shows that the Bogoliubov
formula for weakly interacting ``fundamental'' bosons can be recovered from a purely fermionic model! In our approach,
this result emerges in the simultaneous limit $c^{-1}\to \infty$ (BEC regime), $\hpn \to \infty$ (broad resonance regime).
A similar result has been established by Strinati \emph{et al.} \cite{AAAAStrinati,BBStrinati,CCStrinati,ZStrinati},
who work in a purely fermionic setting, or, in our language, in the broad resonance limit $\hpn\to \infty$ from the outset.

This observation is strengthened by an investigation of the dimensionless density equation (\ref{RenDensConstr}) at $T=0$
in the BEC limit:
\begin{eqnarray}\label{BogoliubovDens}
1 &=&3\pi^2( \frac{\rex}{16\pi\sqrt{-\sigex}} + 2\nRM)\nonumber\\\nonumber
&=& 3\pi^2(2\ZpR\frac{\rex}{\hpn^2} + 2\nRM)\\
&=& \OmRC + \OmRM.
\end{eqnarray}
Here we have dropped the term $\OmC = \mathcal{O}(\hpn^{-2})$ and we may similarly neglect $\OmM$. In the first line
of eq. (\ref{BogoliubovDens}) the first term is the explicit result for $\Omega_{F,0}$. The second line uses the explicit
result for $\ZpR$ (\ref{ZRBEC}) and we recover the definition of $\OmRC$ (\ref{OmCRDef}). This is precisely
the density equation one obtains
when assuming fundamental bosons of mass $2M$. This is an important result: Our $\ZpR$ - renormalization procedure generates
precisely the macrophysics we would have obtained when starting microscopically with a purely bosonic action. In our
approach however, the bosons emerge dynamically.

\section{Gap Equation for the Molecule Propagator}\label{sec:beyond}

The formulation of the problem as an Euclidean functional integral for fermionic and bosonic fields permits the use of
many of the highly developed methods of quantum field theory and statistical physics. It is an ideal starting point
for systematic improvements beyond MFT. We have mentioned in the previous sections that one possible alternative to the standard
one loop approximation could be to first integrate out the fermions and then perform the remaining $\hat{\phi}$-integral
in one loop order. This procedure has, however, some problems.

Let us consider the dimensionless renormalized inverse bosonic propagator (\ref{Pphifull}) ($\omega_B=0$) in the
approximation
\begin{eqnarray}\label{MFTtildeMass}
\mathcal{P}_\phi(\vec{\tilde{q}}) = \ApR \qn^2\delta_{ab} + m^2_{\phi , ab}.
\end{eqnarray}
The exact mass matrix
\begin{eqnarray}
m_{\phi, ab}^2 = \frac{\partial^2 \tilde{u}}{\partial \phi_a\partial\phi_b}\Big|_{\phi=0}
\end{eqnarray}
is diagonal but in general non-degenerate in the $\phi_1,\phi_2$ basis. It
vanishes at the critical temperature of a second order phase transition. Already after the first step of the $\hat\psi$ -
integration $\mathcal{P}_\phi$ differs from the classical inverse molecule propagator. The molecule propagator will appear in the
molecule fluctuations as described by the partition function
\begin{eqnarray}\label{Intermediate}
Z = \int \mathcal{D} \hat{\phi} e^{-S_{MFT}[\hat{\phi}]},
\end{eqnarray}
where $S_{MFT}$ is the intermediate action resulting from the Gaussian integration of the fermion fields. It is composed of
the classical boson piece and a loop contribution from the fermion fields,
\begin{eqnarray}
S_{MFT} &=& \int d^4x \big\{\hat{\phi}^*\big(Z_\phi[\hat{\phi}]\partial_\tau -\Apb[\hat{\phi}]\triangle + \bar{\nu}- 2\sigma\big)\hat{\phi} \nonumber\\
&&+U_1\hF[\hat{\phi}^*\hat{\phi}] + ... \big\}.
\end{eqnarray}
At this stage the contribution from the fermion loop still depends on the fluctuating boson field.
For example, the formula for the dressed molecule density in the symmetric phase
\begin{eqnarray}\label{MolPro}
\nRM = \int\frac{d^3q}{(2\pi)^3}\Big(\mathrm{e}^{P_\phi(\vec{\qn})/\Tn} - 1\Big)^{-1}= -\frac{1}{2} \frac{\partial U_1\hB}{\partial\sigex}
\end{eqnarray}
involves $P_\phi$ (\ref{MFTtildeMass}). It generalizes \footnote{Eq. (\ref{MolPro}) approximates the exact eq. (\ref{MolProExact}) in the limit where corrections to
the dependence of $P_\phi$ on the Matsubara frequencies can be neglected, see app. \ref{app:WFR}. The second part can be
viewed as a Schwinger-Dyson equation for the $\sigma$ - dependence of $U$.}
eq. (\ref{BosNumber}) with $\mu$ replaced by $\sigma$. We see that the inverse propagator $P_\phi$ appears
in the bosonic fluctuation contribution $U_1\hB$ to the effective potential and influences the
field equation for $\phi$.

For a simple one loop evaluation of the functional integral (\ref{Intermediate}) one would have to use a propagator with
$m_\phi^{(F) \,2}$ instead of $m_\phi^2$ in eq. (\ref{MFTtildeMass}). This has an important shortcoming. Due to
the difference between $U_{MFT}$ and $U$
the masslike term $m_\phi^{(F)\,2}$ vanishes at some temperature different from $T_c$. As a consequence of this
mismatch one observes a first order phase transition for sufficiently large values of $\hpn$. Clearly such a first order
phase transition may be suspected to arise from an insufficient approximation. (A similar
fake first order transition has been observed for relativistic scalar theories. There it is well understood
\cite{ATetradis93,CTetradis92} that an appropriate resummation (e.g. by renormalization group methods) cures the disease and
the true phase transition is second order. For the crossover problem the second order nature of the phase transition
has recently been established by functional renormalization group methods for the whole range of concentrations,
cf. \cite{Diehl:2007th})

In order to improve this situation we use for the bosonic fluctuations Schwinger-Dyson type equations where the inverse
propagator $P_\phi$ involves the second derivative of the full effective potential $U$ (rather than $U_{MFT}$).
It is obvious that this is needed for a reliable estimate of $\nRM$ since the exact expression (\ref{ExactDens}) and therefore
also (\ref{MolPro}) involves the full molecule propagator including the contribution of the molecule fluctuations.
\begin{figure}[t!]
\begin{center}
\setlength{\unitlength}{1mm}
\begin{picture}(85,40)
      \put (0,0){
     \makebox(84,38){
     \begin{picture}(84,38)
      \put(0,0){\epsfxsize84mm \epsffile{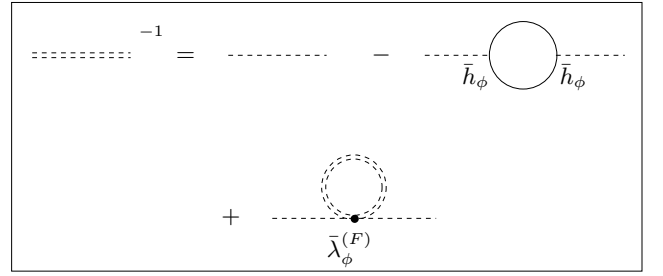}}
      \put(42,2){$\lpb\hF$}
      \put(17,30){$^{-1}$}
      \put(22,28){$=$}
      \put(48,28){$-$}
      \put(28,6.5){$+$}
      \put(60,25){$\hpb$}
      \put(73,25){$\hpb$}
     \end{picture}
      }}
   \end{picture}
\end{center}
\vspace*{-0.25ex} \caption{Lowest order Schwinger-Dyson equation for the inverse molecule propagator (double dashed line)
in the symmetric phase. The first two terms on the rhs denote the ``mean field inverse propagator'' after integrating out
the fermionic fluctuations with a dashed line for the classical inverse molecule propagator and a solid line for the fermion
propagator. The third term on the rhs accounts for the molecule fluctuations. Here $\lpb\hF$ is the molecule
self-interaction induced by the fermion fluctuations.}
\label{SDEGraphic}
\end{figure}
Since our functional integral treats the molecule fluctuations exactly on the same footing as the fermionic atom
fluctuations we can derive the lowest order Schwinger-Dyson or gap equation for $m_\phi^2$ in the standard way. For
the symmetric phase it is graphically represented in fig. \ref{SDEGraphic} and involves the full propagator and therefore
$m_{\phi, ab}^2$.

\subsection{Symmetric phase}

We will consider a Taylor expansion of $\tilde{u} = 2Mk_F^{-5} U$ in terms of the invariant
\begin{eqnarray}
\rho =k_F^{-3} \ZpR \bar{\phi}^*\bar{\phi}.
\end{eqnarray}
Correspondingly we use a renormalized Yukawa coupling and four-boson vertex,
\begin{eqnarray}
\hpR = 2M\hpb/(k_F \ZpR)^{1/2}, \quad \lambda_\phi = \tilde{\lambda}_\phi/\ZpR^2.
\end{eqnarray}
For the symmetric phase we expand
\begin{eqnarray}
\tilde{u} &=& \mpR \rhoR +\frac{1}{2} \lpR \rhoR^2 + ...\\\nonumber
\tilde{u}_{MFT} &=& \mpR \rhoR + \frac{1}{2} \lpR\hF \rhoR^2 + ...
\end{eqnarray}

The gap equation for $m_\phi^2$ takes the form
\begin{eqnarray}
m_\phi^2&=& m_\phi^{(F)\,2} + m_\phi^{(B)\,2},\\
m_\phi^{(B)\,2} &=&  2\sum\limits_n\Tn \int\frac{d^3\qn}{(2\pi)^3}\lpR\hF(\qn,\tilde{\omega}_n)
P_\phi^{-1}(\qn,\tilde{\omega}_n).\nonumber\\
\end{eqnarray}
Here $\lpR\hF$ is the effective vertex involving four bosonic fields $\sim (\phi^*\phi)^2$, as induced by the
fermion fluctuations. It depends on $\qn^2$ and $\tilde{\omega}_n$. Neglecting the $\tilde{\omega}_n$ - dependence,
i.e. replacing $\lpR\hF(\qn,\tilde{\omega}_n) \to \lpR\hF(\qn) \equiv\lpR\hF(\qn,\tilde{\omega}_n = 0)$, one
can perform the Matsubara sum
\begin{eqnarray}\label{BosonMassIntegral}
m_\phi^{(B)\,2} &=&  \int\frac{d^3\qn}{(2\pi)^3}\lpR\hF(\qn) \coth \big( \frac{\ApR \qn^2 +
m_\phi^2}{2\Tn}\big).\nonumber\\
\end{eqnarray}
We have not yet computed the momentum dependence of $\lpR\hF(\qn)$ but a simple qualitative consideration of the
relevant diagram shows that for large $\qn^2$ one has a fast decay $\lpR\hF(\qn) \propto \qn^{-4}$. This makes the
momentum integral (\ref{BosonMassIntegral}) ultraviolet finite. It will be dominated by small values of $\qn^2$. For our
purpose we consider a crude approximation where we replace $\lpR\hF(\qn) \to \lpR\hF \equiv \lpR\hF
(q=0)$. Of course, we have now to restrict the momentum integration to low momenta. This can be done efficiently by
subtracting the leading UV divergence similar as in the computation of $\nRM$, i.e. replacing $\coth x$ by
$\coth x -1$. This procedure yields
\begin{eqnarray}\label{BMIUV}
m_\phi^{(B)\,2}&=&  2\lpR\hF \int\frac{d^3\qn}{(2\pi)^3} \Big(\exp2\alpha - 1\Big)^{-1}.
\end{eqnarray}

We recognize on the rhs of eq. (\ref{BMIUV}) the expression for the number density of dressed molecules and obtain the
gap equation
\begin{eqnarray}\label{MBU}
m_\phi^2 &=& m_\phi^{(F)\,2} + \frac{\lpR\hF\Omega_M}{3\pi^2}
\end{eqnarray}
where we recall that $\Omega_M$ depends on $\mpR$. Our gap equation has a simple interpretation: The bosonic
contribution to $\mpR$ vanishes in the limit where only very few dressed molecules play a role ($\Omega_M\to 0$)
or for vanishing coupling. We are aware that our treatment of the suppression of the high momentum contributions is
somewhat crude. It accounts, however, for the relevant physics and a more reliable treatment would require a quite involved
computation of $\lpR\hF(\qn,\tilde{\omega})$. This complication is an inherent problem of gap equations which often require
the knowledge of effective couplings over a large momentum range. As an alternative method one may employ functional
renormalization \cite{Tetradis} for the crossover problem \cite{Diehl:2007th}, where only the knowledge of couplings
in a narrow momentum interval is required at every renormalization step.

The fermionic contribution to the mass term $m_\phi^{(F)\,2}$ (cf. eq. (\ref{MFTMass})) reads explicitly
\begin{eqnarray}\label{FermSYMMass}
m_\phi^{(F)\,2} &=& (\ZpR\epsilon_F)^{-1} \mpbF = \frac{\tilde{\nu} - 2\sigex}{\ZpR} +
\frac{\partial \tilde{u}_1^{(F)}}{\partial \rho}\\\nonumber
&=& \frac{\tilde{\nu} - 2\sigex}{\ZpR} - \frac{h_\phi^2}{4\Tn} \int\frac{d^3\qn}{(2\pi)^3}
\Big[\frac{1}{\gamma_\phi}\tanh\gamma_\phi -\frac{2\Tn}{\qn^2}\Big].
\end{eqnarray}
The expression in the last line has to be evaluated with $\gamma_\phi =\gamma$ in the symmetric phase.
The r.h.s. of the gap equation (\ref{MBU}) involves the coupling $\lpR\hF$ for the molecule-molecule
interactions
\begin{eqnarray}\label{147}
\lpR\hF &=& \frac{2Mk_F}{\ZpR^2}\lpb\hF =\frac{\partial^2 \tilde{u}_1\hF}{\partial \rho^2} \\\nonumber
&=& \frac{\hpR^4}{32\Tn^3}\int\frac{d^3\qn}{(2\pi)^3}
\big\{\gamma_\phi^{-3}\tanh\gamma_\phi - \gamma_\phi^{-2}\cosh^{-2}\gamma_\phi \big\}.
\end{eqnarray}
Again, in the symmetric phase $\lpR\hF$ has to be evaluated at $\rhoR =0$, i.e. $\gamma_\phi =\gamma$.

The molecule fluctuations give a positive contribution
to $\mpR$, opposite to the fermionic fluctuations. This has a simple interpretation. The fermion fluctuations
induce a self-interaction between the molecules $\sim \lpR\hF$. In turn, the fluctuations of the molecules behave
similarly to interacting fundamental bosons and modify the
two point function for the molecules. The quantum corrections to the fermionic and bosonic fluctuations to the inverse
molecule propagator are represented graphically in fig. \ref{SDEGraphic}. We emphasize that an additional microscopic
molecule interaction could now easily be incorporated by adding to the mean field value for $\lpR^{(F)}$ a
``classical part'' $\lpR^{(cl)}$. In this case the renormalization of $\bar{\nu}_\Lambda$ discussed in sect.
\ref{sec:renormalization} would be modified. For a constant $\lpR^{(cl)}$ the UV - divergent part would contribute
to $m_\phi^{2}(T=0,n=0)$. One would again end up with a contribution of the form (\ref{BMIUV}), now with $\lpR\hF$
replaced by $\lpR^{(cl)}+\lambda_\phi\hF$. Our approximation (\ref{BMIUV}) therefore treats the interactions of
dressed molecules similar to fundamental interacting bosons.


Actually, the Schwinger-Dyson equation for the molecule propagator (fig. \ref{SDEGraphic}) also describes the contribution
of molecule fluctuations to the momentum dependent part encoded in $\ApR$. In the limit of a momentum independent
$\lpb\hF$ the contribution of the boson loop to $\Apb$ vanishes in the symmetric phase.

\subsection{Superfluid phase}
\begin{figure}[t!]
\begin{minipage}{\linewidth}
\begin{center}
\setlength{\unitlength}{1mm}
\begin{picture}(85,54)
      \put (0,0){
     \makebox(80,49){
     \begin{picture}(80,49)
      \put(0,0){\epsfxsize80mm \epsffile{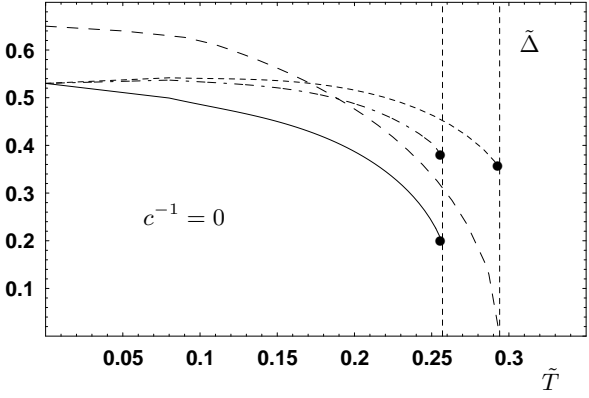}}
      \put(73,-2){$\Tn$}
      \put(70,43){$\tilde{\Delta}$ }
      \put(20,20){$c^{-1} =0$}
      \put(58.7,28.5){$\bullet$}
      \put(66.3,27){$\bullet$}
      \put(58.7,17){$\bullet$}
      \end{picture}
      }}
  \end{picture}
\end{center}
\vspace*{-1.25ex} \caption{Temperature dependence of the gap $\tilde{\Delta}=\sqrt{\rex}$ at the resonance. The role of molecule
fluctuations and the uncertainties in their treatment for $T\to T_c$ are demonstrated by four choices of $\lambda_\phi$
in the gap and density equation.  The critical
temperatures are indicated by vertical dashed lines, with values $\Tn_c = 0.259, 0.292$.
}
\label{Tdependence}
\end{minipage}
\end{figure}
In the superfluid phase we choose an expansion of $\tilde{u}$ around the minimum at
\begin{eqnarray}
\rhob_0 = r/\hpb^2=\frac{k_F^3\rex}{\hpn^2},\quad \rhoR_0 = \rex/\hpR^2,
\end{eqnarray}
namely
\begin{eqnarray}
\tilde{u} = \frac{1}{2}\lpR(\rhoR - \rhoR_0)^2 + ... 
\end{eqnarray}
and define correspondingly
\begin{eqnarray}\label{MLamNu}
\hat{m}_\phi^{(F)\,2} &=& \frac{\partial \tilde{u}_{MFT}}{\partial\rhoR}\Big|_{\rhoR_0},\quad
\lpR^{(F)} = \frac{\partial^2 \tilde{u}_{MFT}}{\partial\rhoR^2}\Big|_{\rhoR_0}.\nonumber\\
\end{eqnarray}
Our truncation of $\tilde{u}_1\hB$ approximates
\begin{eqnarray}
\tilde{u}_1\hB = \hat{m}_\phi^{(B)\,2}(\rhoR - \rhoR_0) + \frac{1}{2}\lpR\hB (\rhoR - \rhoR_0)^2 + ...
\end{eqnarray}

The location of the minimum $\rhoR_0$ is determined by the condition
\begin{eqnarray}\label{Rho0Cond}
\hat{m}_\phi^{(F)\,2} +\hat{m}_\phi^{(B)\,2} =0, \quad \hat{m}_\phi^{(B)\,2} = \frac{\partial \tilde{u}_1^{(B)}}
{\partial\rhoR}
\Big|_{\rhoR=\rhoR_0}.
\end{eqnarray}
This defines the gap equation for $\rhoR_0$, which is the equivalent of eq. (\ref{MBU}) for the superfluid phase. The
computation of the bosonic contribution $\hat{m}_\phi^{(B)\, 2}$ encounters the same problems as for $m_\phi^{(B)\, 2}$
in the symmetric phase. Again we replace $\lpR\hF(q)$ by a constant $\lpR\hF$ evaluated for $q=0$ and
subtract the leading UV divergence of the momentum integral. This results in
\begin{eqnarray}
\hat{m}_{\phi}^{(B)\,2} 
&=& 2\lpR\hF\int\frac{d^3\qn}{(2\pi)^3} \frac{\ApR\qn^2+ \lpR\rhooR/2}{\sqrt{\ApR\qn^2(\ApR\qn^2 +2\lpR\rhooR)}}\\
&&\times \big(\exp \sqrt{\ApR\qn^2(\ApR\qn^2 +2\lpR\rhooR)}/\Tn -1 \big)^{-1}.\nonumber
\end{eqnarray}
In terms of the shorthands $\alpha, \kappa,\alpha_\phi$ (\ref{DefAlpha},\ref{DefKappa},\ref{DefAlphaPhi}) we arrive at
the gap equation for $\rho_0$
\begin{eqnarray}\label{MBUB}
m_\phi^{(F)\,2} &+& 2\lpR\hF\int\frac{d^3\qn}{(2\pi)^3}
\frac{\alpha + \kappa/2}{\alpha_\phi}\big(\exp 2\alpha_\phi -1 \big)^{-1}= 0.\nonumber\\
\end{eqnarray}

The quantity $\alpha_\phi$ contains a mass term $2\lambda_\phi\rhooR$ which involves the ``full'' vertex $\lpR = \lpR\hF +
\lpR\hB$. We have computed the Schwinger-Dyson eq. for $\lpR$ in app. \ref{app:SDE}. For zero momentum $\qn\to 0$ we find
that $\lpR$ vanishes in the superfluid phase. This is, however, only part of the story since the gap equation involves a
momentum dependent vertex $\lpR(\qn)$. In order to demonstrate the uncertainty arising from our lack of knowledge of
$\lpR(\qn)$ we present in fig. \ref{Tdependence} our results for different choices of $\lpR$ in the gap eq. (\ref{MBUB}).

In detail, we show in fig. \ref{Tdependence} four approximation scenarios: (i) the ``standard'' BCS gap equation
(long dashed) neglects the molecule fluctuations, i.e.
$\lambda_\phi = \lambda_\phi^{(F)} =0$ in both the gap and the density equation. This yields a second order
transition but disagrees with QMC results for $T\to 0$. (ii) Bogoliubov density (short dashed) with $\lambda_\phi\hF$ in
the density equation, while the molecule fluctuations in the gap equation are neglected.
This improves the behavior for $T\to 0$, but induces a fake first order phase transition for $T\to T_c$.
(iii) Neglection of molecule fluctuations in the effective coupling $\lambda_\phi$ (dashed-dotted), i.e. we use
$\lambda_\phi = \lambda_\phi\hF$ in the density and gap equation. (iv) Our best estimate (solid line) includes also
corrections from molecule fluctuations for $\lambda_\phi$. As described in this section we use $\lambda_\phi$ in the propagator
of the diagram in the gap equation, whereas the coefficient multiplying the diagram is given by $\lambda_\phi\hF$. (We use
$\lambda_\phi\hF$ in the density equation.) The first order nature is weaker than in (iii), but still present. In a
recent functional renormalization group treatment we have established the second order nature of the phase transition
throughout the crossover, indicating that we now control the universal long range physics governing the phase
transition \cite{Diehl:2007th}.


\subsection{Phase transition}

For $T=T_c$ the gap equations (\ref{MBU}) and (\ref{MBUB}) match since $\rhooR = 0$, $\alpha_\phi=\alpha$,
$\kappa=0$. Also the expression for the molecule density becomes particularly simple
\begin{eqnarray}\label{UBDens}
\nRM = k_F^{3}\frac{\Gamma(3/2) \zeta(3/2)}{4\pi^2}\Big(\frac{\Tn}{\ApR}\Big)^{3/2}.
\end{eqnarray}

\section{Conclusions}
\label{sec:conclusions}

Our functional integral investigation of ultracold fermionic atoms strengthens the picture of a smooth crossover between
Bose-Einstein condensation of molecules and BCS type superfluidity. One and the same field $\phi$ can both describe molecules
and collective atom excitations of the type of Cooper pairs. The two different pictures correspond to two regions in the
space of microscopic couplings and external parameters ($T,n$). In dependence on a concentration parameter $c$, which is
related to the magnetic field $B$, one
can continuously change from one to the other region. Away from the critical temperature the ``macroscopic observables''
are typically analytic in $c^{-1}$ despite the divergence of the scattering length for two-atom scattering at $c^{-1}=0$.
Only for $T\to T_c$ one expects to encounter the non-analytic critical behavior which is characteristic for a second order
phase transition in the universality class of $O(2)$ - Heisenberg magnets.

For small and moderate Yukawa couplings $\hpn$ the physical picture of the crossover for $T>T_c$ is rather simple. Far on the BEC side the low temperature physics is
dominated by tightly bound molecules. As the concentration increases, the molecule waves start to mix with collective
di-atom states. More precisely, the molecule state with momentum $\vec{p}$ mixes with pairs of atoms with momenta
$\vec{q}_1,\vec{q}_2$ that are correlated in momentum space such that $\vec{q}_1 + \vec{q}_2 = \vec{p}$ (Cooper pairs).
On the BCS side the relevant state becomes dominantly a Cooper pair. Our formalism uses only one field $\phi$ for both
the molecule and Cooper pair states. Nevertheless, the relative importance of the microscopic molecule versus the collective
Cooper pair is reflected in the propagator of $\phi$, i.e. the wave function renormalization $\ZpR$ and the gradient
coefficient $\Apb$.

On the BEC side the propagation of $\phi$ corresponds to free molecules and $\Apb$ takes the classical
value $\Apb^{(cl)} = 1/4M$, and similarly for the wave function renormalization, $\ZpR^{(cl)} =1$. The increasing mixing in
the crossover region results in $\Apb$ and $\ZpR$ growing larger than $\Apb^{(cl)},\ZpR^{(cl)}$. Indeed, the
diagram responsible for the increase of $\Apb,\ZpR$ precisely corresponds to a molecule changing to a virtual atom pair and
back. Finally, far in the BCS regime one has $\Apb \gg \Apb^{(cl)}, \ZpR \gg \ZpR^{(cl)}$ such that the classical contribution is negligible.
The existence of a microscopic molecule is not crucial anymore. At long distances the physics looses memory of the
microscopic details - the approximation of a pointlike effective interaction between atoms becomes valid.

For a broad Feshbach resonance the situation is, in principle, similar. However, the region where $\ZpR$ and
$\Apb/\Apb^{(cl)}$ are large covers now both sides of the Feshbach resonance. The difference between the bare
(microscopic) and dressed molecules is a crucial feature. For a broad Feshbach resonance
the microscopic molecules are irrelevant for the whole crossover region. For this limit we have demonstrated in sect.
\ref{sec:renconstZR} that our formalism is equivalent to a purely fermionic model with a pointlike interaction. This
demonstrates that in the broad resonance limit, the use of the two channel model is rather a question of computational
ease than a physical issue -- it describes the same physics as a single channel model.

On the other hand, the two channel
model is more general and also covers the case of narrow resonances. This issue is discussed in more detail in
\cite{Diehl:2005an}. It remains to be seen if it will be possible to investigate the narrow resonances also experimentally
in the future.

The universality of the low temperature physics and the crossover can be traced back to the ``renormalizable couplings''
of an effective non-relativistic quantum field theory for long distances. More precisely, the long distance physics will
only depend on the relevant (or marginal) couplings in the infrared renormalization flow. These are precisely the
dimensionless parameters $c$ and $h_\phi$. Improved approximations will influence the relation between microscopic
atomic and molecular physics and $(c,h_\phi)$. At this point also ``subleading interactions'' like the $\sigma$-exchange
or other interactions contributing to the background scattering length $a_{bg}$ will play a role.

However, universality predicts that the relations
between long-distance (``macroscopic'') quantities should become computable only in terms of the ``renormalizable
couplings'' $m_\phi^2$ (or $c$) and $h_\phi$. In consequence, we find a universal phase diagram in terms of these parameters,
where the ``memory'' of the microscopic physics only concerns the values of $c$ and $h_\phi$, providing for a universal
phase diagram in terms of these parameters. In addition, the universal critical exponents and amplitude ratios for
$T\to T_c$ will be independent of $c$ and $h_\phi$. Furthermore, the BCS limit is independent of $h_\phi$ since only one
parameter ($c$) characterizes the effective pointlike atom interaction. Also the BEC limit does not depend on $h_\phi$
since the fluctuations of unbound atoms become irrelevant. On the other hand, the behavior in the crossover region can
depend on $h_\phi$ as an important universal parameter. This demonstrates that a pointlike approximation for the effective
atom interaction is not always applicable for small $|c^{-1}|$.

The couplings $h_\phi$ and $\hpn$ are related by the multiplicative factor $\ZpR^{1/2}$. In the broad resonance limit
$\hpn\to \infty$ one finds that the renormalized coupling $h_\phi$ is given by an infrared fixed point. It therefore
ceases to be an independent coupling.

For a more general form of the microscopic action we expect that deviations from universality become
most visible far in the BCS regime, where further channels may play a role for the effective interaction, as well
as in the far BEC regime, where more details of the microscopic dispersion relation for the molecules and their
microscopic interactions may become relevant. Going beyond our particular ansatz for the microscopic action
universality should actually work best in the crossover region of small $|c^{-1}|$. On the other hand, for a given
microscopic action the results are most accurate in the BEC and BCS regimes where fluctuation effects are most easily
controlled. It is precisely the presence of strong fluctuation or renormalization effects in the
crossover regime that is responsible for the ``loss of microscopic memory'' and therefore for universality!

The formulation as a Yukawa model solves the problems with a large effective scattering length for the atoms
in the crossover region in a simple way. The large scattering length is simply due to the small ``mass term''
$\bar{m}_\phi^2$ of the molecule field. It does not require a large atom-molecule interaction $\propto \hpn$. As long as
the dimensionless Yukawa or Feshbach coupling $\hpn$ remains small, the region of small and vanishing $|c^{-1}|$ (large or
diverging effective atom interaction) poses no particular problem. In the symmetric phase all quantities can be computed in
a perturbative expansion in $\hpn$. For $\tilde{h}_\phi \to 0$ a nontrivial
exact limiting solution of the functional integral has been found \cite{Diehl:2005an}. Nevertheless, for large $\hpn$ a new
strong interaction appears and the bosonic fluctuations become important. Indeed, the crossover for a broad
Feshbach resonance amounts to a theoretical challenge. One has to deal with a genuinely non-perturbative setting.

In the present paper we have included the molecule fluctuations by the solution for Schwinger-Dyson or gap equations.
For $T=0$ the results are quite satisfactory as shown by the comparison with quantum Monte Carlo simulations
\cite{Carlson03} in fig. \ref{CrossoverSnGap}. However, as $T$ increases towards the critical temperature $T_c$ our
method becomes less reliable since the details of the treatment of the molecule fluctuations play an increasing role,
cf. fig. \ref{Tdependence}. The main problem results from the neglected momentum dependence of the four-boson-vertex
$\lambda_\phi$. This is related to a general difficulty for Schwinger-Dyson equations: the momentum integrals involved
require the knowledge of effective couplings in a large momentum range. A promising alternative may be the use of
functional renormalization \cite{Tetradis}. At every renormalization step only a narrow momentum range plays a role and
the momentum dependence of suitably chosen effective vertices is much less important. First results of a functional
renormalization group treatment can be found in \cite{Diehl:2007th}.

In this light the present paper should be viewed as a starting point for more accurate investigations with further
functional integral techniques. It is well suited for systematic extensions, among which we would like to stress
a more appropriate treatment of the momentum dependence of the couplings, an extended inclusion of the effect
of boson fluctuations, and the modification of the fermion propagator by the renormalization effects originating
from mixed fermion-boson contributions.

It remains to be seen if theoretical improvements, together with a reduction of systematic experimental uncertainties, will
finally lead to an understanding of ultracold fermionic atoms with high quantitative precision.\\\\

\textbf{Acknowledgement} \\\\
We would like to thank T. Gasenzer, H. Stoof and M. Zwierlein for useful discussions.

\begin{appendix}

\section{Partial bosonization and particle density}
\label{sec:partial}

In this appendix we extend the formulation for the functional integral in such a way that we are able to
discuss situations which are (i) inhomogeneous and (ii) beyond
the ``small density approximation'' (SDA). The treatment of inhomogeneities is particularly desirable in the
context of ultracold gases which are prepared in traps of various shapes. In particular, we show how the ``local density
approximation'' emerges from our formalism. Further we consider situations beyond small densities. The price to pay is
the inclusion of a further functional integration over a now fluctuating field $\hat \sigma$. In the SDA, treating the
chemical potential as a source term is appropriate, and we will quantify this statement here.

We will start the discussion for a single fermion field. This describes a situation far off a Feshbach resonance, where
molecules of effective
collective states are unimportant. Actually, the discussion in this appendix
covers a very large class of fermionic systems with effective pointlike interactions. Besides ultracold fermionic atoms,
it may be applied to neutrons (e.g. in neutron stars), dilute gases with short range interactions (e.g. dipole
interactions) and also covers certain aspects of electron gases (where the Coulomb interaction is replaced by a pointlike
repulsion). We then extend the discussion to the case of fermions and bosons in order to account for strong
interactions via a Feshbach resonance. The concept presented here will also be technically useful for the analysis of
strongly interacting ultracold atoms in the frame of the functional renormalization group.

\subsection{Functional integral}
\label{partial2}

For an arbitrary fermionic theory with ``classical'' action
$S_F[\hat\psi]$ we define the partition function as a functional of the local source $J(x)$
\begin{eqnarray}\label{1}
Z_F[J]=\int {\cal D}\hat\psi \exp
\Big\{-S_F[\hat\psi]+\int\hspace{-0.12cm} dx J(x)\hat\psi^\dagger(x)\hat\psi(x)\Big\}.\nonumber\\
\end{eqnarray}
With $W_F[J]=\ln Z_F[J]$ the (relative) particle density becomes
\begin{equation}\label{2}
\bar{n}_\Lambda(x)=\langle\hat\psi^\dagger(x)\hat\psi(x)\rangle=
\frac{\delta W_F}{\delta J(x)}.
\end{equation}
The physical particle density is related to $\bar{n}_\Lambda(x)$ by a constant shift that we have discussed in sect.
\ref{sec:EffActMFT}. The source may be composed \footnote{The split between a constant part in $V_l$ and $\bar{\mu}_\Lambda$
is arbitrary. For interacting atoms in a homogeneous situation ($V_l\equiv 0$), $\bar{\mu}_\Lambda$ is related to the true
chemical potential $\mu$ by a constant shift (see below). For electronic systems $V_l$ typically corresponds to an
electrostatic potential.} of a bare chemical potential $\bar{\mu}_\Lambda$ and a local potential $V_l(x)$ ,
\begin{equation}\label{3}
J(x)=\bar{\mu}_\Lambda-V_l(x).
\end{equation}
For ultracold atoms the local potential $V_l(x)$ represents the trapping potential.

In this section we develop a general functional integral approach for the computation of the density $n(x)$. For this
purpose we introduce a field $\sigma(x)$ which is conjugate to $J(x)$. It will play the role of an effective chemical
potential which differs from eq. (\ref{3}) unless the density is small. Our approach can be used for arbitrary $n$ and is
not restricted to a small density approximation. The effects of density fluctuations can be incorporated by functional
integration over $\sigma$ via partial bosonization.

\subsection{Partial bosonization}
\label{Partial3}
Partial bosonization is achieved by inserting a Gaussian integral which is equivalent to one up to an irrelevant
multiplicative constant (Hubbard-Stratonovich transformation \cite{Stratonovich}, \cite{Hubbard59})
\begin{eqnarray}\label{4}
Z_F[J]&=&\int\hspace{-0.1cm} {\cal D}\hat\psi{\cal D} \hat{\sigma} \exp
\Big\{\hspace{-0.1cm}-S_F[\hat\psi]+\hspace{-0.1cm}\int\hspace{-0.12cm} dx\Big[J(x)\hat\psi^\dagger(x)\hat\psi(x)\nonumber\\
&&-\frac{1}{2}m^2\Big(\hat{\sigma}(x) -J(x)-\frac{1}{m^2}\hat\psi^\dagger(x)\hat\psi(x)\Big)^2\Big].
\end{eqnarray}
The partition function $Z_F[J]$ can now be computed from an equivalent functional
integral 
which also involves bosonic fields $\hat{\sigma}(x)$
\begin{eqnarray}\label{5A}
Z_F[J]&=&\int {\cal D}\hat\psi{\cal D}\hat{\sigma}\exp
\Big\{-S_B[\hat{\sigma},\hat\psi] \nonumber\\
&&+m^2\int\hspace{-0.12cm} dx\Big(J(x)\hat{\sigma}(x)-\frac{1}{2}J^2(x)\Big)\Big\}
\end{eqnarray}
with
\begin{equation}\label{5}
S_B=S_F[\hat\psi]+\int\hspace{-0.12cm} dx\left\{\frac{1}{2}m^2\hat{\sigma}^2-\hat{\sigma}\hat\psi^\dagger\hat\psi
+\frac{1}{2m^2}(\hat\psi^\dagger\hat\psi)^2\right\}.
\end{equation}
Here the expectation value $\sigma(x)=\langle\hat{\sigma}(x)\rangle$ is related to the density $\bar{n}_\Lambda(x)$ by
\begin{eqnarray}\label{6}
\bar{n}_\Lambda(x)&=&\frac{\delta W_F}{\delta J(x)}=m^2\Big(\sigma(x)-J(x)\Big),\nonumber\\
\sigma(x)&=&\langle\hat{\sigma}(x)\rangle=\frac{1}{m^2}
\langle\hat\psi^\dagger(x)\hat\psi(x)\rangle+J(x).
\end{eqnarray}

After partial bosonization the new action $S_B$ contains a mass term for
$\hat{\sigma}$, a Yukawa interaction $\sim\hat{\sigma}\hat\psi^\dagger\hat\psi$ and an additional four-fermion interaction
$\sim(\hat\psi^\dagger\hat\psi)^2/(2m^2)$. The resulting explicit four-fermion vertex in $S_B$ therefore becomes
\begin{eqnarray}
\bar{\lambda}_\psi = \bar{\lambda} + \frac{1}{m^2}.
\end{eqnarray}
We can choose $m^2$ such that $\bar{\lambda}_\psi$ vanishes by requiring
\begin{eqnarray}
\label{BosonCond}
\frac{1}{m^2} = -\bar{\lambda}.
\end{eqnarray}
The limit of non-interacting fermions obtains then for $m^2\rightarrow\infty$. We note that such a cancellation is possible
only for $\bar{\lambda} <0$. It is not compulsory for our formalism an we will not always assume the relation
(\ref{BosonCond}) in the following.


The partition function $Z_F$ (\ref{5A}),(\ref{5}) is closely related to the standard formulation of a
scalar-fermion-model with
\begin{equation}\label{7}
Z_B=\int {\cal D}\hat\psi {\cal D}\hat{\sigma}\exp
\Big\{-S_B[\hat{\sigma},\hat\psi]+\int\hspace{-0.12cm} dxj(x)\hat{\sigma}(x)\Big\}
\end{equation}
by $(W_B=\ln Z_B)$
\begin{equation}\label{8}
j(x)=m^2J(x), \hspace{0.2cm} W_B=W_F+
\frac{m^2}{2}\int\hspace{-0.12cm} dx J^2(x).
\end{equation}
In this formulation we may define the (one particle irreducible) effective action
$\Gamma$ by the usual Legendre transformation
\begin{equation}\label{9}
\Gamma=-W_B+\int\hspace{-0.12cm} dxj(x)\sigma(x), \hspace{0.2cm}\sigma(x)=
\frac{\delta W_B}{\delta j(x)}.
\end{equation}
For a given source $j(x)$ the expectation value $\sigma(x)$ obeys the field equation
\begin{equation}\label{10}
\frac{\delta\Gamma}{\delta\sigma(x)}=j(x).
\end{equation}
The effective action may be decomposed into a classical part
$\Gamma_{cl}[\sigma]=S_B[\hat{\sigma}=\sigma,~\psi=0]$ and a quantum part $\Gamma_q$
\begin{equation}\label{11}
\Gamma=\Gamma_{cl}+\Gamma_q=
\frac{m^2}{2}\int\hspace{-0.12cm} dx \sigma^2(x)+\Gamma_q[\sigma].
\end{equation}

We next turn to the relative particle density $\bar{n}_\Lambda(x)$. It can be computed from $\Gamma$ by
decomposing the field equation
\begin{eqnarray}\label{12}
\sigma(x)&=&J(x)+\sigma_1(x),\nonumber\\
\frac{\delta\Gamma}{\delta\sigma(x)}
&=&m^2\sigma(x)+\frac{\delta\Gamma_q}{\delta\sigma(x)}=m^2J(x)
\end{eqnarray}
as
\begin{equation}\label{13}
\bar{n}_\Lambda(x)=m^2\sigma_1(x)=-
\frac{\delta\Gamma_q}{\delta\sigma(x)}
\Big(J(x)+\sigma_1(x)\Big).
\end{equation}
This establishes the exact general relation between $\bar{n}_\Lambda(x)$ and the $\sigma$ - functional derivative of the quantum part
of the effective action.

The limit $|\sigma_1|\ll|J|$ corresponds (for fixed $J(x)$) to small densities or
small interactions $(m^2\rightarrow \infty)$ according to
\begin{equation}\label{14}
|\bar{n}_\Lambda(x)|\ll m^2|J(x)|.
\end{equation}
In this limit we may expand
\begin{equation}\label{15}
\frac{\delta\Gamma_q}{\delta\sigma(x)}
\Big(J(x)+\sigma_1(x)\Big)
=-n_0(x)+b(x)\sigma_1(x)+\dots
\end{equation}
with
\begin{equation}\label{16}
n_0(x)=-
\frac{\delta\Gamma_q}{\delta\sigma(x)}\Big|_{J(x)}~,~b(x)=
\frac{\delta^2\Gamma_q}{\delta^2\sigma(x)}\Big|_{J(x)}.
\end{equation}
This yields
\begin{equation}\label{17}
\bar{n}_\Lambda(x)=
\frac{n_0(x)}{1+b(x)/m^2}.
\end{equation}
We emphasize, however, that eq. (\ref{13}) remains valid for arbitrary densities.

The explicit computation of $\bar{n}_\Lambda(x)$ needs an evaluation of $\Gamma_q$. In mean field theory the
fluctuation part $\Gamma_q$ is estimated by including only the fermionic fluctuations in the functional integral
(\ref{7}), while keeping $\hat{\sigma}(x)=\sigma(x)$ as a fixed ``background''. In this scheme, the generalization of eq.
(\ref{BarNFMFT}) for the relative particle density reads, according to eq. (\ref{13})
\begin{equation}\label{23}
\int d^3x\bar{n}_\Lambda(x)=-\int d^3x\frac{\delta\Gamma_q}{\delta\sigma(x)} =-\frac{1}{2}\tilde{\text{Tr}}\tanh
\left(\frac{\bar{P}_F-\sigma}{2T}\right)
\end{equation}
where the remaining  trace $\tilde{\text{Tr}}$ is over three dimensional phase space and spinor indices $\alpha$ -
in momentum space it reads $\tilde{\text{Tr}} \hat{=}V_3\int \frac{d^3q}{(2\pi)^3}\sum_\alpha$, with $V_3$ the volume
of (three-dimensional) space. At this point the construction from sect. \ref{sec:renormalization} can be performed,
replacing the relative ($\bar n_\lambda$) by the physical particle number density $\bar n_\lambda + \hat n$. This yields
\begin{equation}\label{26}
N=\int d^3x(\bar{n}_\Lambda+\hat{n})=\tilde{\text{Tr}}
\Big(\exp\Big(\frac{\bar{P}_F-\sigma}{T}\Big)+1\Big)^{-1}.
\end{equation}
This formula has a simple interpretation. The trace $\tilde{\text{Tr}}$ over an operator $\hat{A}$ can be evaluated in a
complete orthonormal system of complex functions $f_m(x)$,
\begin{eqnarray}\label{27}
&&\int d^3xf^*_{m'}(x)f_m(x)=\delta_{m'm},\nonumber\\
&&\int d^3xf^*_{m'}(x)\hat{A}f_m(x)=A_{m'm},\nonumber\\
&&\tilde{\text{Tr}}\hat{A}=\sum_mA_{mm}.
\end{eqnarray}
With $\sigma(x)=\mu-\hat{V}(x)$ (in analogy to eq. (\ref{3})) we can define the Hamilton operator
$\hat{H}=\bar{P}_F+\hat{V}$ (with $P=-\Delta/(2M)$ for nonrelativistic atoms). Choosing the
$f_m$ to be eigenfunctions of the Hamiltonian with eigenvalue $E_m$ eq. (\ref{26}) becomes
\begin{equation}\label{28}
N=\sum_m\left(\exp
\left(\frac{E_m-\mu}{T}\right)+1\right)^{-1}
\end{equation}
and we recognize the well known fermionic occupation number. In the limit of noninteracting atoms ($m^2\rightarrow\infty)$
we can cut the Taylor expansion in eq. (\ref{15}) at the lowest term, $\bar{n}_\Lambda=n_0$. In consequence,
$\delta\Gamma_q/\delta\sigma$ is evaluated with the local potential $V_l$ in eq. (\ref{3}). In this limit one has $\sigma =
\mu - V_l$ and $\hat{V}$ therefore equals the trap potential $V_l$. As it should be the Hamiltonian
$\hat{H}$ reduces to the quantum Hamiltonian of a single atom in a potential and the density becomes independent of $m^2$.

\subsection{Universal field equation for the density}
\label{universalfieldequation}

The crucial advantage of our formalism is the possibility to compute the effective action
$\Gamma[\sigma]$ without any specification of the local potential (trapping potential)
$V_l(x)$. The detailed geometry of the trap only enters at a second step when one solves
the field equation (\ref{10}). This offers the
great advantage that the fluctuation problem can be solved in a homogeneous setting, i.e. one encounters standard
momentum integrals in the loop expressions rather than summations over spherical harmonics or other function systems
adapted to the geometry of the trap. The results are universal and apply to all geometries, provided the inhomogeneity
is sufficiently weak.

For $\sigma(\vec{x})$ mildly varying in space and independent of $\tau$ one may use a derivative expansion
\begin{equation}\label{29}
\Gamma[\sigma]=\int\hspace{-0.12cm} dx\left\{U_\sigma(\sigma)+ \frac{1}{2}A_\sigma(\sigma)\vec{\nabla}\sigma\vec{\nabla}\sigma+\dots\right\}.
\end{equation}
We emphasize that the derivative expansion always applies for situations close enough
to homogeneity. The following results are therefore valid independently of the way how
$\Gamma[\sigma]$ is computed - in particular, they do not rely on MFT. However, for definiteness, let us give the
result for $A_\sigma$ in the latter scheme ($U(\sigma)$ can be read off from eq. (\ref{USigmaPhi}) for
$\gamma_\phi=\gamma$). For this purpose, we proceed along the lines in sect. \ref{GradCoeff} and extract from
$\Gamma_{q,MFT}$ the term quadratic in $\delta\sigma_Q$
\begin{eqnarray}\label{40}
\Gamma^{(2)}_{q,MFT}&=&\frac{V_3}{T}\delta\sigma^*_Q
P_\sigma(Q)\delta\sigma_Q~,\\\nonumber
P_\sigma(Q)&=&
\int\limits_{Q'}\Big\{\frac{1}{(P_F(Q')-\sigma)(P_F(Q'+Q)-\sigma)}\\\nonumber
&& + \frac{1}{(P_F(Q')-\sigma)(P_F(Q'-Q)-\sigma)}\Big\}
\label{41}
\end{eqnarray}
with $\int_{Q'} = \sum_{n'} T \int \frac{d^3q'}{(2\pi)^3}$. Inserting the background $\sigma(X)=
\sigma+\exp(iQX)\delta\sigma_Q + \exp(-iQX)\delta\sigma^*_Q$ into eq. (\ref{29}) yields
\begin{eqnarray}\label{42}
&&A_\sigma(\sigma)=\lim\limits_{q^2\rightarrow 0} \hspace{0.5cm}\frac{\partial}{\partial q^2}P_\sigma(0,q)\\\nonumber
&&=\frac{1}{4MT^2}\int\frac{d^3q}{(2\pi)^3}\left\{\frac{ \tanh~\gamma}{ \cosh ^2\gamma}-\frac{q^2}{9MT}
\frac{2 \cosh ^2\gamma-3}{ \cosh ^4\gamma}\right\}.
\end{eqnarray}
We find that  $A_\sigma(\sigma)$ is ultraviolet finite. The zero temperature limit of this expression reads
\footnote{The prefactor has been determined numerically with an error of less than $0.3\%.$} (for $\sigma >0$)
\begin{eqnarray}\label{Zsigma0}
A_\sigma = \frac{1}{12\pi^2}\sqrt{\frac{2M}{\sigma}},
\end{eqnarray}
and diverges for $\sigma\to 0$.

\subsubsection{Density field equation}
We are interested in time-independent situations and therefore consider $\sigma(x) = \sigma(\vec{x})$ independent of $\tau$.
Variation with respect to $\sigma$ yields the stationary field equation
(with $U_\sigma'=\partial U_\sigma/\partial\sigma$ etc.)
\begin{equation}\label{30}
U_\sigma'(\sigma)-A_\sigma(\sigma)\triangle\sigma-
\frac{1}{2}A'_\sigma(\sigma)\vec{\nabla}\sigma\vec{\nabla}\sigma=m^2(\bar{\mu}_\Lambda-V_l)
\end{equation}
where $\sigma$ and $V_l$ depend on $\vec{x}$.
Once $\sigma(\vec{x})$ is found one can compute the particle density from
\begin{eqnarray}\label{31}
n&=&-U'+A_\sigma\Delta\sigma+\frac{1}{2}
A'_\sigma\vec{\nabla}\sigma\vec{\nabla}\sigma\nonumber\\
&=&m^2\Big(\sigma-\bar{\mu}_\Lambda+V_l+\frac{\hat{n}}{m^2}\Big),\\
U&=&U_\sigma-\frac{1}{2}m^2\sigma^2-\hat{n}\sigma.
\end{eqnarray}
Here we have chosen the definition of $U$ such that the physical particle density from eq. (\ref{25}) is
reproduced in eq. (\ref{31}).

In the second line of eq. (\ref{31}) appears a shifted or ``additively renormalized chemical potential'',
\begin{equation}\label{33}
\mu=\bar{\mu}_\Lambda-\frac{\hat{n}}{m^2}.
\end{equation}
This combination is the true ``physical'' chemical potential as can be seen along the lines of eqs. (\ref{24},\ref{25}).
Indeed, the interaction term in the operator language $\propto (a^\dagger (x) a (x))^2$ is translated to the interaction term
in the functional integral (\ref{1}) for the microscopic action  by the use of
\begin{eqnarray}\label{33X}
\frac{1}{2m^2}\left(a^\dagger(x)a(x)\right)^2 \rightarrow \frac{1}{2m^2} \left(\hat\psi^\dagger(x)\hat\psi(x) + \hat{n}\right)^2.
\end{eqnarray}
The term $(\hat{n}/m^2)\hat\psi^\dagger(x)\hat\psi(x)$ precisely shifts $\mu$ to $\bar{\mu}_\Lambda$ in the functional integral (\ref{1}).
A detailed account for a similar ``chemical potential shift'' for bosonic atoms can be found in \cite{Gollisch01}.

Hence the second part of eq. (\ref{31}) yields a linear relation between the physical particle density $n$ and $\sigma$,
\begin{equation}
\sigma=\mu-V_l+\frac{n}{m^2}.\label{34.2}
\end{equation}
This will be our central equation relating the density $n$, the effective chemical potential $\sigma$, the chemical
potential $\mu$ and the local trap potential $V_l$. We observe that a great part of the ``ultraviolet divergencies''
(for $\Lambda \to \infty$) present in the functional integral description is related to $\bar{\mu}_\Lambda$ and $\hat{n}$.
These divergencies do not appear if the ``physical quantities'' $\mu$ and $n(x)$ are used.

Eq. (\ref{34.2}) can now be used to eliminate $\sigma$ in favor of the physical particle density. Insertion into the first
eq. (\ref{30}) yields the central field equation for the density in an inhomogeneous situation,
\begin{eqnarray}\label{34}
&&n-\frac{1}{m^2}A_\sigma\left(\mu-V_l + \frac{n}{m^2}\right)\triangle n\nonumber\\
&&\qquad\quad+\frac{1}{m^2}A'_\sigma(\mu-V_l + \frac{n}{m^2})
\left(\vec{\nabla}V_l-\frac{1}{2m^2}\vec{\nabla}n\right)\vec{\nabla}n\nonumber\\
&&\qquad=-U'(\mu-V_l + \frac{n}{m^2})
-A_\sigma(\mu-V_l + \frac{n}{m^2})\Delta V_l\nonumber\\
&&\qquad\quad+\frac{1}{2}A'_\sigma(\mu-V_l + \frac{n}{m^2})\vec{\nabla}V_l\vec{\nabla}V_l.
\end{eqnarray}
Once the functions $U'(\sigma)$ and $A_\sigma(\sigma)$ have been computed this
equation permits the determination of $n(\vec{x})$ as a function of $\mu$.
For its solution the boundary conditions of $n(\vec{x})$ have to be specified appropriately
- for trapped atoms $n(\vec{x})$ has to vanish away from the trap. In practice, one
often does not know $\mu$ in a given experimental situation, but rather has information
about the total particle number $N=\int d^3xn(x)$. In this case one may compute $N(\mu)$
by integrating the solution of eq. (\ref{34}). Our setting can be reformulated in a more intuitive form in terms of a
``density functional'' $\hat{\Gamma}[n]$ to which we proceed in the next section.

\subsubsection{Density Functional}
\label{app:DensFunc}
We can write the effective action in a more intuitive form as a functional of the particle density. It is
convenient to add the source explicitly such that the field equation becomes
\begin{eqnarray}
\label{FieldEqGeneral}
\frac{\delta \hat{\Gamma}}{\delta n(x)} = 0.
\end{eqnarray}
Using eq. (\ref{34.2}) one has
\begin{eqnarray}
\label{FreeEnergy}
\hat{\Gamma}[n] 
&=& \Gamma[\sigma] + \int\hspace{-0.12cm} dx [m^2(V_l - \mu) - \hat{n}]\sigma \\\nonumber
&=& \int\hspace{-0.12cm} dx\left\{U( \frac{n}{m^2}+\mu-V_l )+\frac{n^2}{2m^2} \right.\\\nonumber
 && \quad\left.   +\frac{1}{2}A_\sigma(\frac{n}{m^2}+\mu-V_l )\left(\vec{\nabla}(\frac{n}{m^2} - V_l)\right)^2 +...\right\},
\end{eqnarray}
where again a density independent term has been dropped. The explicit form of eq. (\ref{FieldEqGeneral}) is given by eq.
(\ref{34}).

In particular, a homogeneous situation is governed by the effective potential
\begin{eqnarray}
U_n(n)= \frac{n^2}{2m^2} + U(n).
\end{eqnarray}
In this case we can use $n$ as the independent variable, replacing $\mu$. In turn, the chemical potential $\mu (n)$ follows
from the minimum condition $\partial U_n/\partial n =0$. In practice, one first solves for $\sigma (n)$ by inverting
\begin{eqnarray}
\label{DensSelfCons}
U'(\sigma) = - n.
\end{eqnarray}
In MFT this step does not depend on $m^2$. The interaction
strength $m^{-2}$ only enters the determination of $\mu$ through eq. (\ref{34.2}). Similarly, in the local density
approximation we can trade $\mu$ for the density at a given reference location $x_0$, $n_0=n(x_0)$. Using again eq.
(\ref{DensSelfCons}), the computation of the density profile employs
\begin{eqnarray}\label{SigmaDensRelat}
\sigma(x)= \sigma (n_0) + V_l(x_0) - V_l(x) + \frac{n(x)-n_0}{m^2}.
\end{eqnarray}
For a given central density $n_0$ (e.g. $x_0=0$) and given trap potential $V_l(x) - V_l(0)$ we can now evaluate $U'$
as a function of $n$ and determine $n(x)$ from $U'(\sigma (n))=-n$.

On the other hand, if $n(x)$ is known we may use a functional of the variable $\sigma$ for fixed $n(x)$. Using
\begin{eqnarray}
\frac{\delta\Gamma}{\delta\sigma}= m^2\sigma - n + \hat{n}
\end{eqnarray}
we may define
\begin{eqnarray}
\bar{\Gamma} &=& \Gamma + \int dx \Big\{ \big(n-\hat{n}\big) \sigma - \frac{m^2}{2} \sigma^2\Big\},\nonumber\\
\frac{\delta\bar{\Gamma}}{\delta\sigma}&=&0.
\end{eqnarray}
Here $n(x)$ is considered as a fixed function. The corresponding potential
\begin{eqnarray}
\bar{U} = U_\sigma - \frac{m^2}{2}\sigma^2 + \big(n-\hat{n}\big) \sigma = U + n\sigma
\end{eqnarray}
is particularly useful for a homogeneous situation where $n$ is a fixed constant. The solution for $\sigma (n)$ corresponds
to the minimum of $\bar{U}$.

\subsubsection{Small and local density approximations}
In the \emph{small density approximation} (SDA) $\sigma$ is given by
\begin{eqnarray}
\sigma= \mu - V_l(x).
\end{eqnarray}
We can now specify the validity of the small density limit more precisely and replace the condition (\ref{14}) by
\begin{equation}\label{35}
n\ll m^2|\mu-V_l|.
\end{equation}
In lowest order in the small density approximation  one obtains the density of
non-interacting fermions in a local potential as
\begin{eqnarray}\label{36}
n&=&-U'(\mu-V_l)
-A_\sigma(\mu-V_l)\Delta V_l\nonumber\\
&&+\frac{1}{2}A'_\sigma(\mu-V_l)
\vec{\nabla}V_l \vec{\nabla}V_l.
\end{eqnarray}
We emphasize that eq. (\ref{34}) and its low density limit (\ref{36}) are rather universal formulae which determine the
density of fermions in a local potential. Their possible applications reach far beyond the particular problem of ultracold
trapped atoms.

Another useful approximation, the \emph{local density approximation} (LDA), obtains by setting $A_\sigma = 0$ in eq. (\ref{34}),
\begin{eqnarray}\label{LDAEq}
n&=&-U'(\mu_l + \frac{n}{m^2}),
\end{eqnarray}
where we define the ``local chemical potential''
\begin{eqnarray}
\mu_l(x) = \mu - V_l(x).
\end{eqnarray}
Its validity does not require a small density but rather that the derivative terms in eq. (\ref{34}) (or (\ref{29})) are
small as compared to $U'$. For a given size  of $A_\sigma$ this always applies for a sufficiently homogeneous trap.
Indeed, the local character of eq. (\ref{LDAEq}) guarantees that weak changes in $\mu_l(x)$ result in weak changes of
$n(x)$. The error of the LDA can now easily be estimated by an iterative procedure. Inserting the solution of eq.
(\ref{LDAEq}) into eq. (\ref{34}) allows for a simple direct computation of the subsequent terms. Obviously, the error
depends on the size of $A_\sigma$ which is often a rather small quantity (see below).

If both the density is sufficiently small and the trap sufficiently homogeneous one can work with the \emph{small local
density approximation} (SLDA). This results in the simple formulae
\begin{equation}
\label{nUeq}
n = - \frac{\partial U}{\partial\sigma}
\end{equation}
and
\begin{equation}\label{SigMuEq}
\sigma = \mu_l.
\end{equation}
In this approximation we can regard $\sigma(x)$ as a fixed external parameter and compute $n(x)$ by eq. (\ref{nUeq}). We
will find that for realistic ultracold atom systems like $^6\mathrm{Li}$ the LDA is valid whereas eqs. (\ref{nUeq}),
(\ref{SigMuEq}) are oversimplifications (except for the BEC regime far away from the Feshbach resonance).

We emphasize again that our method is valid quite generally and not bound to the case of small density. If necessary,
mean field theory can be improved by including the bosonic fluctuations in the computation of $U$ and $A_\sigma$ by
performing the functional integral over $\hat{\sigma}$. Our method can also be applied in the presence of additional
condensate fields. As a simple application we compute in appendix \ref{sec:Meta} the density at low
$T$ in the mean field approximation. In particular, this shows that the dilute gas of ultracold atoms is a metastable
state, the ground state being a liquid or solid.

\subsection{Fermions and Bosons}

We can proceed in complete analogy to sects. \ref{partial2}, \ref{Partial3} with $Z_F$ replaced by
\begin{eqnarray}\label{ZFM}
Z_{FM}[J]&=&\int\hspace{-0.1cm} {\cal D}\hat\psi{\cal D} \hat{\phi} \exp
\Big\{\hspace{-0.1cm}-S_{FM}[\hat\psi,\hat{\phi}]\\\nonumber
&&+\hspace{-0.1cm}\int\hspace{-0.12cm} dx\Big[J(x)\big(\hat\psi^\dagger(x)\hat\psi(x)+2\hat{\phi}^*(x)\hat{\phi}(x)\big)\Big]\Big\},
\end{eqnarray}
and
\begin{equation}
\frac{\delta \ln Z_{FM}}{\delta J(x)}=n(x),\quad J(x) = \mu +(1-\tilde{\beta})\frac{\hat{n}}{m^2}-V_l(x).
\end{equation}
The classical action reads now
\begin{eqnarray}
S_{FM}\hspace{-0.12cm} &=&\hspace{-0.12cm}\int \hspace{-0.12cm} dx \Big[\hat\psi^\dagger\big(\partial_{\tau}
-\frac{\triangle}{2M} \big)\hat\psi-
 \frac{1}{2}\big(\frac{1}{m^2} - \bar{\lambda}_\psi\big)\big( \hat\psi^\dagger\hat\psi\big)^2\nonumber\\
&&\hspace{-1.5cm}+\hat{\phi}^*\Big(\hspace{-0.05cm}\partial_\tau\hspace{-0.12cm} -\hspace{-0.05cm}\frac{\triangle}{4M}\hspace{-0.05cm}
 +\hspace{-0.05cm} \bar{\nu}_\Lambda +\frac{2\hat{n}}{m^2}\big(1-2\tilde{\beta}+\beta\big) + V_M(x)
-2V_l(x)\Big)\hat{\phi} \nonumber\\
&&\hspace{-1.5cm}-\frac{2\beta}{m^2}\big(\hat{\phi}^*\hat{\phi}\big)^2 -\frac{2\tilde{\beta}}{m^2}\big(\hat\psi^\dagger\hat\psi\big)\hat{\phi}^*\hat{\phi}
-\frac{\hpb}{2}\big(\hat{\phi}^*\hat\psi^T\epsilon\hat\psi - \hat{\phi}\hat\psi^\dagger\epsilon\hat\psi^*\big) \Big]\nonumber\\
\end{eqnarray}
and the partition function describes the coupled system of atoms in a local potential $V_l$ and molecules
in a ``molecule potential'' $V_M$. (The term $\propto -2V_l\hat{\phi}^*\hat{\phi}$ in $S_{FM}$ cancels the corresponding term
from $-2J\hat{\phi}^*\hat{\phi}$.) The atoms are coupled to the molecules by the Yukawa coupling $\hpb$.
Again, the bare parameters $\bar{\nu}_\Lambda$, $\hpb$ and $m^2$ have to be fixed by appropriate
observable ``renormalized'' parameters. We have also included a possible local self-interaction of the molecules $\sim
2\beta/m^2$ and between free atoms and molecules $\sim 2\tilde{\beta}/m^2$. We concentrate on $V_M= 2V_l$ and $\beta=\tilde{\beta}=1$. This simply counts the molecules
as two atoms as far as the local interactions are concerned, e.g. the local self-interaction is $\sim (\nF+ 2\nM)^2$ and
the energy in the trap potential is $\sim \nF + 2\nM$. We note that in this particular case $\hat{n}$ drops out.
(The corrections to $\mu$ are computed similar  to eqs. (\ref{33},\ref{33X}).)

Replacing in the Hubbard-Stratonovich transformation (\ref{4}) $\hat\psi^\dagger\hat\psi\to \psi^\dagger\psi + 2\hat{\phi}^*
\hat{\phi}$ all steps in sec. \ref{universalfieldequation} can be performed in complete analogy, with the only difference that
$Z_B$ in (\ref{7}) involves now also an integration over $\hat{\phi}$ and an appropriate source for $\hat{\phi}$, and
$S_B$ reads (for $m^{-2} = \bar{\lambda}_\psi$)
\begin{eqnarray}\label{YukawaAction}
S_B\hspace{-0.15cm}&=&\hspace{-0.15cm}\int \hspace{-0.12cm} dx \Big[\psi^\dagger\big(\partial_{\tau}
-\frac{\triangle}{2M} -\hat{\sigma}\big)\psi\nonumber\\
&&+\hat{\phi}^*\big(\partial_\tau -\frac{\triangle}{4M} + \bar{\nu}_\Lambda - 2\hat{\sigma}\big)\hat{\phi} \nonumber\\
&&\hspace{-0.12cm}-\frac{\hpb}{2}\Big(\hat{\phi}^*\psi^T\epsilon\psi - \hat{\phi}\psi^\dagger\epsilon\psi^*\Big) + \frac{m^2}{2}\hat{\sigma}^2\Big].
\end{eqnarray}
This is a simple model for fermions with Yukawa coupling to scalar fields $\hat{\phi}$ and $\hat{\sigma}$.

\subsubsection{Effective action}
The effective action $\Gamma
[\sigma,\bar{\phi}]$ obtains 
again by a Legendre transform similar to eq. (\ref{9}) and obeys now 
\begin{eqnarray}
\frac{\delta\Gamma}{\delta\sigma}= m^2\mu_l = m^2\sigma- n.
\end{eqnarray}
In particular, a homogeneous situation is characterized by an extremum of $\bar{U} = U_\sigma - (m^2/2)\sigma^2 + n\sigma$,
\begin{eqnarray}
\label{Ubar}
\bar{U}=n \sigma - \frac{\hpb^2M}{4\pi a} \bar{\phi}^*\bar{\phi} - 2\sigma\bar{\phi}^*\bar{\phi} +U_1 = U + n\sigma .
\end{eqnarray}
Similar to sect. (\ref{sec:ContribMolFluct}) we may proceed beyond the Gaussian functional integration for $\hat\psi$ (MFT) by
adding to $U_1$ a piece from the one loop contribution from the $\hat\phi$-fluctuations
\begin{eqnarray}
U_1&=& U_1^{(F)}+U_1^{(B)}.\label{BosonicPot}
\end{eqnarray}

The contributions from dressed uncondensed molecules is now given by
\begin{eqnarray}\label{nRenorm}
\frac{\partial U_1^{(B)}}{\partial\sigma}=-2\nRM
\end{eqnarray}
Indeed, eq. (\ref{nRenorm}) follows from eq. (\ref{DressedMol}) if $\mu$ is replaced in the classical bosonic propagator by the effective
chemical potential $\sigma$.

\subsubsection{Field equations}
Collecting the different pieces the extremum of $\bar{U}$ (eq. (\ref{Ubar})) occurs for
\begin{eqnarray}
\frac{\partial \bar{U}}{\partial \sigma} = 0 = n - 2\bar{\phi}^*\bar{\phi} - n_{F,0} - 2\nRM.
\end{eqnarray}
This is precisely the relation (\ref{RenDensConstr}) as it should be. (Recall $n_{F,0} = -\partial U_1\hF/\partial \sigma$.)
In other words, the density obeys
\begin{equation}
n= - \frac{\partial U }{\partial \sigma}
\end{equation}
With $J=\mu$ we may derive
the relation between the effective chemical potential $\sigma$ and the chemical potential $\mu$
\begin{eqnarray}
\mu =\sigma - n/m^2
\end{eqnarray}
directly from the identity
\begin{eqnarray}
Z_B^{-1}\int\mathcal{D}\hat\psi\mathcal{D}\hat\sigma\mathcal{D}\hat{\phi} \frac{\delta}{\delta\hat\sigma}\exp
\big\{-S_B + m^2 \int J\hat\sigma\big\} = 0.\nonumber\\
\end{eqnarray}
The evaluation of $U(\sigma,
\phi)$ is now sufficient for the computation of the total atom density $n$. For the homogeneous setting one can therefore
determine $\sigex$ by
\begin{equation}\label{DensHom}
 \frac{\partial \tilde{u} }{\partial \sigex} = -\frac{1}{3\pi^2}.
\end{equation}

\subsubsection{Effective chemical potential}
In summary, our problem is now formulated as a functional integral for a Yukawa model. In the small density approximation
we can treat $\sigma (x) = \mu + V_l(x)$ as an external parameter. The partition function becomes
\begin{eqnarray}\label{ZDecoup}
Z = \int \mathcal{D}\hat\psi\mathcal{D}\hat{\phi} \exp \big[- S_B[\hat\psi, \hat{\phi}; \sigma] + \mathrm{ source \, terms}\big]
\end{eqnarray}
where $S_B$ is given by eq. (\ref{YukawaAction}) and the density obtains from eq. (\ref{DensHom}). Beyond the small density
approximation $\sigma$ is treated as a field and the partition function involves an additional integration over
$\hat{\sigma}$. In the limit where the $\hat{\sigma}$ - fluctuations can be neglected we may consider the effective
chemical potential $\sigma$ instead of $\mu$ as a free ``external parameter''. In particular, this offers for the
homogeneous case the advantage that we are not bound to the validity of the small density approximation (which may not be
accurate for realistic systems as $^6\mathrm{Li}$). It is sufficient to compute the value of $\sigex (\Tn)$ for a given
density or given $k_F$.

For the homogeneous case we can both deal with situations at fixed effective chemical potential $\sigma$ or at fixed
density $n$. For a fixed $\sigma$ we may choose an arbitrary fiducial $\bar{k}_F$ (not determined by $n$) and do all the
rescalings with $\bar{k}_F$ (instead of $k_F$). One can then work with a fixed value of $\sigex$ and compute $n$ or $k_F$
by the relation
\begin{eqnarray}
\frac{\partial\tilde{u}}{\partial \sigex} = -\frac{1}{3\pi^2} \frac{k_F^3}{\bar{k}_F^3}.
\end{eqnarray}
Many experimental settings can be idealized, however, by a fixed value of $n$. Then the choice $\bar{k}_F = k_F$ and the
determination of $\sigex$ via eq. (\ref{DensHom}) provides directly all results for fixed $n$. For an inhomogeneous
setting it is sensible to use a suitable fiducial $\bar{k}_F$.

\section{Metastability}
\label{sec:Meta}

At low $T$ the thermal equilibrium state of atoms is a liquid or solid, with density essentially determined by the ``size''
of the atoms (typically set by the van der Waals length). When we deal with a dilute gas of atoms at ultracold
temperature we obviously do not consider the stable
thermal equilibrium state which minimizes the free energy. Indeed, we will see in this section that the metastability can
be captured within our formalism if the weak cutoff dependence $\mathcal{O}(\Lambda^{-1})$ is not neglected. For this
purpose, we consider a homogeneous system and neglect the effects from molecules.

Under the above circumstances the field equation (\ref{34}) reduces to a simple self-consistency relation for the density,
\begin{eqnarray}\label{43}
n&=&-U'\left(\mu+\frac{n}{m^2}\right)\nonumber\\
&=&2\int\frac{d^3q}{(2\pi)^3}\frac{1}{\mathrm{e}^{(q^2/(2M)-\mu-n/m^2)/T}+1}
\end{eqnarray}
For the gross features we first consider the $T\to 0$ limit where the last equation reduces to
\begin{equation}\label{44}
n=\frac{1}{3\pi^2}\left\{2M\left(\mu+\frac{n}{m^2}\right)\right\}^{3/2}.
\end{equation}
The resulting cubic equation for $y=n/(\mu m^2)$
\begin{equation}\label{45}
(y+1)^3-\kappa y^2=0~,~\kappa=\frac{9\pi^4m^4}{8M^3\mu}
\end{equation}
may have several solutions, depending on $\kappa$. We concentrate on $\mu>0$ and consider first $\kappa\gg 1$. The solution
with small $y\approx\kappa^{-1/2}$,
\begin{equation}\label{46}
n_1=\frac{1}{3\pi^2}(2M\mu)^{3/2}
\end{equation}
describes a dilute gas of atoms. In lowest order the relation between $n$ and $\mu$
is indeed independent of $m^2$ and $\Lambda$. For large $\kappa$ the second solution with positive $y$ occurs
for $y\approx\kappa$
\begin{equation}\label{47}
n_2=\frac{9\pi^4m^6}{8M^3}
\end{equation}
while the third solution has negative $y$ and should be discarded. As $\kappa$ decreases
(i.e. $\mu$ increases) the two solutions approach each other and melt for
$\kappa_c=27/4~,~y_c=2$ or
\begin{equation}\label{48}
n_c=2\mu
m^2=\frac{\pi^4m^6}{3M^3},
\end{equation}
which delimits the metastable gas phase. No solution with $y>0$ exists for $\kappa<\kappa_c$. The stability of the solution
depends on the second derivative of $U_n$ with respect to $n$, i.e. for $T=0$
\begin{equation}\label{49}
\frac{\partial^2U_n}{\partial n^2}=\frac{1}{m^2}
\left\{1-\frac{3}{2}\left(\frac{1+y}{\kappa}\right)^{1/2}\right\}.
\end{equation}
The solution $n_1$ (with small $y$) turns out to be stable $(\partial^2U_n/\partial n^2>0)$ whereas the solution $n_2$
(with $y\approx\kappa)$ is unstable. For $\kappa_c$ one has $\partial^2U_n/\partial n^2=0$.

What happens for $n>n_c$? In order to study this question we cannot neglect the
ultraviolet cutoff $\Lambda$ anymore. In the evaluation of eq. (\ref{43}) the upper
bound of the momentum space integral is actually given by $q_{max}=\Lambda$ 
instead of infinity. For $\Lambda<(2M(\mu+n/m^2))^{1/2}$ this multiplies the r.h.s of eq. (\ref{44}) by an additional
factor $\Lambda^{3}/(2M(\mu+n/m^2))^{3/2}$ such that
\begin{equation}\label{50}
n=\frac{\Lambda^3}{3\pi^2} \hspace{0.2cm} \text{for} \hspace{0.2cm}
\frac{n}{m^2}+\mu>\frac{\Lambda^2}{2M}.
\end{equation}
This solution is again stable with $\partial^2U_n/\partial n^2=1/m^2$. We associate it with the liquid or solid thermal
equilibrium state. Indeed, the free energy for $n_{gs}=\Lambda^3/(3\pi^2)$ is much lower than for the solution $n_1$. Also
the density is determined by the effective size of the atoms $\sim\Lambda^{-1}$. Inversely, we can define our cutoff by
the density
$n_{gs}$ of the liquid or solid ground state for $T=0$, $\Lambda = (3\pi^2n_{gs})^{1/3}$. Of course, in this region our
approximations are no longer valid, but the detailed properties of the liquid or solid state are not the purpose of this
paper.

Beyond the
limit $T\to 0$, a computation of the phase boundary for the existence of a metastable gas for arbitrary temperature requires
the solution of the condition for the melting of the stable and unstable extrema,
\begin{eqnarray}
\label{ncrit}
0&\stackrel{!}{=}&\frac{\partial^2U_n}{\partial n^2}=\frac{1}{m^4}M_\sigma^2
\end{eqnarray}
with
\begin{eqnarray}
M_\sigma^2 &=& m^2 + U''\\\nonumber
           &&  m^2 - \frac{1}{2T}\int\frac{d^3q}{2\pi^3}\cosh^{-2}\gamma.
\end{eqnarray}
One of the dependent variables, $\mu$ or $n$, has to be eliminated via the equation of state (\ref{43}). For the range of
$T$ considered in this paper the influence of the temperature is negligible and the qualitative result (\ref{47}) is not
affected.


\section{Boson Propagator in MFT}
\label{app:WFR}

In mean field theory the gradient terms for the fields $\sigma$ and $\bar{\phi}$ obtain by evaluating the fermion loop in an
inhomogeneous background field. We write the inverse fermion propagator as $ \mathcal{P}+\mathcal{F}$ with $\mathcal{P}$ the
part independent of $\sigma$ and $\bar{\phi}$. We decompose the fields in a position independent part and a harmonic (single
momentum mode) small local perturbation,
\begin{eqnarray}
\mathcal{F}(X,Y)&=&\mathcal{F}_{h} +\mathcal{F}_{inh} \\\nonumber
&=&  \left(
\begin{array}{cc}
  {-\epsilon_{\alpha\beta}\hpb\bar{\phi}^*(X)} & {\delta_{\alpha\beta}\sigma(X)} \\
  {-\delta_{\alpha\beta}\sigma(X)} & {\epsilon_{\alpha\beta}\hpb\bar{\phi}(X)}
\end{array}
\right)\delta (X-Y),\nonumber
\end{eqnarray}
where $\epsilon_{\alpha\beta}$ is the 2 dimensional totally antisymmetric symbol and
\begin{eqnarray}\label{ansatzWFR}
\bar{\phi}(X)&=&\bar{\phi}+ \delta\bar{\phi} \exp(\mathrm{i}K X),\\\nonumber
\sigma(X)&=& \sigma + \delta\sigma \exp(\mathrm{i}K X)+\delta\sigma^* \exp(-\mathrm{i}K X).
\end{eqnarray}
The one-loop fermionic fluctuations can now be expanded around the homogeneous fields,
\begin{eqnarray}\label{InhomExp}
\Gamma_q^{(MFT)}\hspace{-0.25cm}&=& \hspace{-0.15cm}- \frac{1}{2}\Tr\ln(\mathcal{P}+\mathcal{F}_{h} +\mathcal{F}_{in})\\
\hspace{-0.25cm}&=& \hspace{-0.15cm}- \frac{1}{2}\Tr\ln(\mathcal{P}\hspace{-0.05cm}+\mathcal{F}_{h})
                 -\frac{1}{2}\Tr\ln(1\hspace{-0.12cm}+\mathcal{F}_{in}(\mathcal{P}\hspace{-0.05cm}+\mathcal{F}_{h})^{-1})\nonumber\\
\hspace{-0.25cm}&=&\hspace{-0.15cm}V U_{MFT} 
+     \frac{1}{4}\Tr([\mathcal{F}_{in}(\mathcal{P}+\mathcal{F}_{h})^{-1}]^2) + \mathcal{O}(\mathcal{F}_{in}^4).\nonumber
\end{eqnarray}
The terms with odd powers in $\mathcal{F}_{inh}$ vanish due to translation invariance.

For a practical computation, we switch to Fourier space where
\begin{eqnarray}
\mathcal{P}(Q_1,Q_2)=
\left(
\begin{array}{cc}
  {\hspace{-0.15cm}0} & {\hspace{-0.3cm}-P_F(-Q_1)} \\
  {\hspace{-0.15cm}P_F(Q_1)} & {\hspace{-0.3cm}0}
\end{array}
\right) \delta_{\alpha\beta}\delta(Q_1 - Q_2),\nonumber\\
\end{eqnarray}
and
\begin{eqnarray}
&&(\mathcal{P} +  \mathcal{F}_h)^{-1}(Q_1,Q_2)=\mathcal{R}(Q_1)\delta (Q_1- Q_2),\\\nonumber
&&\mathcal{R}(Q_1)=\left(
\begin{array}{cc}
  {\epsilon_{\alpha\beta}\hpb\bar{\phi}} & {\delta_{\alpha\beta}(P_F(-Q_1)-\sigma)} \\
  {-\delta_{\alpha\beta}(P_F(Q_1)-\sigma)} & {-\epsilon_{\alpha\beta}\hpb\bar{\phi}^*}
\end{array}
\right) \\ \nonumber
&&\qquad \times \frac{1}{[P_F(Q_1)-\sigma][P_F(-Q_1)-\sigma] + \hpb^2\bar{\phi}^*\bar{\phi}}.
\end{eqnarray}
We concentrate first on the gradient term for the molecules ($\Apb$) and set $\delta\sigma =0$. Then the inhomogeneous part
reads in momentum space
\begin{eqnarray}
\mathcal{F}_{in}(Q_1,Q_2) =
\hpb\epsilon_{\alpha\beta}\left(
\begin{array}{cc}
  {\hspace{-0.25cm}-\delta\bar{\phi}^*\delta_{Q_1,Q_2 + K} } & {\hspace{-0.25cm}0} \\
  {\hspace{-0.25cm}0} & {\hspace{-0.25cm}\delta\bar{\phi}\delta_{Q_1,Q_2 - K}}
\end{array}
\right)\nonumber\\
\end{eqnarray}
($\delta_{R,S}=\delta(R-S)$). This yields
\begin{eqnarray}
&&\frac{1}{4}\Tr([\mathcal{F}_{in}(\mathcal{P}+\mathcal{F}_{h})^{-1}]^2)\\\nonumber
&=&\frac{1}{4}\mathrm{tr}\int\limits_{Q_1,Q_2}\mathcal{F}_{in}(Q_1,Q_2)\mathcal{R}(Q_2)
\mathcal{F}_{in}(Q_2,Q_1)\mathcal{R}(Q_1)
\end{eqnarray}
where $\mathrm{tr}$ is the trace over the ``internal'' $4\times 4$ matrix. One obtains
\begin{eqnarray}\label{BosPropGen}
&&\Gamma_q^{MFT}=\\\nonumber
&&-\frac{V_3}{T} \hpb^2\delta\bar{\phi}^*\delta\bar{\phi} \int\limits_{Q'} \frac{P_F(-Q')-\sigma}{[P_F(Q')-\sigma]
[P_F(-Q')-\sigma] + \hpb^2\bar{\phi}^*\bar{\phi}}\\\nonumber
&&\times\frac{P_F(Q' - K)-\sigma}{[P_F(K- Q' )-\sigma][P_F(Q' - K)-\sigma] + \hpb^2\bar{\phi}^*\bar{\phi}},
\end{eqnarray}
Insertion of the ansatz (\ref{ansatzWFR}) into the effective action \footnote{Note the slight abuse of conventions. Here,
$\bar{\mathcal{P}}$ stands for the 11 - entry of the inverse propagator matrix in the $\phi^*,\phi$ basis, instead of the
full matrix.}
\begin{eqnarray}\label{SlightAbuse}
\Gamma = \frac{V_3}{T} \delta\bar{\phi}^*\delta\bar{\phi} \bar{\mathcal{P}}_\phi(K)
\end{eqnarray}
yields eq. (\ref{Zphi}). We note the simpler form for $\bar{\phi}=0$ (symmetric phase) where
\begin{eqnarray}\label{ZphiSYM}
\bar{\mathcal{P}}_\phi(K)&=& 2\pi \mathrm{i} n T +\frac{q^2}{4M} \\\nonumber
   &&\hspace{-1.2cm}-\int\limits_{Q} \frac{\hpb^2}{[P_F(Q)-\sigma][P_F(K-Q)-\sigma]}.
\end{eqnarray}
The quantum corrections differ from $P_\sigma$ (eq. (\ref{40})) by the overall factor $-\hpb^2/2$ and the different momentum
structure in the denominator.

We note that in the superfluid phase the symmetries would also be consistent with a gradient term of the form
$\int_Q \bar{\phi}^*\delta\bar{\phi}(Q)\bar{\phi}^*\delta\bar{\phi}(-Q)$. The coefficient of this term cannot be computed with the ansatz
(\ref{ansatzWFR}) - it would require a more general ansatz $\bar{\phi}(X) = \bar{\phi} + \delta\bar{\phi}_+\exp(\mathrm{i}K X) +
\delta\bar{\phi}_-\exp(-\mathrm{i}K X)$. We will omit this term in the present paper.

\subsection{Momentum dependence}
For spacelike momenta the loop integral depends on the square of the external momentum only and we define $\Apb(q)$ and
 $\Apb$ by eqs. (\ref{Zphi1},\ref{Zphi2}). The fluctuation correction to $P_\phi$ at $n=0$ is given by
\begin{eqnarray}\label{DeltaPfull}
\Delta \bar{P}_\phi &=& -\hpb^2 T\sum\limits_m \int \frac{d^3q'}{(2\pi)^3}\frac{f(q')}{[((2m+1)\pi T)^2 + f^2(q') + r]}\nonumber\\
&&\,\, \times \frac{f(q' -q)}{[((2m+1)\pi T)^2 + f^2(q' -q) + r]}\\\nonumber
&=& -\frac{\hpb^2}{2}\hspace{-0.1cm}\int\hspace{-0.1cm}
 \frac{d^3q'}{(2\pi)^3}\frac{f(q')f(q' -q)}{f(q')-f(q'-q)}\\\nonumber
&& \,\,\times\Big\{\frac{1}{\sqrt{b}}\tanh\frac{\sqrt{b}}{2T} -
\frac{1}{\sqrt{a}}\tanh \frac{\sqrt{a}}{2T}\Big\}
\end{eqnarray}
with $f(q)=q^2/2M - \sigma$, $a=f^2(q')+r$, $b=f^2(q' -q)+r$.
The momentum dependent gradient coefficient reads
\begin{eqnarray}
\Apb (q^2) &=& \frac{1}{4M} + \frac{\Delta \bar{P}_\phi(q^2) -\Delta \bar{P}_\phi(0)}{q^2} \nonumber\\
&=& \Apb^{(cl)} + \Delta \Apb(q^2).
\end{eqnarray}
As argued in sects. \ref{sec:MolFrac} and \ref{EffAtDens}, the physically relevant quantity is $\Apb/\ZpR$. We have plotted
the momentum dependence of $\Apb (q)/\ZpR$ in the broad resonance limit $\hpn^2 \gg 1$ in fig. \ref{Zqtot}. At large momenta,
the gradient coefficient slowly tends to zero. This has no impact on observables, since the thermal distribution
functions are suppressed at lower momenta already (see below, also fig. \ref{nqtot}). It might be an artefact of the
neglected momentum dependence of $\ZpR$.

In the present paper we neglect for the numerical results the momentum dependence of $\Apb$ and approximate $\Apb (q) =
\Apb\equiv\Apb (q=0)$. Let us discuss here the validity of this approximation in the broad resonance regime at the critical
temperature. The impact of the
momentum dependence of $\Apb (q)/\ZpR$ is most clearly seen for the Bose distribution in fig. \ref{nqtot}. There we explicitly
compare the results with a momentum dependent $\Apb (q)/\ZpR$ and an approximation of constant $\Apb/\ZpR$. In the
BEC regime the error is very small with $\Apb (q)/\ZpR$ close to the classical value $1/2$ for all $q$. The main difference from the
classical propagator concerns the renormalization of the mass term $\mpb$. In the crossover regime the approximation
of constant $\Apb/\ZpR$ underestimates the number of molecules with large $q^2$, but the error remains acceptable. In
contrast, the deviation from the result with the classical molecule propagator is already substantial. In the BCS regime
the underestimate of $N_M(q)$ for the dominant range in $q$ is quite substantial. Though the overall effect of the boson
fluctuations is small in the BEC regime, this may affect the quantitative results for the number density of molecules and
the condensate. In view of fig. \ref{nqtot} (c) the estimates of $\OmM$ and $\OmC$ and $\Omega_M,\Omega_C$ in the present
paper are most likely too small.

We next discuss $\Apb = \Apb (q=0)$ more explicitly.
With $\Apb = 1/4M + \Delta \Apb$ we can compute $\Delta \Apb$ as the term linear in $q^2$ in the Taylor expansion
of $\Delta \bar{P}_\phi$ (\ref{DeltaPfull}). Using $\Apn= 2M \Apb$ we find the result
\begin{eqnarray}\label{Aphi1}
\Apn &=&\frac{1}{2} + \frac{\hpn^2}{288\Tn^3}\int
\frac{d^3\qn}{(2\pi)^3}\,\,\qn^2\gamma_\phi^{-7}\big[3(5\gamma^4 - 5\gamma^2\gamma_\phi^2+2\gamma_\phi^4)\nonumber\\
&&\times[\tanh\gamma_\phi -\gamma_\phi\cosh^{-2}\gamma_\phi]+ 2\gamma^2\gamma_\phi(\gamma^2 -\gamma_\phi^2)\nonumber\\
&& \times[\gamma_\phi \cosh^{-4}\gamma_\phi - 6\tanh\gamma_\phi -2\gamma_\phi\tanh^2\gamma_\phi\big],
\end{eqnarray}
simplifying in the symmetric phase $\gamma_\phi=\gamma$ to
\begin{eqnarray}\label{AphiSYM}
\Apn &=&\frac{1}{2} + \frac{\hpn^2}{48\Tn^3}\int
\frac{d^3\qn}{(2\pi)^3}\,\,\qn^2\gamma^{-3}\big[\tanh\gamma -\gamma\cosh^{-2}\gamma\big].\nonumber\\
\end{eqnarray}
The loop correction to the gradient coefficient is strictly positive and monotonically growing for
$\tilde{\sigma}$ increasing from negative values (BEC side) to its saturation value $\tilde{\sigma}=1$ on the BCS side. For
the BEC regime it vanishes in the limit $\tilde{\sigma}\to -\infty$. However, the physical quantity $\ApR$
=$\Apn/\ZpR$ approaches the finite value $1/2$ in the BEC regime - this is an indicator of the emergence of an
effective bosonic theory. For $\tilde{\sigma}\approx 1$ instead, $\Apn/\ZpR$ is much larger than the
value for elementary pointlike bosons, $1/2$. Indeed the integral is dominated by modes peaked around $\tilde{\sigma}$,
explaining the strong increase as $\tilde{\sigma} \to 1$.

In the limiting BEC and BCS cases, $\Apn$ can be approximated by much simpler formulae. The BEC result is given in
app. \ref{AnalBEC}, eq. (\ref{ZphiBEC}). In the BCS regime, the critical temperature is very small ($\Tn\lesssim 0.1$) and
$0.4\lesssim\sigex\lesssim 1$. We then find an approximate behavior for the $\rex$-dependent
gradient coefficient
\begin{eqnarray}\label{Zphi02}
\Apn &=& \frac{1}{2}+\frac{7\zeta(3)}{12\pi^4}\frac{\tilde{h}_\phi^2\sigex^{3/2}}{4\Tn^2 + \rex}.
\end{eqnarray}
In the temperature range of interest the classical contribution to $\Apb$ is small ($\Apn^{(cl)}=1/2$). Neglecting
it and restricting to the symmetric phase ($\bar{\phi} =0$), this is consistent with the symmetric BCS result in
\cite{CMelo93}, and we see how the condensate regulates the divergence for $T\to 0$.

\subsection{Frequency dependence}
Similarly the loop correction can be evaluated as a function of external Matsubara frequency $\omega_m$
($Q = (\omega_m,\vec{q}\,)$) for vanishing external spacelike momentum $q$. This amounts to the renormalization
correction to the operator $\phi^*\partial_\tau\phi$. In momentum space this corresponds to the part
$\sim \mathrm{i}\omega_n$ in the inverse molecule propagator. Hence we only need to consider the imaginary part of the
loop integral. The denominator in the integrand in eq. (\ref{BosPropGen}) is real and the imaginary part of the numerator
becomes $2\pi \mathrm{i} n T f(q')$. This yields ($\omega = 2\pi n T$)\footnote{As in eq. (\ref{SlightAbuse}),
$\bar{\mathcal{P}}$ stands for the 11 - entry of the inverse propagator matrix in the $\phi^*,\phi$ basis}
\begin{eqnarray}
\mathrm{Im} \Delta \bar{\mathcal{P}}_\phi(\omega, q=0) &=&  \omega \hpb^2 T \sum\limits_m \int\frac{d^3q'}{(2\pi)^3}
f(q') \\\nonumber
&&\hspace{-0.5cm}\times\{((2m+1)\pi T)^2 + f^2(q') +r \}^{-1}\\\nonumber
&&\hspace{-0.5cm}\times\{((2m+1)\pi T + \omega)^2 + f^2(q') +r \}^{-1}\hspace{-0.1cm}.
\end{eqnarray}
Obviously $\mathrm{Im}\Delta P_\phi$ vanishes for $n=0$ ($\omega=0$). For $n\neq 0$ we can perform the Matsubara sum
(we suppress the argument of $f$),
\begin{eqnarray}
\mathrm{Im}\Delta \bar{\mathcal{P}}_\phi = \frac{\omega \hpb^2}{4} \int\frac{d^3q'}{(2\pi)^3}\frac{\tanh (\sqrt{f^2 +r}/2T)}
{(\omega/2)^2 + f^2 + r}.
\end{eqnarray}
We may define the coefficient of the Matsubara frequencies as ($\omega_1 =2\pi T$)
\begin{eqnarray}
Z_{\phi,\tau}(\omega) &=& \frac{\mathrm{Im} \big(\bar{\mathcal{P}}_\phi(\omega,0)\big)}{\omega},\\\nonumber
Z_{\phi,\tau}(\omega_1) &=& \mathrm{Im} \frac{\bar{\mathcal{P}}_\phi(\omega_1,0)}{\omega_1}\\\nonumber
&=&  1 + \frac{\hpb^2}{4} \int\frac{d^3q'}{(2\pi)^3}\frac{\tanh (\sqrt{f^2 +r}/2T)}{(\pi T)^2 + f^2 + r}.
\end{eqnarray}

We can study the bosonic propagator as a function of $m$. The loop corrected imaginary part of
the inverse boson propagator can be brought into the form
\begin{eqnarray}
\mathrm{i}\omega_m(1 + c(q^2, T,\sigma,m^2)), \quad \omega_m =2\pi m.
\end{eqnarray}
Each Matsubara mode is renormalized by an $m$-dependent quantity. In the present paper we neglect these corrections, i.e.
we take $c(q^2,T, \sigma, m^2)=0$.

\section{Analytical results in the BEC regime}
\label{AnalBEC}
We exploit the fact that in the BEC regime $\sigex/2\Tn \to -\infty$, which means that we can replace the functions
$\tanh\gamma \to 1, \gamma \cosh^{-2}\gamma\to 0$. In the superfluid phase, we use additionally $\rex/|\sigex| \to 0$.
The loop integrals can then be evaluated analytically. They are temperature independent. Furthermore, their values
coincide in the symmetric and superfluid phase, if terms of $\mathcal{O}(\rex/\sigex)$ or higher order in $\sigex^{-1}$
are ignored. We find
\begin{eqnarray}
\Delta\tilde{m}_\phi^{(F)\,2} &=& \frac{\hpn^2\sqrt{-\sigex}}{8\pi},\label{MassBCS}\\
\tilde{\lambda}_\phi^{(F)} &=& \frac{\hpn^4}{128\pi\sqrt{-\sigex}^3},\\
\Delta \Apn &=& \frac{\hpn^2}{64\pi\sqrt{-\sigex}},\label{ZphiBEC}\\
\Delta \ZpR &=& \frac{\hpn^2}{32\pi\sqrt{-\sigex}}\label{ZRBEC}.
\end{eqnarray}
For the fermionic particle density contribution we find a term $\mathcal{O}(\rex/\sqrt{-\sigex})$,
\begin{eqnarray}
n_{F,0} &=& k_F^3\frac{\rex}{16\pi\sqrt{-\sigex}}.
\end{eqnarray}
The BCS gap equation is solved by using (\ref{MassBCS}),
\begin{eqnarray}
c^{-1} =\sqrt{-\sigex},
\end{eqnarray}
independently of the value of $\rex$.
Hence in the BEC limit, the relation between binding energy and scattering length \cite{Diehl:2005an} is independent of the
density scale $k_F$. Indeed, many body effects should be unimportant in this regime. Approaching the resonance, the
impact of the pairing gap $\rex$ becomes important and $\sigex \neq \epsilon_M /\epsilon_F$.

In the limit of large Yukawa couplings, we can then evaluate the gradient coefficient of the effective Bose distribution
(cf. eqs. (\ref{SymmDens},\ref{SuperFlDens})):
\begin{eqnarray}
\ApR=\frac{\Apn}{\ZpR} = \frac{1}{2}.
\end{eqnarray}
Similarly, the fermionic contribution to the four-boson coupling evaluates to
\begin{eqnarray}
\lpR^{(F)}  =\frac{\tilde{\lambda}_\phi^{(F)}}{\ZpR^2} =\frac{8\pi}{\sqrt{-\sigex}}.
\end{eqnarray}
Using the relation $\lambda_p = 4\pi a_p/M_p$ between coupling strength
and scattering length, the molecular scattering length in the BEC limit is given by
\begin{eqnarray}
a_M  = 2 a_R
\end{eqnarray}
where we have changed back to dimensionful quantities. This reproduces the Born approximation for the scattering of particles of
mass $M_p = 2M$. In approaching the resonance for the fermionic scattering length (crossover regime) $c^{-1} = 0$, the
bosonic scattering length however remains finite. Note that both $\ApR$ and $\lpR^{(F)}$ are effectively independent of
$\hpn$ in the broad resonance limit.

\section{Schwinger-Dyson Equations for the molecule couplings}
\label{app:SDE}

In this appendix we provide details of our computation of the effective molecule-molecule interaction $\lambda_\phi$.
We work in dimensionless renormalized units. In the symmetric phase we expand the effective potential $u$ and the
mean field effective potential $\tilde{u}_{MFT}$ in powers of $\rho$,
\begin{eqnarray}
\tilde{u} &=& m_\phi^2\rho + \frac{1}{2}\lambda_\phi\rho^2 
+ ...\\\nonumber
\tilde{u}_{MFT} &=& m_\phi^{(F)\,2}\rho + \frac{1}{2}\lambda_\phi^{(F)}\rho^2 
+ ...
\end{eqnarray}
The contribution from the boson loop reads
\begin{eqnarray}\label{U1B}
\tilde{u}_1^{(B)} &=& m_\phi^{(B)\,2}\rho + \frac{1}{2}\lambda_\phi^{(B)}\rho^2 + ...\\\nonumber
          &=& \big(m_\phi^2 - m_\phi^{(F)\,2}\big)\rho + \frac{1}{2}\big(\lambda_\phi - \lambda_\phi^{(F)}
              \big)\rho^2 + ...
\end{eqnarray}
(see below).
In the symmetric phase we evaluate $\tilde{u}_1^{(B)}$ in an approximation where we truncate in
quadratic order in $\rho$ (\ref{U1B}).
\begin{eqnarray}\label{MLambda}
m_\phi^2 &=&  m_\phi^{(F)\,2}+ m_\phi^{(B)\,2},\nonumber\\
\lambda_\phi &=& \frac{\partial^2 U}{\partial \rho^2}\Big|_{\rho=0}= \lambda_\phi^{(F)} + \lambda_\phi^{(B)}.
\end{eqnarray}
We determine the coupling $\lambda_\phi$ from the ``Schwinger-Dyson'' equation
\begin{eqnarray}\label{LBU}
\lambda_\phi\hspace{-0.1cm} &=& \hspace{-0.1cm}\lambda_\phi\hF + \frac{\partial^2 U_1\hB}{\partial \rho^2}\Big|_{\rho=0}\\\nonumber
            &=& \lpR\hF - \frac{3\lpR\hF
\lpR}{2\Tn}\int\hspace{-0.15cm}\frac{d^3\qn}{(2\pi)^3}\alpha^{-1}
\big[(\exp 2\alpha - 1\big)^{-1}\nonumber\\
 &&\qquad + 2 \alpha \sinh^{-2}\alpha \big]\nonumber\\\nonumber
 &=& \lambda_\phi\hF + \lambda_\phi \cdot I_\lambda.
\end{eqnarray}
which has the solution
\begin{eqnarray}
\lambda_\phi = \frac{\lambda_\phi\hF}{1- I_\lambda}.
\end{eqnarray}
For $m_\phi^2 \to 0$ the last term in eq. (\ref{LBU}) becomes infrared divergent. Divergences of this type of quantum
corrections to quartic couplings are familiar from quantum field theory and statistical physics of critical phenomena.
Indeed, the point $m_\phi^2 = 0$ corresponds to the critical line (or hypersurface) for the phase transition to superfluidity
- for negative $m_\phi^2$ the symmetric phase becomes unstable. The remedy to this infrared problem has been well understood
from the solution of functional renormalization group equations: the strong fluctuation effects drive $\lambda_\phi$ to
zero at the critical line \cite{ATetradis93,BTetWet94,CTetradis92}.

Our gap equations recover this important feature in a direct way. As $m_\phi^2$ approaches zero the negative last term in eq.
(\ref{LBU}) becomes more and more important as compared to $\lambda_\phi$ on the left hand side. The solution to eq.
(\ref{LBU}) implies
\begin{eqnarray}\label{LambdaLimit}
\lim\limits_{m_\phi \to 0} \lambda_\phi(m_\phi) \to 0.
\end{eqnarray}
For small values of $m_\phi^2$ in the vicinity of the phase transition we can expand the integral in eq. (\ref{LBU}) as
\begin{eqnarray}
I_\lambda = -15\Tn \lambda_\phi\hF \int \frac{d^3\qn}{(2\pi)^3}\big(\ApR \qn^2 +m_\phi^2\big)^{-2}.
\end{eqnarray}
One infers $\lambda_\phi\propto m_\phi$ according to
\begin{equation}\label{FixedPoint}
\lambda_\phi  = \frac{8\pi}{15 \Tn}\ApR^{3/2} m_\phi.
\end{equation}

In the superfluid phase, we expand the effective potential around $\rho_0$
\begin{eqnarray}\label{TruncPotSSB}
\tilde{u} = m_\phi^2(\rho - \rho_0) + \frac{\lambda_\phi}{2}(\rho-\rho_0)^2 .
\end{eqnarray}
We choose again a basis of real renormalized fields $\phi_1, \phi_2$ according to
\begin{eqnarray}
\phi = \frac{1}{\sqrt{2}}\big(\phi_1 + \mathrm{i}\phi_2\big) , \quad \rho = \frac{1}{2}\big( \phi^2_1 + \phi_2^2\big).
\end{eqnarray}
Without loss of generality we may consider a background of real $\phi$, i.e. $\phi_{1,0}^2 =2\rho_0, \phi_{2,0}=0$.
With $\phi_1' = \phi_1 - \phi_{1,0}$ the potential (\ref{TruncPotSSB}) becomes
\begin{eqnarray}\label{TruncPotSSB2}
&\tilde{u}&= \frac{1}{2}m_\phi^2\phi^2_2  +\frac{1}{2}(m_\phi^2 + 2\lambda_\phi \rho_0)\phi_1'^2 +
\lambda_\phi\sqrt{\rho_0/2}\phi_1'\phi_2^2 \nonumber\\
&+&\frac{\lambda_\phi}{8}\phi^4_2 + \frac{\lambda_\phi}{4} \phi_2^2\phi_1'^2 +
\frac{\lambda_\phi}{8} \phi_1'^4+ ...\nonumber\\
\end{eqnarray}
We can associate $m_\phi^2$ and $\lambda_\phi$ with the terms quadratic and quartic in $\phi_2$. The dots denote cubic and
quintic terms $\propto \phi_1'^3, \phi_2^4\phi_1', \phi_2^2\phi_1'^3$ that will not contribute in our approximation and
we neglect terms of $\mathcal{O}(\phi^6)$. We use the Schwinger-Dyson equations for the $\phi_2^2$
and $\phi_2^4$ vertices which result in eq. (\ref{MBUB}) and
\begin{eqnarray}\label{LBUB}
\lambda_\phi \hspace{-0.2cm}&=&\hspace{-0.1cm}\lpR\hF \hspace{-0.1cm}- \frac{3\lpR\hF
\lpR}{2\Tn}\hspace{-0.2cm}\int\hspace{-0.2cm}\frac{d^3\qn}{(2\pi)^3}\alpha_\phi^{-3}\big[ \big(\alpha -\kappa\big)^2
\big(\exp 2\alpha_\phi - 1\big)^{-1}  \nonumber\\
&&+ 2\big(\alpha + \kappa/2\big)^2 \alpha_\phi \sinh^{-2}\alpha_\phi\big].
\end{eqnarray}
Again, we observe in eq. (\ref{LBUB}) the appearance of infrared divergences in the contribution $\propto \lambda_\phi\hF
\lambda_\phi$
from the Goldstone fluctuations. If we would define $\lambda_\phi(\rho)$ by the $\phi_2^4$ - vertex evaluated at some value
$\rho > \rho_0$ they would be regulated. Taking the limit $\rho\to \rho_0$ we obtain similar to eq. (\ref{LambdaLimit})
$\lambda_\phi (\rho\to \rho_0)\to 0$. The Goldstone boson fluctuations renormalize $\lambda_\phi$ to zero, as found in
\cite{ATetradis93,BTetWet94,CTetradis92}. In our approximation $\lambda_\phi$ vanishes for all $T < T_c$ in the superfluid
phase. In consequence, the mass term $2\lambda_\phi \rho_0$ of the radial mode vanishes for all $T <T_c$. However, as we have
discussed in the main text, this vanishing of $\lambda_\phi$ concerns only the effective vertex at zero external momentum,
whereas the loop integrals often involve vertices at nonzero momentum.

Furthermore, the contribution from the fluctuations of the radial mode in eqs. (\ref{LBU}) and (\ref{LBUB}) are not treated
very accurately. First, the $\phi_1'^2\phi_2^2$ vertex contains in principle a contribution $\propto \nu_\phi\rho_0$
($\nu_\phi$ the coefficient of the contribution $\propto (\rho - \rho_0)^3$) which shifts $\lambda_\phi\hF \to
\lambda_\phi\hF + 2\nu_\phi\hF \rho_0$ and is neglected here.
Second, the structure of the inverse propagator of the radial mode is actually not simply $\Apb q^2$ with constant
$\Apb$. Indeed, the effective quartic coupling $\lambda_\phi$ only vanishes for zero external momentum, with a typical
momentum dependence $\lambda_\phi\propto |q|$. For a definition of a mass term at $q=0$ this is consistent with
$\lambda_\phi\rho_0\to 0$. However, this effect will then become visible as an infrared divergence of the gradient coefficient for the radial mode, $\bar{A}_{\phi, r} \propto \rho_0 |q|^{-1}$ (which differs from $\Apb$ for the Goldstone
mode). In consequence, one has $\bar{P}_\phi(q\to 0) \propto |q|$ and the ``radial mode contribution'' to the
Schwinger-Dyson equation is not infrared divergent. We note, however, that the radial mode contribution is subleading
as compared to the Goldstone mode contribution such that our approximation catches the dominant
features for the behavior of $\lambda_\phi$.

Finally, in the superfluid phase the potential (\ref{TruncPotSSB2}) also contains a cubic term $\propto \phi_1'\phi_2^2
\propto \lambda \rho_0^{1/2}$. This term contributes to the Schwinger-Dyson equation for $m_\phi^{(B)\, 2}$.
The coefficient of this contribution $\propto \lambda_\phi\hF \lambda_\phi\rho_0$ vanishes for $\lambda_\phi=0$.
Nevertheless, for a momentum dependent $\lambda_\phi$ one has to take it into account, as well as similar corrections
to the Schwinger-Dyson equation for $\lambda_\phi$ which involve quintic couplings.

\section{Numerical Procedures}
\label{app:numerics}

In this section we give a short summary which equations are actually used for numerical solutions. All quantities are given
in dimensionless renormalized units. We first give a complete list relating dimensionful, dimensionless and dimensionless
renormalized parameters, couplings and fields.\\
(i) Relations between dimensionful and dimensionless parameters (the Fermi energy is given by $\epsilon_F = k_F^2/2M$)
\begin{eqnarray}
\qn &=& q/k, \quad \Tn = T/\epsilon_F,  \quad \tilde{\nu}=\bar{\nu}/\epsilon_F,\quad c = \bar{a} k_F\nonumber,\\\nonumber
\sigex &=& \sigma/\epsilon_F,\quad  \hpn = 2M\hpb/k_F^{1/2}, \\\nonumber
\Apn &=& 2M \Apb, \quad \tilde{m}_\phi^2 = \bar{m}_\phi^2/\epsilon_F, \quad
\tilde{\lambda}_\phi = 2M k_F \bar{\lambda}_\phi\\\nonumber
\tilde{\psi} &=&k_F^{-3/2}\psi,\quad \tilde{\phi} = k_F^{-3/2}\bar{\phi}, \\
\tilde{\rho} &=&\tilde{\phi}^*\tilde{\phi}= k_F^{-3}\bar{\rho}, \quad \rex= \hpn^2\tilde{\rho} = r/\epsilon_F^2.
\end{eqnarray}
(ii) Relations between dimensionless and dimensionless renormalized parameters
\begin{eqnarray}
\ApR &=& \Apn/\ZpR , \quad m_\phi^2 = \tilde{m}_\phi^2/\ZpR,\quad \nu = \tilde{\nu}/\ZpR,\nonumber\\
\lambda_\phi &=& \tilde{\lambda}_\phi/\ZpR^2,\quad h_\phi = \hpn /\ZpR^{1/2}, \quad \rho = \ZpR  \tilde{\rho}.
\end{eqnarray}
All other parameters, couplings and fields are invariant under a rescaling with $\ZpR$. In particular, note the invariance of
the concentration $c$ and the dimensionless superfluid gap parameter $\rex$. In our approximation, we neglect details of
the renormalization coefficients for the Matsubara frequencies and use
\begin{eqnarray}
Z_{\phi,\tau}/\ZpR = 1.
\end{eqnarray}

The input parameters are the detuning $\nun$, the temperature $\Tn$ and the Yukawa coupling $\hpn$, with $c$ following from
eq. (\ref{NuMuC}). Alternatively, we could choose $c^{-1},\hpn,\Tn$. We have to solve the field equations for the effective
chemical potential $\sigex$ (density equation) and the field expectation value $\phi$. The latter has the general form
\begin{eqnarray}
\frac{\partial u}{\partial\rho}(\rho_0) \cdot \phi_0 = 0
\end{eqnarray}
In the superfluid phase it determines the expectation value $\phi_0$.
In the normal or symmetric phase the field expectation value vanishes, $\phi_0=0$. The field equation for $\phi$
is now replaced by the gap equation for the boson mass term $m_\phi^2= \partial u(0)/\partial\rho$. The two equations
provide a closed system, whose solution for $\sigex$ and $\rho_0$ or $m_\phi^2$ determines all observables of the
crossover problem.

In the superfluid or symmetry-broken phase we have $\phi_0 \neq 0$ such that $ \partial u(\rho_0)/\partial\rho =0$ must
hold. The density equation
and the latter condition are then solved for $\sigex$ and $\rex = \hpn^2\tilde{\phi}^*\tilde{\phi}$. However, the full
four-boson coupling $\lambda_\phi$ enters both the density equation and the gap equation. Therefore, an
additional gap equation for $\lambda_\phi$ is needed, cf. sect. \ref{sec:beyond} and app. \ref{app:SDE}. Hence the solution
is given by $\sigex, \rex, \lambda_\phi$. In the ``zero momentum approximation'' we find $\lambda_\phi=0$.

All other quantities of interest can then be reconstructed from the solution. For example, we can obtain the bare
noncondensed molecule fraction and the condensate fraction by
\begin{eqnarray}\label{barOm}
\bar{\Omega}_M = 6\pi^2 k_F^{-3} \bar{n}_M,\quad \bar{\Omega}_C = 6\pi^2 \frac{\rex}{\hpn^2}
\end{eqnarray}
where $\bar{n}_M$ is given by eqs. (\ref{SymmDens}). The dressed noncondensed molecule fraction and
condensate fraction are the given by
\begin{eqnarray}
\Omega_M = \ZpR \bar{\Omega}_M ,\quad \Omega_C = \ZpR \bar{\Omega}_C.
\end{eqnarray}

Let us now give the explicit formulae used in the different regions of the phase diagram.\\
(i) Normal phase.\\
The full boson mass is determined by the condition
\begin{eqnarray}\label{Cross1SYM}
m_\phi^2 = m_\phi^{(F)\, 2} + \frac{1}{3\pi^2}\lambda_\phi\hF\Omega_M.
\end{eqnarray}
Here $\Omega_M$ depends on the full boson mass term $m_\phi^2$, and $m_\phi^{(F)\, 2},\lambda_\phi\hF$ (eqs.
(\ref{FermSYMMass}, \ref{147})) have to be evaluated at $\rex=0$.

The field equation for $\sigex$ is equivalent to the condition
\begin{eqnarray}\label{DensSYM}
1 = \Omega_F + \Omega_M
\end{eqnarray}
which uses the dressed density fractions. Here
\begin{eqnarray}
\Omega_F &=& 6\pi^2\int \frac{d^3\qn}{(2\pi)^3}\big(\mathrm{e}^{ 2\gamma} + 1\big)^{-1},\label{OmegF}\\
\Omega_M &=& 6\pi^2\int\frac{d^3\qn}{(2\pi)^3} \Big(\mathrm{e}^{(\ApR \qn^2 + m_\phi^2)/\Tn} - 1\Big)^{-1}\nonumber\\
      &=& \frac{3\Gamma(3/2)}{2}\Big(\frac{\Tn}{\ApR}\Big)^{3/2}\mathrm{Li}_{3/2} \big(e^{-m_\phi^2/\Tn}\big)\label{OmegM}.
\end{eqnarray}
(ii) Superfluid phase.\\
Now $\rex$ and $\sigex$ are determined by the equations
\begin{eqnarray}\label{Cross1SSB}
m_\phi^{(F)\, 2} + \frac{1}{3\pi^2}\lambda_\phi\hF \Omega_M &=&0,\\
\Omega_{F,0} + \Omega_M + \OmC &=& 1.
\end{eqnarray}
These equations are still coupled to a gap equation for $\lambda_\phi$ (cf. app. \ref{app:SDE} for details). Here
$\Omega_M$ is given by eq. (\ref{SymmDens})($\Omega_M = 6\pi^2k_F^{-3}n_M$), $\bar{\Omega}_C$ by eq. (\ref{barOm}) and
\begin{eqnarray}
\Omega_{F,0} = -3\pi^2\int \frac{d^3\qn}{(2\pi)^3}\big(\frac{\gamma}{\gamma_\phi}\tanh\gamma_\phi -1\big).
\end{eqnarray}
(iii) Phase transition.\\
At the critical line $T=T_c$ we have $m_\phi^2=\phi_0 =0$ and we solve
\begin{eqnarray}
m_\phi^{(F)\, 2} + \frac{1}{3\pi^2}\lambda_\phi\hF\Omega_M  =0,\\
\Omega_F + \Omega_M = 1
\end{eqnarray}
at $\rex =0$ for $\sigex$ and $T_c$. Here $\Omega_M$ is evaluated at $m_\phi^2 = 0$ and reads
\begin{eqnarray}
\Omega_M = \frac{3\Gamma(3/2)\zeta(3/2)}{2}\Big(\frac{\Tn}{\ApR}\Big)^{3/2}\label{OmegMPT}.
\end{eqnarray}

\end{appendix}

\bibliographystyle{apsrev}
\bibliography{Citations}

\end{document}